\documentclass[preprint,12pt]{elsarticle}
\usepackage{amsmath}
\usepackage[dvipdfmx]{}
\usepackage[dvipdfmx]{color}
\usepackage{url}
\usepackage{graphicx}
\usepackage{bm}
\usepackage{here}
\usepackage{caption}
\usepackage[subrefformat=parens]{subcaption}
\newcounter{bla}

\begin{document}
\renewcommand{\include}[1]{}
\renewcommand\documentclass[2][]{}
\setcounter{tocdepth}{2}
\thispagestyle{empty}

\begin{frontmatter}



\title{Code O-SUKI-N 3D: Upgraded Direct-Drive Fuel Target 3D Implosion Code in Heavy Ion Inertial Fusion}


\author[a]{H. Nakamura}
\author[a]{K. Uchibori}
\author[a]{S. Kawata\corref{author}}
\author[b]{T. Karino}
\author[a]{R. Sato}
\author[c]{A. I. Ogoyski}

\cortext[author] {Corresponding author.\\\textit{E-mail address:} kwt@cc.utsunomiya-u.ac.jp, s.kwta.g@gmail.com}
\address[a]{Graduate School of Engineering, Utsunomiya University, Utsunomiya 321-8585, Japan}
\address[b]{Collaborative Laboratories for Advanced Decommissioning Science, Japan Atomic Energy Agency, Fukushima 970-8026, Japan}
\address[c]{Department of Physics, Varna Technical University, Varna 9010, Bulgaria}

\begin{abstract}

The Code O-SUKI-N 3D is an upgraded version of the 2D Code O-SUKI (Comput. Phys. Commun. 240, 83 (2019)). Code O-SUKI-N 3D is an integrated 3-dimensional (3D) simulation program system for fuel implosion, ignition and burning of a direct-drive nuclear-fusion pellet in heavy ion beam (HIB) inertial confinement fusion (HIF).The Code O-SUKI-N 3D consists of the three programs of Lagrangian fluid implosion program, data conversion program, and Euler fluid implosion, ignition and burning program. The Code O-SUKI-N 3D can also couple with the HIB illumination and energy deposition program of OK3 (Comput. Phys. Commun. 181, 1332 (2010)). The spherical target implosion 3D behavior is computed by the 3D Lagrangian fluid code until the time just before the void closure of the fuel implosion. After that, all the data by the Lagrangian implosion code are converted to the data for the 3D Eulerian code. In the 3D Euler code, the DT fuel compression at the stagnation, ignition and burning are computed. The Code O-SUKI-N 3D simulation system provides a capability to compute and to study the HIF target implosion dynamics. 

\end{abstract}


\end{frontmatter}



\noindent
{\bf Program summary}

\begin{small}
\noindent
{\em Program Title:}  O-SUKI-N 3D                                        \\
{\em Licensing provisions: CC BY NC 3.0 }                                   \\
{\em Programming language:} C++                                 \\
{\em Computer:} Workstation (Xeon, 2 GHz or higher recommended)\\
{\em RAM:} 120GBytes minimum\\
{\em Operating system:} UNIX, Linux (For example: CentOS 6.4, Ubuntu 18.04.1 LTS)\\
{\em Journal reference of previous version: Code O-SUKI: 2D version}                  \\
{\em Nature of problem:}     
Nuclear fusion energy would be energy source for society. In this paper we focus on heavy ion beam (HIB) inertial confinement fusion (HIF). A spherical mm-radius deuterium (D) - tritium (T) fuel pellet is irradiated by HIBs to be compressed to about a thousand times of the solid density. The DT fuel temperature reaches $\sim$5-10KeV for the ignition to release the fusion energy. The typical HIBs total input energy is several MJ, and the HIBs pulse length is about a few tens of ns. The O-SUKI-N 3D code system provides an integrated tool to simulate the HIF DT fuel pellet implosion, ignition and burning in 3 dimensions (3D). The O-SUKI-N 3D code system is an upgraded version of the Code O-SUKI (Comput. Phys. Commun. 240, 83 (2019)) which is a 2D implosion simulation system in HIF. The DT fuel is compressed to the high density, and so the DT fuel spatial deformation may be serious at the DT fuel stagnation. Therefore, the O-SUKI and O-SUKI-N 3D systems employ a Lagrangian fluid code first to simulate the DT fuel implosion phase until just before the stagnation. Then all the simulation data from the Lagrangian code are converted to them for the Euler fluid code, in which the DT fuel ignition and burning are simulated. \\
{\em Solution method:}     
In the two fluid codes (Lagrangian and Euler fluid codes) in the O-SUKI-N 3D system the three-temperature fluid model (J. Appl. Phys. 60, 898 (1986)) is employed to simulate the pellet dynamics in HIF. 
\\
{\em Additional comments including Restrictions and Unusual features:} The Lagrange code is weak against the spatial mesh deformation from nature of its numerical algorithm. When short-wavelength perturbations are imposed near the poles of the spherical target, the spatial meshes might crash and the computation run may stop. \\
   \\

\end{small}

\section{Introduction}\label{sec:1}

Code O-SUKI-N 3D (3 dimension) is an upgraded 3D version of our Code O-SUKI \cite{CPC-O-SUKI}, and it provides a capability to simulate a deuterium (D) - tritium (T) fuel target implosion, ignition and burning in 3D in heavy ion beam (HIB) inertial confinement fusion (ICF). 

In ICF, DT fuel target implosion, ignition and burning are essentially important to release a sufficient fusion energy output. In ICF a few mg DT in a fuel pellet is compressed to about a thousand times the solid density by an input driver energy, for example, lasers or heavy ion beams (HIBs) or pulse power. In addition, the ion temperature of the compressed DT must reach $\sim$5-10 KeV \cite{ICFBook}. In order to compress the DT fuel stably to the high density, the implosion non-uniformity should be less than a few percent\cite {kwtANDniu} The key issues of the fuel implosion in ICF include how to realize the uniform implosion. The O-SUKI-N 3D code system provides an integrated computer simulation tool to study the DT fuel implosion, ignition and burning in heavy ion inertial confinement fusion (HIF) \cite{Kawata1, Kawata2}. 
         
The DT fuel implosion is simulated until just before the void closure time by the Lagrangian code, which can couple with the OK3 code to include the time-dependent HIBs energy deposition profile in the target energy absorber layer. For example, the detail HIBs illumination on a HIF DT target can be computed by a computer code of OK3 \cite{ogoyski1, ogoyski2, ogoyski3}. The Lagrange code data are converted to the data imported to the Euler code. The Euler code is robust against the target fuel deformation. The DT fuel ignition and burning are simulated further by the 3D Euler fluid code. The O-SUKI-N 3D code system simulates the 3D HIF target implosion dynamics, and would contribute to release the fusion energy stably for society.

\section{O-SUKI-N 3D code algorithm description}
\par

\subsection{O-SUKI-N 3D code structure}
     The O-SUKI-N 3D code system consists of three parts: The Lagrangian fluid code \cite{Schulz}, the data conversion code from the Lagrangian code to the Euler code, and Euler code. The fluid model is the three-temperature model in Ref. \cite{Tahir}. The Lagrangian fluid code, the data conversion code and the Euler code are described below in detail. 
     
     In the Lagrangian fluid code the spatial meshes move together with the fluid motion \cite{Schulz}. However, the Lagrange meshes can not follow the fluid large deformation. On the other hand, the Euler meshes are fixed to the space, and the fluid moves through the meshes. Therefore, just before the void closure time, that is, the stagnation phase, the Lagrangian code is used to simulate the DT fuel implosion. After the void closure time, the Euler code is employed to simulate the DT fuel further compression, ignition and burning. Between the Lagrangian code and the Euler code the data should be converted by the data conversion code. 

	All the simulation process is performed in its integrated way by using the script of "CodeO-SUKI-N-fusion-start.sh". The processes executed by this shell script are as follows: \\
1. Make the stack size infinite.\\
2. Remove all output data file and make the new output files.\\
3. Change the permission of shell scripts to executable. \\
4. Compile the main function of the Lagrangian code and execute it.\\
5. If any problems do not appear during the calculation of the Lagrangian code, compile the main function of the data conversion code and execute it.\\
6. If there is no problem during the data conversion, compile the main function of the Euler code and execute it.\\

\subsection{Steps in Lagrangian code}\par
     The Lagrangian code has the following steps: 

\begin{enumerate}
\item Initialize the variables and calculation of total input energy. \par
\item Calculation of time step size.\par
\item Calculation of coordinates.\par
\item Solve equation of motion. \par
\item Solve density by equation of continuity.\par
\item Calculation of artificial viscosity.\par
\item Transfer the data to the OK3. \par
\item Calculation of energy deposition distribution in code OK3. For details of the OK3, see the refs.\cite{ogoyski1,ogoyski2,ogoyski3}. \par
\item Solve energy equations\par
\item Calculation of heat conduction\par
\item Calculation of temperature relaxation among three temperatures.\par
\item Solve equation of state\par
\item Save the results.\par
\item End the Lagrangian calculation right before the void closure.\par
\item Transfer the data to converting code. \par
\end {enumerate}

\subsection{Data Conversion code from Lagrangian fluid code to Euler fluid code}

\begin {enumerate}
\item Read variables saved in Lagrangian code.\par
\item Generate the Eulerian mesh.\par
\item Calculate the interpolation of the physical quantity to them on the Eulerian mesh.\par
\item Write the converted data to the Eulerian code.\par
\end {enumerate}

\subsection{Steps in Eulerian code}

\begin {enumerate}
\item Read the mesh number from the converted data and define the each matrices.\par
\item Initialize the variables.\par
\item Calculation of time step size.\par
\item Solve equation of motion. \par
\item Track the material boundaries of DT, Al and Pb.\par
\item Linearly interpolate the boundary lines and transcribe them on the Eulerian code. \par
\item Discriminate the materials by using the transferred boundary line. \par
\item Solve density by equation of continuity.\par
\item Calculate artificial viscosity.\par
\item Solve energy equations\par
\item Calculation of fusion reaction.\par
\item Calculation of heat conduction\par
\item Calculation of temperature relaxation among three temperatures.\par
\item Solve equation of state.\par
\item Save the results.\par
\item End.
\end{enumerate}

\section{Files included}

The logical coordinates in the Lagrangian code are identified by the mesh number of $( i, j, k )$. One Lagrange mesh is shown in Fig. \ref{Lmesh}. The discretization method in Ref. \cite{Schulz} is employed in the Lagrangian fluid code. 
We use the spatial coordinate of $ {\bm R} = (x(i, j, k), y(i, j, k), z(i, j, k))$. The vector $ {\bar {\bm R}} $ is nomal to $ {\bm R}$. 
\begin{figure}[H]
	\centering
	\includegraphics[height=8cm]{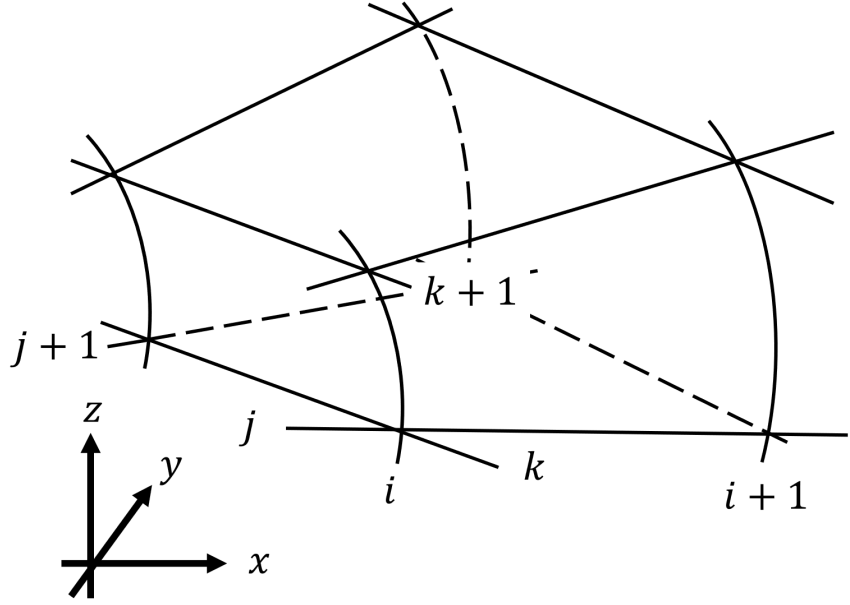}
	\caption{Lagrangian Mesh.}\label{Lmesh}
\end{figure}

The definition points of the discretized physical quantities in the Lagrange and Euler codes are presented in Figs. \ref {definition_La} and \ref{definition_Eu}, respectively. The subscripts $i$, $j$ and $k$ correspond to the positions in space, and the subscript $n$ corresponds to time $n\times dt$. The displacements in the $i$, $j$ and $l$ directions are defined as follows: 
	\begin{eqnarray*}
		&&\begin{cases}
			dx^n_{i+\frac{1}{2},j,k} \equiv dxi_{i+\frac{1}{2},j,k}=x^n_{i+1,j,k}-x^n_{i,j,k}\\
			dy^n_{i+\frac{1}{2},j,k} \equiv dyi_{i+\frac{1}{2},j,k}=y^n_{i+1,j,k}-y^n_{i,j,k}\\
			dz^n_{i+\frac{1}{2},j,k} \equiv dzi_{i+\frac{1}{2},j,k}=z^n_{i+1,j,k}-z^n_{i,j,k}
		\end{cases}\\
		&&\begin{cases}
			dx^n_{i,j+\frac{1}{2},k} \equiv dxj_{i,j+\frac{1}{2},k}=x^n_{i,j+1,k}-x^n_{i,j,k}\\
			dy^n_{i,j+\frac{1}{2},k} \equiv dyj_{i,j+\frac{1}{2},k}=y^n_{i,j+1,k}-y^n_{i,j,k}\\
			dz^n_{i,j+\frac{1}{2},k} \equiv dzj_{i,j+\frac{1}{2},k}=z^n_{i,j+1,k}-z^n_{i,j,k}
		\end{cases}\\
		&&\begin{cases}
			dx^n_{i,j,k+\frac{1}{2}} \equiv dxk_{i,j,k+\frac{1}{2}}=x^n_{i,j,k+1}-x^n_{i,j,k}\\
			dy^n_{i,j,k+\frac{1}{2}} \equiv dyk_{i,j,k+\frac{1}{2}}=y^n_{i,j,k+1}-y^n_{i,j,k}\\
			dz^n_{i,j,k+\frac{1}{2}} \equiv dzk_{i,j,k+\frac{1}{2}}=z^n_{i,j,k+1}-z^n_{i,j,k}
		\end{cases}\\
	\end{eqnarray*}	

\begin{figure}[H]
	\centering
	\includegraphics[height=8cm]{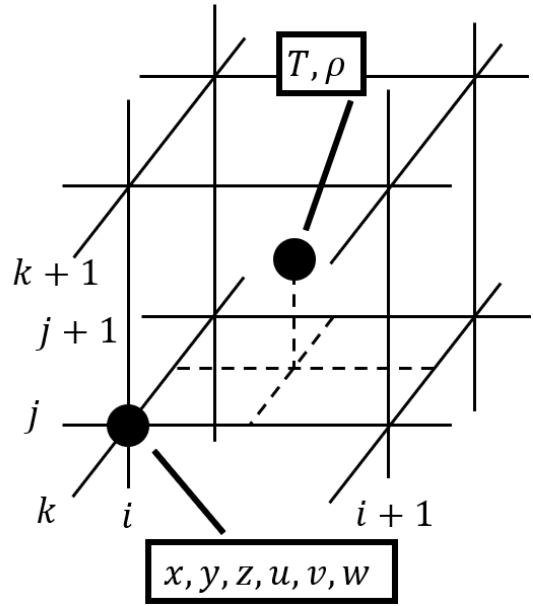}
	\caption{Definition points of discretized physical quantities in the Lagrangian code.}\label{definition_La}
	\end{figure}
	
\begin{figure}[H]
	\centering
	\includegraphics[height=8cm]{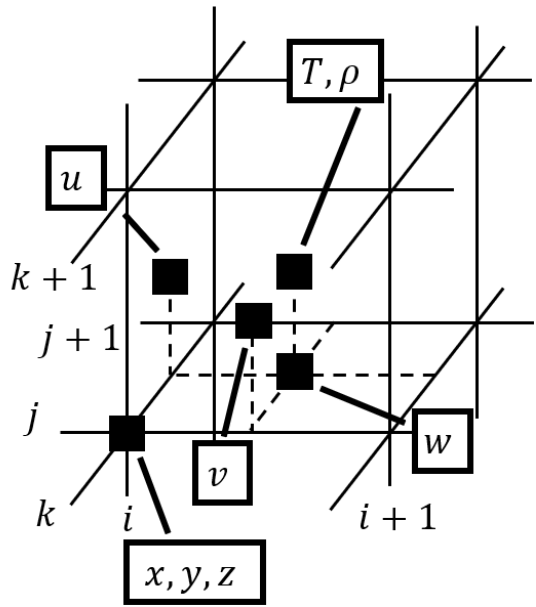}
	\caption{Definition points of discretized physical quantities in the Eulerian code.}\label{definition_Eu}
	\end{figure}

In the Lagrange code, the derivatives in $ x $, $ y $ and $ z $ are as follows:
\begin{equation}
\begin{split}
\frac{\partial }{\partial x}
&=\frac{\partial}{\partial i}\frac{\partial i}{\partial x}+\frac{\partial}{\partial j}\frac{\partial j}{\partial x}+\frac{\partial}{\partial k}\frac{\partial k}{\partial x}\\
&=\frac{1}{J}\left[(\frac{\partial y}{\partial j}\frac{\partial z}{\partial k}-\frac{\partial y}{\partial k}\frac{\partial z}{\partial j})\frac{\partial}{\partial i}+(\frac{\partial y}{\partial k}\frac{\partial z}{\partial i}-\frac{\partial y}{\partial i}\frac{\partial z}{\partial k})\frac{\partial}{\partial j}+(\frac{\partial y}{\partial i}\frac{\partial z}{\partial j}-\frac{\partial y}{\partial j}\frac{\partial z}{\partial i})\frac{\partial}{\partial k}\right]\\
& \equiv \frac{1}{J}\left[Dix\frac{\partial}{\partial i}+Djx\frac{\partial}{\partial j}+Dkx\frac{\partial}{\partial k}\right]
\end{split}
\end{equation}
\begin{equation}
\begin{split}
\frac{\partial }{\partial y}&
=\frac{\partial}{\partial i}\frac{\partial i}{\partial y}+\frac{\partial}{\partial j}\frac{\partial j}{\partial y}+\frac{\partial}{\partial k}\frac{\partial k}{\partial y}\\
&=\frac{1}{J}\left[(\frac{\partial z}{\partial j}\frac{\partial x}{\partial k}-\frac{\partial z}{\partial k}\frac{\partial x}{\partial j})\frac{\partial}{\partial i}+(\frac{\partial z}{\partial k}\frac{\partial x}{\partial i}-\frac{\partial z}{\partial i}\frac{\partial x}{\partial k})\frac{\partial}{\partial j}+(\frac{\partial z}{\partial i}\frac{\partial x}{\partial j}-\frac{\partial z}{\partial j}\frac{\partial x}{\partial i})\frac{\partial}{\partial k}\right]\\
& \equiv \frac{1}{J}\left[Diy\frac{\partial}{\partial i}+Djy\frac{\partial}{\partial j}+Dky\frac{\partial}{\partial k}\right]
\end{split}
\end{equation}
\begin{equation}
\begin{split}
\frac{\partial }{\partial z}&
=\frac{\partial}{\partial i}\frac{\partial i}{\partial z}+\frac{\partial}{\partial j}\frac{\partial j}{\partial z}+\frac{\partial}{\partial k}\frac{\partial k}{\partial z}\\
&=\frac{1}{J}\left[(\frac{\partial x}{\partial j}\frac{\partial y}{\partial k}-\frac{\partial x}{\partial k}\frac{\partial y}{\partial j})\frac{\partial}{\partial i}+(\frac{\partial x}{\partial k}\frac{\partial y}{\partial i}-\frac{\partial x}{\partial i}\frac{\partial y}{\partial k})\frac{\partial}{\partial j}+(\frac{\partial x}{\partial i}\frac{\partial y}{\partial j}-\frac{\partial x}{\partial j}\frac{\partial y}{\partial i})\frac{\partial}{\partial k}\right]\\
& \equiv \frac{1}{J}\left[Diz\frac{\partial}{\partial i}+Djz\frac{\partial}{\partial j}+Dkz\frac{\partial}{\partial k}\right]
\end{split}
\end{equation}
The generalized formula is as follows: 
\begin{equation}\label{ijktrans}
\frac{\partial}{\partial \bm r}=\frac{1}{J}\left[Di\bm r\frac{\partial}{\partial i}+Dj\bm r\frac{\partial}{\partial j}+Dk\bm r\frac{\partial}{\partial k}\right]
\end{equation}
\begin{equation}
	Di\bm r \equiv \left(
	\begin{array}{c}
	Dix\\
	Diy\\
	Diz
	\end{array}
	\right), 
	Dj\bm r \equiv \left(
	\begin{array}{c}
	Djx\\
	Djy\\
	Djz
	\end{array}
	\right),  
	Dk\bm r \equiv \left(
	\begin{array}{c}
	Dkx\\
	Dky\\
	Dkz
	\end{array}
	\right)
\end{equation}



\subsection{Lagrangian code and OK3}
\begin{enumerate}
\item {\bf BC\_LC.cpp}\\
The boundary conditions are included in the procedure.
\item {BeamMaking.cpp}\\
The function calculates the total input energy.
\item {\bf CONSTANT.h}\\
The file contains the definition of constant values and normalization factors.
\item {\bf Derf.c}\\
The file contains the error function in the double precision.
\item {\bf HIFScheme.h}\\
The file contains 1, 2, 3, 6, 12, 20, 32, 60 and 120-beam irradiation schemes. (see also Refs. \cite{ogoyski1, ogoyski2, ogoyski3}.)
\item {\bf IMOK.cpp}\\
The file contains a procedure to transfer the Lagrange mesh date. The file sets the initial target surface numerically.

\item {\bf InitMesh\_LC.cpp}\\
The file initializes the Lagrangian coordinates and determines the number of the target layer. The number of the layers can be selected from 1 to 5 layers. The user must set the mesh number of each layer in this file.
\item {\bf InputOK3.h}\\
The input data file contains the target parameters, the HIB parameters. 
\item {\bf Insulation.cpp}\\
The file contains a procedure to calculate the adiabat $\alpha$ to evaluate the fuel preheating \cite{CPC-O-SUKI, ICFBook}. 
\item {\bf Lagrange\_set.cpp}
This function performs auxiliary calculations for spatial differentiation and thermal conductivity calculations in the Lagrange code. 
\item {\bf Legendre.cpp}\\
The procedure performs the mode analyses based on the spherical harmonics in order to find the implosion non-uniformity. The analysis results are also output in this procedure. 
\item {\bf Lr\_LC.cpp}\\
A procedure to calculate the Rosseland mean free path (see Ref. \cite {Zeldovich}). 
\item {\bf MS.cpp}\\
A function to solve matrix by the Gauss elimination method. 
This function is optimized specifically for Langnge calculations.  
\item {\bf MS\_TDMA.cpp}\\
A function to solve matrix by TDMA (TriDiagonal-Matrix Algorithm).
\item {\bf OK3code.cpp}\\
The file is the main routine of the HIBs illumination code of OK3 and contains the following procedures\cite{ogoyski1,ogoyski2,ogoyski3}. The details for each procedure relating to the HIBs illumination code OK3 are found in Ref. \cite{CPC-O-SUKI, ogoyski1, ogoyski2, ogoyski3}. The relating procedures are listed here: Irradiation(), InitEdp1(), Focus(), fDis(), Divider(), kBunch(), PointC(), PointF(), PointAlpha(), BeamCenterRot(), BeamletRot( ), Rotation() and StoppingPower1.cpp. The procedure of StoppingPower1.cpp contains a function Stop1. This function serves a heart of the OK1 code \cite{ogoyski1} and describes the energy deposition model. It calculates the stopping power from the projectile ions into the solid target. The one-ion stopping power is considered to be a sum of the deposition energy in the target nuclei, the target bound and free electrons and the target ions\cite{mehlhorn}. 

\item {\bf PelletSurface.h}\\
The file sets the initial target surface numerically.
\item {\bf RMS.cpp}\\
The procedure in this file calculates the root-mean-square (RMS) deviation in target non uniformity.
\item {\bf ResultIMP.cpp}\\
This file contains a procedure to calculate the implosion velocity.
\item {\bf SLC.cpp}\\
This file contains the procedure to output the time history of each physical quantity obtained by cutting one each in $\theta$ and $\phi$ directions. The positions of $\theta$ and $\phi$ are changed in "input\_LC.h". 

\item {\bf Acceleration.cpp}\\
A procedure for calculating the target acceleration. 

\item {\bf artv\_LC.cpp}\\
This file contains a procedure calculate the artificial viscosity. When dealing with shock waves propagating in a compressive fluid at a supersonic speed in fluid dynamics simulations, it is impossible to employ sufficient number of multiple meshes to describe the real shock front structure, because its thickness is very thin. As a method, we introduce the following artificial viscosity devised by Von Neumann and Richtmyer\cite{artv}.\\

     The three-dimensional artificial viscosity is written:
 
\begin{equation}
		q_A=\rho c^2_1\left|\displaystyle\frac{\partial }{\partial i}\left(\frac{\partial u}{\partial i}\right)^A_{\_}\right|\left(\frac{\partial u}{\partial i}\right)^A_{\_}\\
		\label{artv_LC1}
\end{equation}
\begin{equation}
		q_B=\rho c^2_1\left|\displaystyle\frac{\partial }{\partial j}\left(\frac{\partial u}{\partial j}\right)^B_{\_}\right|\left(\frac{\partial u}{\partial j}\right)^B_{\_}\\
		\label{artv_LC2}
\end{equation}
\begin{equation}
		q_C=\rho c^2_1\left|\displaystyle\frac{\partial }{\partial k}\left(\frac{\partial u}{\partial k}\right)^C_{\_}\right|\left(\frac{\partial u}{\partial k}\right)^C_{\_}
		\label{artv_LC3}
	\end{equation}
	
	\begin{eqnarray*}
		&&\left(\frac{\partial u}{\partial i}\right)^A_{\_}=\min{\left[\left(\frac{\partial u}{\partial i}\right)^A,0\right]}\nonumber\\
		&&\left(\frac{\partial u}{\partial j}\right)^B_{\_}=\min{\left[\left(\frac{\partial u}{\partial j}\right)^B,0\right]}\nonumber\\
		&&\left(\frac{\partial u}{\partial k}\right)^C_{\_}=\min{\left[\left(\frac{\partial u}{\partial k}\right)^C,0\right]}\nonumber\\
		&&\left(\frac{\partial u}{\partial i}\right)^A=\frac{\bar{\bm R}_i\cdot{\bm u}_i}{\left|\bar{\bm R}_i\right|}\nonumber\\
		&&\left(\frac{\partial u}{\partial j}\right)^B=\frac{\bar{\bm R}_j\cdot{\bm u}_j}{\left|\bar{\bm R}_j\right|}\nonumber\\
		&&\left(\frac{\partial u}{\partial k}\right)^C=\frac{\bar{\bm R}_k\cdot{\bm u}_k}{\left|\bar{\bm R}_k\right|}\nonumber
	\end{eqnarray*}

Here $\bar{\bm R_i}, \bar{\bm R_j}$ and $\bar{\bm R_k}$ are the normal vectors to the $i, j, k$ directions, respectively. $q_A$, $q_B$ and $q_C$ are the artificial viscosities in the directions of $\bar{\bm R_i}$, $\bar{\bm R_j}$ and $\bar{\bm R_k}$, respectively. Equations (\ref{artv_LC1}), (\ref{artv_LC2}) and (\ref{artv_LC3}) are discretized as follows: 
	\begin{eqnarray}
\begin{split}
		&{q_A}^{n}_{i+\frac{1}{2},j+\frac{1}{2},k+\frac{1}{2}}=\\
&\left({q_{A1}}^{n}_{i+\frac{1}{2},j+\frac{1}{2},k+\frac{1}{2}}+{q_{A2}}^{n}_{i+\frac{1}{2},j+\frac{1}{2},k+\frac{1}{2}}+{q_{A3}}^{n}_{i+\frac{1}{2},j+\frac{1}{2},k+\frac{1}{2}}+{q_{A4}}^{n}_{i+\frac{1}{2},j+\frac{1}{2},k+\frac{1}{2}}\right)
\end{split}\\
\begin{split}
		&{q_B}^{n}_{i+\frac{1}{2},j+\frac{1}{2},k+\frac{1}{2}}=\\
&\left({q_{B1}}^{n}_{i+\frac{1}{2},j+\frac{1}{2},k+\frac{1}{2}}+{q_{B2}}^{n}_{i+\frac{1}{2},j+\frac{1}{2},k+\frac{1}{2}}+{q_{B3}}^{n}_{i+\frac{1}{2},j+\frac{1}{2},k+\frac{1}{2}}+{q_{A4}}^{n}_{i+\frac{1}{2},j+\frac{1}{2},k+\frac{1}{2}}\right)
\end{split}\\
\begin{split}
		&{q_C}^{n}_{i+\frac{1}{2},j+\frac{1}{2},k+\frac{1}{2}}=\\
&\left({q_{C1}}^{n}_{i+\frac{1}{2},j+\frac{1}{2},k+\frac{1}{2}}+{q_{C2}}^{n}_{i+\frac{1}{2},j+\frac{1}{2},k+\frac{1}{2}}+{q_{C3}}^{n}_{i+\frac{1}{2},j+\frac{1}{2},k+\frac{1}{2}}+{q_{A4}}^{n}_{i+\frac{1}{2},j+\frac{1}{2},k+\frac{1}{2}}\right)
\end{split}
	\end{eqnarray}
	
	Here, the expression appeared are summarized below: 
	\begin{eqnarray}
		{q_{A1}}^{n}_{i+\frac{1}{2},j+\frac{1}{2},k+\frac{1}{2}}=\rho^{n}_{i+\frac{1}{2},j+\frac{1}{2},k+\frac{1}{2}}c_1^2ddiV^{A1}_{i+\frac{1}{2},j+\frac{1}{2},k+\frac{1}{2}}diV^{A1}_{i+\frac{1}{2},j+\frac{1}{2},k+\frac{1}{2}}\\
		{q_{A2}}^{n}_{i+\frac{1}{2},j+\frac{1}{2},k+\frac{1}{2}}=\rho^{n}_{i+\frac{1}{2},j+\frac{1}{2},k+\frac{1}{2}}c_1^2ddiV^{A2}_{i+\frac{1}{2},j+\frac{1}{2},k+\frac{1}{2}}diV^{A2}_{i+\frac{1}{2},j+\frac{1}{2},k+\frac{1}{2}}\\
		{q_{A3}}^{n}_{i+\frac{1}{2},j+\frac{1}{2},k+\frac{1}{2}}=\rho^{n}_{i+\frac{1}{2},j+\frac{1}{2},k+\frac{1}{2}}c_1^2ddiV^{A3}_{i+\frac{1}{2},j+\frac{1}{2},k+\frac{1}{2}}diV^{A3}_{i+\frac{1}{2},j+\frac{1}{2},k+\frac{1}{2}}\\
		{q_{A4}}^{n}_{i+\frac{1}{2},j+\frac{1}{2},k+\frac{1}{2}}=\rho^{n}_{i+\frac{1}{2},j+\frac{1}{2},k+\frac{1}{2}}c_1^2ddiV^{A4}_{i+\frac{1}{2},j+\frac{1}{2},k+\frac{1}{2}}diV^{A4}_{i+\frac{1}{2},j+\frac{1}{2},k+\frac{1}{2}}\\
		{q_{B1}}^{n}_{i+\frac{1}{2},j+\frac{1}{2},k+\frac{1}{2}}=\rho^{n}_{i+\frac{1}{2},j+\frac{1}{2},k+\frac{1}{2}}c_1^2ddjV^{B1}_{i+\frac{1}{2},j+\frac{1}{2},k+\frac{1}{2}}djV^{B1}_{i+\frac{1}{2},j+\frac{1}{2},k+\frac{1}{2}}\\
		{q_{B2}}^{n}_{i+\frac{1}{2},j+\frac{1}{2},k+\frac{1}{2}}=\rho^{n}_{i+\frac{1}{2},j+\frac{1}{2},k+\frac{1}{2}}c_1^2ddjV^{B2}_{i+\frac{1}{2},j+\frac{1}{2},k+\frac{1}{2}}djV^{B2}_{i+\frac{1}{2},j+\frac{1}{2},k+\frac{1}{2}}\\
		{q_{B3}}^{n}_{i+\frac{1}{2},j+\frac{1}{2},k+\frac{1}{2}}=\rho^{n}_{i+\frac{1}{2},j+\frac{1}{2},k+\frac{1}{2}}c_1^2ddjV^{B3}_{i+\frac{1}{2},j+\frac{1}{2},k+\frac{1}{2}}djV^{B3}_{i+\frac{1}{2},j+\frac{1}{2},k+\frac{1}{2}}\\
		{q_{B4}}^{n}_{i+\frac{1}{2},j+\frac{1}{2},k+\frac{1}{2}}=\rho^{n}_{i+\frac{1}{2},j+\frac{1}{2},k+\frac{1}{2}}c_1^2ddjV^{B4}_{i+\frac{1}{2},j+\frac{1}{2},k+\frac{1}{2}}djV^{B4}_{i+\frac{1}{2},j+\frac{1}{2},k+\frac{1}{2}}\\
		{q_{C1}}^{n}_{i+\frac{1}{2},j+\frac{1}{2},k+\frac{1}{2}}=\rho^{n}_{i+\frac{1}{2},j+\frac{1}{2},k+\frac{1}{2}}c_1^2ddkV^{C1}_{i+\frac{1}{2},j+\frac{1}{2},k+\frac{1}{2}}dkV^{C1}_{i+\frac{1}{2},j+\frac{1}{2},k+\frac{1}{2}}\\
		{q_{C2}}^{n}_{i+\frac{1}{2},j+\frac{1}{2},k+\frac{1}{2}}=\rho^{n}_{i+\frac{1}{2},j+\frac{1}{2},k+\frac{1}{2}}c_1^2ddkV^{C2}_{i+\frac{1}{2},j+\frac{1}{2},k+\frac{1}{2}}dkV^{C2}_{i+\frac{1}{2},j+\frac{1}{2},k+\frac{1}{2}}\\
		{q_{C3}}^{n}_{i+\frac{1}{2},j+\frac{1}{2},k+\frac{1}{2}}=\rho^{n}_{i+\frac{1}{2},j+\frac{1}{2},k+\frac{1}{2}}c_1^2ddkV^{C3}_{i+\frac{1}{2},j+\frac{1}{2},k+\frac{1}{2}}dkV^{C3}_{i+\frac{1}{2},j+\frac{1}{2},k+\frac{1}{2}}\\
		{q_{C4}}^{n}_{i+\frac{1}{2},j+\frac{1}{2},k+\frac{1}{2}}=\rho^{n}_{i+\frac{1}{2},j+\frac{1}{2},k+\frac{1}{2}}c_1^2ddkV^{C4}_{i+\frac{1}{2},j+\frac{1}{2},k+\frac{1}{2}}dkV^{C4}_{i+\frac{1}{2},j+\frac{1}{2},k+\frac{1}{2}}
	\end{eqnarray}
\small
	\begin{eqnarray*}
		&&diV^{A1}_{i+\frac{1}{2},j+\frac{1}{2},k+\frac{1}{2}}=\min{\left[\frac{\bar{\bm Ri}^{n+\frac{1}{2}}_{i+\frac{1}{2},j+\frac{1}{2},k+\frac{1}{2}}\cdot{\frac{\partial \bm u}{\partial i}}^{n+\frac{1}{2}}_{i+\frac{1}{2},j,k}}{\left|\bar{\bm Ri}^n_{i+\frac{1}{2},j+\frac{1}{2},k+\frac{1}{2}}\right|},0\right]}\\
		&&diV^{A2}_{i+\frac{1}{2},j+\frac{1}{2},k+\frac{1}{2}}=\min{\left[\frac{\bar{\bm Ri}^{n+\frac{1}{2}}_{i+\frac{1}{2},j+\frac{1}{2},k+\frac{1}{2}}\cdot{\frac{\partial \bm u}{\partial i}}^{n+\frac{1}{2}}_{i+\frac{1}{2},j+1,k}}{\left|\bar{\bm Ri}^n_{i+\frac{1}{2},j+\frac{1}{2},k+\frac{1}{2}}\right|},0\right]}\\
		&&diV^{A3}_{i+\frac{1}{2},j+\frac{1}{2},k+\frac{1}{2}}=\min{\left[\frac{\bar{\bm Ri}^{n+\frac{1}{2}}_{i+\frac{1}{2},j+\frac{1}{2},k+\frac{1}{2}}\cdot{\frac{\partial \bm u}{\partial i}}^{n+\frac{1}{2}}_{i+\frac{1}{2},j+1,k+1}}{\left|\bar{\bm Ri}^n_{i+\frac{1}{2},j+\frac{1}{2},k+\frac{1}{2}}\right|},0\right]}\\
		&&diV^{A4}_{i+\frac{1}{2},j+\frac{1}{2},k+\frac{1}{2}}=\min{\left[\frac{\bar{\bm Ri}^{n+\frac{1}{2}}_{i+\frac{1}{2},j+\frac{1}{2},k+\frac{1}{2}}\cdot{\frac{\partial \bm u}{\partial i}}^{n+\frac{1}{2}}_{i+\frac{1}{2},j,k+1}}{\left|\bar{\bm Ri}^n_{i+\frac{1}{2},j+\frac{1}{2},k+\frac{1}{2}}\right|},0\right]}\\
		&&djV^{B1}_{i+\frac{1}{2},j+\frac{1}{2},k+\frac{1}{2}}=\min{\left[\frac{\bar{\bm Rj}^{n+\frac{1}{2}}_{i+\frac{1}{2},j+\frac{1}{2},k+\frac{1}{2}}\cdot{\frac{\partial \bm u}{\partial j}}^{n+\frac{1}{2}}_{i,j+\frac{1}{2},k}}{\left|\bar{\bm Rj}^n_{i+\frac{1}{2},j+\frac{1}{2},k+\frac{1}{2}}\right|},0\right]}\\
		&&djV^{B2}_{i+\frac{1}{2},j+\frac{1}{2},k+\frac{1}{2}}=\min{\left[\frac{\bar{\bm Rj}^{n+\frac{1}{2}}_{i+\frac{1}{2},j+\frac{1}{2},k+\frac{1}{2}}\cdot{\frac{\partial \bm u}{\partial j}}^{n+\frac{1}{2}}_{i+1,j+\frac{1}{2},k}}{\left|\bar{\bm Rj}^n_{i+\frac{1}{2},j+\frac{1}{2},k+\frac{1}{2}}\right|},0\right]}\\
		&&djV^{B3}_{i+\frac{1}{2},j+\frac{1}{2},k+\frac{1}{2}}=\min{\left[\frac{\bar{\bm Rj}^{n+\frac{1}{2}}_{i+\frac{1}{2},j+\frac{1}{2},k+\frac{1}{2}}\cdot{\frac{\partial \bm u}{\partial j}}^{n+\frac{1}{2}}_{i+1,j+\frac{1}{2},k+1}}{\left|\bar{\bm Rj}^n_{i+\frac{1}{2},j+\frac{1}{2},k+\frac{1}{2}}\right|},0\right]}\\
		&&djV^{B4}_{i+\frac{1}{2},j+\frac{1}{2},k+\frac{1}{2}}=\min{\left[\frac{\bar{\bm Rj}^{n+\frac{1}{2}}_{i+\frac{1}{2},j+\frac{1}{2},k+\frac{1}{2}}\cdot{\frac{\partial \bm u}{\partial j}}^{n+\frac{1}{2}}_{i,j+\frac{1}{2},k+1}}{\left|\bar{\bm Rj}^n_{i+\frac{1}{2},j+\frac{1}{2},k+\frac{1}{2}}\right|},0\right]}\\
		&&dkV^{C1}_{i+\frac{1}{2},j+\frac{1}{2},k+\frac{1}{2}}=\min{\left[\frac{\bar{\bm Rk}^{n+\frac{1}{2}}_{i+\frac{1}{2},j+\frac{1}{2},k+\frac{1}{2}}\cdot{\frac{\partial \bm u}{\partial k}}^{n+\frac{1}{2}}_{i,j,k+\frac{1}{2}}}{\left|\bar{\bm Rk}^n_{i+\frac{1}{2},j+\frac{1}{2},k+\frac{1}{2}}\right|},0\right]}\\
		&&dkV^{C2}_{i+\frac{1}{2},j+\frac{1}{2},k+\frac{1}{2}}=\min{\left[\frac{\bar{\bm Rk}^{n+\frac{1}{2}}_{i+\frac{1}{2},j+\frac{1}{2},k+\frac{1}{2}}\cdot{\frac{\partial \bm u}{\partial k}}^{n+\frac{1}{2}}_{i+1,j,k+\frac{1}{2}}}{\left|\bar{\bm Rk}^n_{i+\frac{1}{2},j+\frac{1}{2},k+\frac{1}{2}}\right|},0\right]}
	\end{eqnarray*}
	\begin{eqnarray*}	
		&&dkV^{C3}_{i+\frac{1}{2},j+\frac{1}{2},k+\frac{1}{2}}=\min{\left[\frac{\bar{\bm Rk}^{n+\frac{1}{2}}_{i+\frac{1}{2},j+\frac{1}{2},k+\frac{1}{2}}\cdot{\frac{\partial \bm u}{\partial k}}^{n+\frac{1}{2}}_{i+1,j+1,k+\frac{1}{2}}}{\left|\bar{\bm Rk}^n_{i+\frac{1}{2},j+\frac{1}{2},k+\frac{1}{2}}\right|},0\right]}\\
		&&dkV^{C4}_{i+\frac{1}{2},j+\frac{1}{2},k+\frac{1}{2}}=\min{\left[\frac{\bar{\bm Rk}^{n+\frac{1}{2}}_{i+\frac{1}{2},j+\frac{1}{2},k+\frac{1}{2}}\cdot{\frac{\partial \bm u}{\partial k}}^{n+\frac{1}{2}}_{i,j+1,k+\frac{1}{2}}}{\left|\bar{\bm Rk}^n_{i+\frac{1}{2},j+\frac{1}{2},k+\frac{1}{2}}\right|},0\right]}\\
		&&ddiV^{A*}_{i,j+\frac{1}{2},k+\frac{1}{2}}=\left|diV^{A*}_{i+\frac{1}{2},j+\frac{1}{2},k+\frac{1}{2}}-diV^{A*}_{i-\frac{1}{2},j+\frac{1}{2},k+\frac{1}{2}}\right|\\
		&&ddiV^{A*}_{i+\frac{1}{2},j+\frac{1}{2},k+\frac{1}{2}}=\frac{1}{2}\left(ddiV^{A*}_{i,j+\frac{1}{2},k+\frac{1}{2}}+ddiV^{A*}_{i+1,j+\frac{1}{2},k+\frac{1}{2}}\right)\\
		&&ddjV^{B*}_{i+\frac{1}{2},j,k+\frac{1}{2}}=\left|djV^{B*}_{i+\frac{1}{2},j+\frac{1}{2},k+\frac{1}{2}}-djV^{B*}_{i+\frac{1}{2},j-\frac{1}{2},k+\frac{1}{2}}\right|\\
		&&ddjV^{B*}_{i+\frac{1}{2},j+\frac{1}{2},k+\frac{1}{2}}=\frac{1}{2}\left(ddjV^{B*}_{i+\frac{1}{2},j,k+\frac{1}{2}}+ddjV^{B*}_{i+\frac{1}{2},j+1,k+\frac{1}{2}}\right)\\
		&&ddkV^{C*}_{i+\frac{1}{2},j+\frac{1}{2},k}=\left|dkV^{C*}_{i+\frac{1}{2},j+\frac{1}{2},k+\frac{1}{2}}-dkV^{C*}_{i+\frac{1}{2},j+\frac{1}{2},k-\frac{1}{2}}\right|\\
		&&ddkV^{C*}_{i+\frac{1}{2},j+\frac{1}{2},k+\frac{1}{2}}=\frac{1}{2}\left(ddkV^{C*}_{i+\frac{1}{2},j+\frac{1}{2},k}+ddkV^{C*}_{i+\frac{1}{2},+\frac{1}{2}j,k+1}\right)\\
	\end{eqnarray*}
\normalsize
For the normal vectors to the $ i, j$ and $k$ directions, the outer products are used at  each side as shown in Fig. \ref{fig:qAbm}, and the averaged values are obtained:  
\small
\begin{eqnarray*}\label{vector}
	&&\bar{\bm Ri}^n_{i,j+\frac{1}{2},k+\frac{1}{2}}=\frac{\bar{\bm R_{A1}}^n_{i,j+\frac{1}{2},k+\frac{1}{2}}+\bar{\bm R_{A2}}^n_{i,j+\frac{1}{2},k+\frac{1}{2}}+\bar{\bm R_{A3}}^n_{i,j+\frac{1}{2},k+\frac{1}{2}}+\bar{\bm R_{A4}}^n_{i,j+\frac{1}{2},k+\frac{1}{2}}}{4}\\	
	&&\bar{\bm Rj}^n_{i+\frac{1}{2},j,k+\frac{1}{2}}=\frac{\bar{\bm R_{B1}}^n_{i+\frac{1}{2},j,k+\frac{1}{2}}+\bar{\bm R_{B2}}^n_{i+\frac{1}{2},j,k+\frac{1}{2}}+\bar{\bm R_{B3}}^n_{i+\frac{1}{2},j,k+\frac{1}{2}}+\bar{\bm R_{B4}}^n_{i+\frac{1}{2},j,k+\frac{1}{2}}}{4}\\
	&&\bar{\bm Rk}^n_{i+\frac{1}{2},j+\frac{1}{2},k}=\frac{\bar{\bm R_{C1}}^n_{i+\frac{1}{2},j+\frac{1}{2},k}+\bar{\bm R_{C2}}^n_{i+\frac{1}{2},j+\frac{1}{2},k}+\bar{\bm R_{C3}}^n_{i+\frac{1}{2},j+\frac{1}{2},k}+\bar{\bm R_{C4}}^n_{i+\frac{1}{2},j+\frac{1}{2},k}}{4}
\end{eqnarray*}
\normalsize
\begin{figure}[H]
	\centering
	\includegraphics[width=10cm]{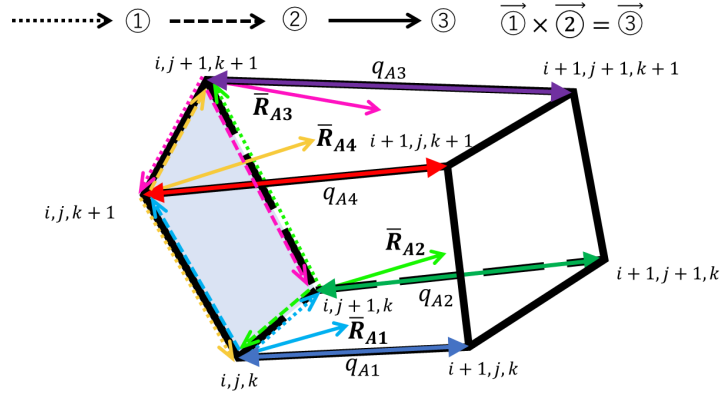}
	\caption{The nomal vector to the $i$ direction}\label{fig:qAbm}
\end{figure}


\item {\bf coc\_LC.cpp}\\
The file calculates the Lagrangian mesh dynamics. The Lagrangian meshes move together with the fluid motion. The new position coordinates for each mesh point are renewed at $n+1$. 


\item {\bf cotc3D\_e.cpp, cotc3D\_r.cpp}\\
For calculation of the heat conduction, the following basic equation is used\cite{Christiansen}.
\begin{eqnarray}
		&&C_{V_k}\frac{DT}{Dt}=\frac{1}{\rho}{\bm \nabla}\cdot(\kappa_k{\bm \nabla}T_k)\ \ \ \ \ \ \ (k=e,r)\\
		&&\ \ \ \ \kappa_e=1.83\times10^{-10}T^{5/2}_{e}(\log{\Lambda})^{-1}Z^{-1}\ \ [{\rm W/mK}]\nonumber\\
		&&\ \ \ \ \kappa_r=\frac{16}{3}\sigma L_RT^3_r\ \ [{\rm W/mK}]\nonumber
	\end{eqnarray}
	
\begin{align*}
\kappa&: Heat conductivity\\
T_k&: Ion,\ electron,\ radiation\ temperature[\rm K]\\
\log{\Lambda}&: Coulomb\ logarithm\\
m&: Mass\\
Z&: Ionization\ degree\\
\sigma&: Stefan-Boltzmann\ constant\\
L_R&: \rm {Rosseland\ mean\ free\ path}\\
\end{align*}

Here,  If $\bm S$ is the area vector of each surface of the mesh, the basic equation is transformed as follows: 
	\begin{eqnarray}\label{trans2}
		C_v\frac{DT}{Dt}&=&\frac{1}{\rho}{\bm \nabla}\cdot(\kappa{\bm \nabla}T)\nonumber\\
		\iiint \rho C_v\frac{DT}{Dt}dV&=&\iiint{\bm \nabla}\cdot(\kappa{\bm \nabla}T)dV\nonumber\\
		MC_v\frac{DT}{Dt}&=&\sum_i \left(\kappa{\bm \nabla} T\right)\cdot \bm{S_i}\nonumber\\
		\frac{DT}{Dt}&=&\frac{1}{MC_v}\sum_i 	\bm{S_i} \cdot \left[\frac{\kappa}{J}Di\bm r\frac{\partial T}{\partial i}+\frac{\kappa}{J}Dj\bm r\frac{\partial T}{\partial j}+\frac{\kappa}{J}Dk\bm r\frac{\partial T}{\partial k}\right]\nonumber\\
	\end{eqnarray}
	
Equation (\ref{trans2}) is discretized as follows:
\small
	\begin{eqnarray}\label{trans3}
		\frac{T^{n+1}_{i+\frac{1}{2},j+\frac{1}{2},k+\frac{1}{2}}-T^{n}_{i+\frac{1}{2},j+\frac{1}{2},k+\frac{1}{2}}}{Dt^{n+\frac{1}{2}}}&=&\frac{1}{MC_v}\nonumber\\
		&\times&\left[A_2\Bigl(T^{n+1}_{i+\frac{3}{2},j+\frac{1}{2},k+\frac{1}{2}}-T^{n+1}_{i+\frac{1}{2},j+\frac{1}{2},k+\frac{1}{2}}\Bigr)\right.\nonumber\\
		&-&A_1\Bigl(T^{n+1}_{i+\frac{1}{2},j+\frac{1}{2},k+\frac{1}{2}}-T^{n+1}_{i-\frac{1}{2},j+\frac{1}{2},k+\frac{1}{2}}\Bigr)\nonumber\\
		&+&B_2\Bigl(T^{n+1}_{i+\frac{1}{2},j+\frac{3}{2},k+\frac{1}{2}}-T^{n+1}_{i+\frac{1}{2},j+\frac{1}{2},k+\frac{1}{2}}\Bigr)\nonumber\\
		&-&B_1\Bigl(T^{n+1}_{i+\frac{1}{2},j+\frac{1}{2},k+\frac{1}{2}}-T^{n+1}_{i+\frac{1}{2},j-\frac{1}{2},k+\frac{1}{2}}\Bigr)\nonumber\\
		&+&C_2\Bigl(T^{n+1}_{i+\frac{1}{2},j+\frac{1}{2},k+\frac{3}{2}}-T^{n+1}_{i+\frac{1}{2},j+\frac{1}{2},k+\frac{1}{2}}\Bigr)\nonumber\\
		&-&\left.C_1\Bigl(T^{n+1}_{i+\frac{1}{2},j+\frac{1}{2},k+\frac{1}{2}}-T^{n+1}_{i+\frac{1}{2},j-\frac{1}{2},k-\frac{1}{2}}\Bigr)\right]\nonumber\\
	\end{eqnarray}
\normalsize
Here, the coefficients in Eq. (\ref{trans3}) are listed: 
\small
\begin{eqnarray*}
&&A_2=\Bigl|{Di\bm r}^n_{i+1,j+\frac{1}{2},k+\frac{1}{2}}\Bigr|{\Bigl|\bm{Si}^n_{i+1,j+\frac{1}{2},k+\frac{1}{2}}\Bigr|}\frac{\kappa^n_{i+1,j+\frac{1}{2},k+\frac{1}{2}}}{J^n_{i+1,j+\frac{1}{2},k+\frac{1}{2}}}\\
&&A_1=\Bigl|{Di\bm r}^n_{i,j+\frac{1}{2},k+\frac{1}{2}}\Bigr|{\Bigl|\bm{Si}^n_{i,j+\frac{1}{2},k+\frac{1}{2}}\Bigr|}\frac{\kappa^n_{i,j+\frac{1}{2},k+\frac{1}{2}}}{J^n_{i+1,j+\frac{1}{2},k+\frac{1}{2}}}\\
&&B_2=\Bigl|{Dj\bm r}^n_{i+\frac{1}{2},j+1,k+\frac{1}{2}}\Bigr|{\Bigl|\bm{Sj}^n_{i+\frac{1}{2},j+1,k+\frac{1}{2}}\Bigr|}\frac{\kappa^n_{i+\frac{1}{2},j+1,k+\frac{1}{2}}}{J^n_{i+\frac{1}{2},j+1,k+\frac{1}{2}}}\\
&&B_1=\Bigl|{Dj\bm r}^n_{i+\frac{1}{2},j,k+\frac{1}{2}}\Bigr|{\Bigl|\bm{Sj}^n_{i+\frac{1}{2},j,k+\frac{1}{2}}\Bigr|}\frac{\kappa^n_{i+\frac{1}{2},j,k+\frac{1}{2}}}{J^n_{i+\frac{1}{2},j,k+\frac{1}{2}}}\\
&&C_2=\Bigl|{Dk\bm r}^n_{i+\frac{1}{2},j+\frac{1}{2},k+1}\Bigr|{\Bigl|\bm{Sk}^n_{i+\frac{1}{2},j+\frac{1}{2},k+1}\Bigr|}\frac{\kappa^n_{i+\frac{1}{2},j+\frac{1}{2},k+1}}{J^n_{i+\frac{1}{2},j+\frac{1}{2},k+1}}\\
&&C_1=\Bigl|{Dk\bm r}^n_{i+\frac{1}{2},j+\frac{1}{2},k}\Bigr|{\Bigl|\bm{Sk}^n_{i+\frac{1}{2},j+\frac{1}{2},k}\Bigr|}\frac{\kappa^n_{i+\frac{1}{2},j+\frac{1}{2},k}}{J^n_{i+\frac{1}{2},j+\frac{1}{2},k}}\\
\end{eqnarray*}
\normalsize

\item {\bf define\_LC.h}\\
It contains the procedure declarations for the Lagrangian code and Ok3. 

\item {\bf dif\_LC.cpp}\\
The following Lagrangian equation of motion is used.
\begin{equation}\label{dme}
		\rho\frac{D\bm u}{Dt}=-{\bm\nabla}(P+q)
	\end{equation}
Equation (\ref {dme}) is expressed as follows: 
	\begin{equation}
\begin{split}\label{dmeu}
\frac{D\bm u}{Dt}\bigg|^n_{i,j,k}&
=-\frac{1}{\rho}\frac{\partial}{\partial \bm r} (P+q)\\&
=-\frac{1}{M^n_{i,j,k}}\left[Di\bm r\frac{\partial}{\partial i}(P+q_A)+Di\bm r\frac{\partial}{\partial j}(P+q_B)+Di\bm r\frac{\partial}{\partial k}(P+q_C)\right]^n_{i,j,k}\\&
\end{split}
\end{equation}
Equation (\ref{dmeu}) is discretized as follows: 

\small
	\begin{eqnarray}
		\bm u^{n+\frac{1}{2}}_{i,j,k}=\bm u^{n-\frac{1}{2}}_{i,j,k} - \frac{Dt^n}{M^n_{i,j,k}}\left[Di\bm r\frac{\partial}{\partial i}(P+q_A)+Di\bm r\frac{\partial}{\partial j}(P+q_B)+Di\bm r\frac{\partial}{\partial k}(P+q_C)\right]^n_{i,j,k}
		\label{dmeu2}
	\end{eqnarray}
\normalsize

Here, each term is shown: 
\begin{equation}
\begin{split}
Di\bm r\frac{\partial}{\partial i}(P+q_A)\bigg|^n_{i,j,k}&=
\frac{1}{4}\left[Di\bm r\frac{\partial}{\partial i}(P+q_{A1})\right]_{i,j+\frac{1}{2},k+\frac{1}{2}}\\&
+\frac{1}{4}\left[Di\bm r\frac{\partial}{\partial i}(P+q_{A2})\right]_{i,j-\frac{1}{2},k+\frac{1}{2}}\\&
+\frac{1}{4}\left[Di\bm r\frac{\partial}{\partial i}(P+q_{A3})\right]_{i,j-\frac{1}{2},k-\frac{1}{2}}\\&
+\frac{1}{4}\left[Di\bm r\frac{\partial}{\partial i}(P+q_{A4})\right]_{i,j+\frac{1}{2},k-\frac{1}{2}}\\
\end{split}
\end{equation}

\begin{equation}
\begin{split}
Dj\bm r\frac{\partial}{\partial j}(P+q_B)\bigg|^n_{i,j,k}&=
\frac{1}{4}\left[Dj\bm r\frac{\partial}{\partial j}(P+q_{B1})\right]_{i+\frac{1}{2},j,k+\frac{1}{2}}\\&
+\frac{1}{4}\left[Dj\bm r\frac{\partial}{\partial j}(P+q_{B2})\right]_{i-\frac{1}{2},j,k+\frac{1}{2}}\\&
+\frac{1}{4}\left[Dj\bm r\frac{\partial}{\partial j}(P+q_{B3})\right]_{i-\frac{1}{2},j,k-\frac{1}{2}}\\&
+\frac{1}{4}\left[Dj\bm r\frac{\partial}{\partial j}(P+q_{B4})\right]_{i+\frac{1}{2},j,k-\frac{1}{2}}\\
\end{split}
\end{equation}

\begin{equation}
\begin{split}
Dk\bm r\frac{\partial}{\partial k}(P+q_C)\bigg|^n_{i,j,k}&=
\frac{1}{4}\left[Dk\bm r\frac{\partial}{\partial k}(P+q_{C1})\right]_{i+\frac{1}{2},j+\frac{1}{2},k}\\&
+\frac{1}{4}\left[Dk\bm r\frac{\partial}{\partial k}(P+q_{C2})\right]_{i-\frac{1}{2},j+\frac{1}{2},k}\\&
+\frac{1}{4}\left[Dk\bm r\frac{\partial}{\partial k}(P+q_{C3})\right]_{i-\frac{1}{2},j-\frac{1}{2},k}\\&
+\frac{1}{4}\left[Dk\bm r\frac{\partial}{\partial k}(P+q_{C4})\right]_{i+\frac{1}{2},j-\frac{1}{2},k}\\
\end{split}
\end{equation}
	
\item {\bf dt\_LC.cpp}\\
This procedure calculates and controls the time step to satisfy the numerical stability condition. The time step $ \Delta t $ in the calculation must satisfy the following conditions.
	\begin{equation}
	\label{eq:Courant}
		\Delta t=\frac{\Delta r}{C_{S}+V_{max}}
	\end{equation}
The time step for the Lagrangian method $ Dt ^ {n + \frac {1}{2}} $ is represented by the following expression.
	\begin{equation}
		Dt^{n+\frac{1}{2}}=\alpha\displaystyle\frac{dr_{min}}{C_S+V_{max}}
	\end{equation}
		\begin{tabbing}
		12345678901234567890123\=789\=12345\=\kill\\
		\>\>$\alpha$\>: Numerical coefficient constant $(\alpha\leq1)$\\
		\>\>$dr_{min}$\>: the minimum grid spacing\\
		\>\>$C_S$\>: Sound speed\\
		\>\>$V_{max}$\>: the maximum flow speed\\
	\end{tabbing}
	
\item {\bf eoenergy\_LC.cpp}\\
The file contains a procedure for calculation of the energy equation. The following Lagrangian energy equation is used except for the heat conductions terms.
	\begin{eqnarray}
	\label{EoEkiso}
	\begin{cases}
		\frac{DT_i}{Dt}=-\frac{k_B}{C_{V_i}}\left[B_{T_i}\frac{D\rho}{Dt}+\frac{p_i+q}{M}\frac{DJ}{Dt}\right]\\
		\frac{DT_e}{Dt}=-\frac{k_B}{C_{V_e}}\left[B_{T_e}\frac{D\rho}{Dt}+\frac{p_e}{M}\frac{DJ}{Dt}\right]\\
		\frac{DT_r}{Dt}=-\frac{k_B}{C_{V_r}}\left[B_{T_r}\frac{D\rho}{Dt}+\frac{p_r}{M}\frac{DJ}{Dt}\right]
	\end{cases}
	\end{eqnarray}
Equation (\ref{EoEkiso}) is discretized as follows: 
	\small \begin{eqnarray}
		T^{n+1}_{i+\frac{1}{2},j+\frac{1}{2},k+\frac{1}{2}}&=&T^n_{i+\frac{1}{2},j+\frac{1}{2},k+\frac{1}{2}}\nonumber\\
		&-&\frac{1}{{C_V}^{n+\frac{1}{2}}_{i+\frac{1}{2},j+\frac{1}{2}k+\frac{1}{2}}}\Bigg[{B_T}^{n+1}_{i+\frac{1}{2},j+\frac{1}{2},k+\frac{1}{2}}(\rho^{n+1}_{i+\frac{1}{2},j+\frac{1}{2}k+\frac{1}{2}}-\rho^n_{i+\frac{1}{2},j+\frac{1}{2},k+\frac{1}{2}})\nonumber\\
		&+&\frac{P^{n+\frac{1}{2}}_{i+\frac{1}{2},j+\frac{1}{2},k+\frac{1}{2}}+q^{n+\frac{1}{2}}_{i+\frac{1}{2},j+\frac{1}{2},k+\frac{1}{2}}}{M_{i+\frac{1}{2},j+\frac{1}{2},k+\frac{1}{2}}}(J^{n+1}_{i+\frac{1}{2},j+\frac{1}{2}k+\frac{1}{2}}-J^n_{i+\frac{1}{2},j+\frac{1}{2}k+\frac{1}{2}})\Bigg]\nonumber
	\end{eqnarray}
	
\normalsize	
\item {\bf eos.cpp}\\
The file contains the procedures to calculate the equation of state. The equation of state for ions is the ideal one. For the equation of state for electrons and the ionization, we use the equation of state based on the Thomas-Fermi model shown in Ref. \cite{Bell}. Users can select the Thomas-Fermi model or the ideal equation of state in the header file of "input\_LC.h". For the equation of state for the radiation, we use the equilibrium blackbody equations \cite{Zeldovich}. 
	
\item {\bf init\_LC.h}\\
It contains the initial conditions such as the initial target temperature and so on. 

\item {\bf init\_matrix\_LC.cpp}
The file get the matrix.

\item {\bf input\_LC.h}\\
The input data for Lagrangian code contains radius, $\theta$ and $\phi$ direction mush number, each layers mesh number,HIB number, beam pulse parameters, fuel target structure,  output date step, etc. 

\item {\bf jacobian\_LC.cpp}\\
 The volume of each mesh is calculated. The Jacobian $J$ is expressed by the following formula. 
\begin{eqnarray}\label{jaco}
\begin{split}
	J
	&=\frac{\partial (x,y,z)}{\partial (i,j,k)}\\
	&=\left[
	\begin{array}{rrr}
	\frac{\partial x}{\partial i} & \frac{\partial x}{\partial j} & \frac{\partial x}{\partial k} \\
	\frac{\partial y}{\partial i} & \frac{\partial y}{\partial j} & \frac{\partial y}{\partial k} \\
	\frac{\partial z}{\partial i} & \frac{\partial z}{\partial j} & \frac{\partial z}{\partial k} 
	\end{array}
	\right]\\
	&=\frac{\partial x}{\partial i}\frac{\partial y}{\partial j}\frac{\partial z}{\partial k}+\frac{\partial x}{\partial j}\frac{\partial y}{\partial k}\frac{\partial z}{\partial i}+\frac{\partial x}{\partial k}\frac{\partial y}{\partial i}\frac{\partial z}{\partial j}
	-\frac{\partial x}{\partial k}\frac{\partial y}{\partial j}\frac{\partial z}{\partial i}-\frac{\partial x}{\partial j}\frac{\partial y}{\partial i}\frac{\partial z}{\partial k}-\frac{\partial x}{\partial i}\frac{\partial y}{\partial k}\frac{\partial z}{\partial j}
\end{split}
\end{eqnarray}
	From Eq. (\ref {jaco}), the Jacobian is expressed as follows: 
	\begin{eqnarray}\label{Ja_R}
	J^n_{k+\frac{1}{2},l+\frac{1}{2},m+\frac{1}{2}}&=&
\left(\frac{\partial x}{\partial i}\right)^n_{i+\frac{1}{2},j+\frac{1}{2},k+\frac{1}{2}}
\left(\frac{\partial y}{\partial j}\right)^n_{i+\frac{1}{2},j+\frac{1}{2},k+\frac{1}{2}}
\left(\frac{\partial z}{\partial k}\right)^n_{i+\frac{1}{2},j+\frac{1}{2},k+\frac{1}{2}}\nonumber\\
&+&\left(\frac{\partial x}{\partial j}\right)^n_{i+\frac{1}{2},j+\frac{1}{2},k+\frac{1}{2}}
\left(\frac{\partial y}{\partial k}\right)^n_{i+\frac{1}{2},j+\frac{1}{2},k+\frac{1}{2}}
\left(\frac{\partial z}{\partial i}\right)^n_{i+\frac{1}{2},j+\frac{1}{2},k+\frac{1}{2}}\nonumber\\
&+&\left(\frac{\partial x}{\partial k}\right)^n_{i+\frac{1}{2},j+\frac{1}{2},k+\frac{1}{2}}
\left(\frac{\partial y}{\partial i}\right)^n_{i+\frac{1}{2},j+\frac{1}{2},k+\frac{1}{2}}
\left(\frac{\partial z}{\partial j}\right)^n_{i+\frac{1}{2},j+\frac{1}{2},k+\frac{1}{2}}\nonumber\\
&-&\left(\frac{\partial x}{\partial k}\right)^n_{i+\frac{1}{2},j+\frac{1}{2},k+\frac{1}{2}}
\left(\frac{\partial y}{\partial j}\right)^n_{i+\frac{1}{2},j+\frac{1}{2},k+\frac{1}{2}}
\left(\frac{\partial z}{\partial i}\right)^n_{i+\frac{1}{2},j+\frac{1}{2},k+\frac{1}{2}}\nonumber\\
&-&\left(\frac{\partial x}{\partial j}\right)^n_{i+\frac{1}{2},j+\frac{1}{2},k+\frac{1}{2}}
\left(\frac{\partial y}{\partial i}\right)^n_{i+\frac{1}{2},j+\frac{1}{2},k+\frac{1}{2}}
\left(\frac{\partial z}{\partial k}\right)^n_{i+\frac{1}{2},j+\frac{1}{2},k+\frac{1}{2}}\nonumber\\
&-&\left(\frac{\partial x}{\partial i}\right)^n_{i+\frac{1}{2},j+\frac{1}{2},k+\frac{1}{2}}
\left(\frac{\partial y}{\partial k}\right)^n_{i+\frac{1}{2},j+\frac{1}{2},k+\frac{1}{2}}
\left(\frac{\partial z}{\partial j}\right)^n_{i+\frac{1}{2},j+\frac{1}{2},k+\frac{1}{2}}\nonumber
\end{eqnarray}
	Here, each term is obtained: 
\begin{eqnarray}
\begin{cases}
&(\frac{\partial x_{i,j,k}}{\partial i})^n_{i+\frac{1}{2},j+\frac{1}{2},k+\frac{1}{2}}=\frac{\Delta_i x^n_{i+\frac{1}{2},j+1,k}+\Delta_i x^n_{i+\frac{1}{2},j,k+1}+\Delta_i x^n_{i+\frac{1}{2},j+1,k+1}+\Delta_i x^n_{i+\frac{1}{2},j,k}}{4}\\
&(\frac{\partial x_{i,j,k}}{\partial j})^n_{i+\frac{1}{2},j+\frac{1}{2},k+\frac{1}{2}}=\frac{\Delta_j x^n_{i+1,j+\frac{1}{2},k}+\Delta_j x^n_{i,j+\frac{1}{2},k+1}+\Delta_j x^n_{i+1,j+\frac{1}{2},k+1}+\Delta_j x^n_{i,j+\frac{1}{2},k}}{4}\\
&(\frac{\partial x_{i,j,k}}{\partial k})^n_{i+\frac{1}{2},j+\frac{1}{2},k+\frac{1}{2}}=\frac{\Delta_k x^n_{i+1,j,k+\frac{1}{2}}+\Delta_k x^n_{i,j+1,k+\frac{1}{2}}+\Delta_k x^n_{i+1,j+1,k+\frac{1}{2}}+\Delta_k x^n_{i,j,k+\frac{1}{2}}}{4}\\
&(\frac{\partial y_{i,j,k}}{\partial i})^n_{i+\frac{1}{2},j+\frac{1}{2},k+\frac{1}{2}}=\frac{\Delta_i y^n_{i+\frac{1}{2},j+1,k}+\Delta_i y^n_{i+\frac{1}{2},j,k+1}+\Delta_i y^n_{i+\frac{1}{2},j+1,k+1}+\Delta_i y^n_{i+\frac{1}{2},j,k}}{4}\\
&(\frac{\partial y_{i,j,k}}{\partial j})^n_{i+\frac{1}{2},j+\frac{1}{2},k+\frac{1}{2}}=\frac{\Delta_j y^n_{i+1,j+\frac{1}{2},k}+\Delta_j y^n_{i,j+\frac{1}{2},k+1}+\Delta_j y^n_{i+1,j+\frac{1}{2},k+1}+\Delta_j y^n_{i,j+\frac{1}{2},k}}{4}\\
&(\frac{\partial y_{i,j,k}}{\partial k})^n_{i+\frac{1}{2},j+\frac{1}{2},k+\frac{1}{2}}=\frac{\Delta_k y^n_{i+1,j,k+\frac{1}{2}}+\Delta_k y^n_{i,j+1,k+\frac{1}{2}}+\Delta_k y^n_{i+1,j+1,k+\frac{1}{2}}+\Delta_k y^n_{i,j,k+\frac{1}{2}}}{4}\\
&(\frac{\partial z_{i,j,k}}{\partial i})^n_{i+\frac{1}{2},j+\frac{1}{2},k+\frac{1}{2}}=\frac{\Delta_i z^n_{i+\frac{1}{2},j+1,k}+\Delta_i z^n_{i+\frac{1}{2},j,k+1}+\Delta_i z^n_{i+\frac{1}{2},j+1,k+1}+\Delta_k z^n_{k+\frac{1}{2},l,m}}{4}\\
&(\frac{\partial z_{i,j,k}}{\partial j})^n_{i+\frac{1}{2},j+\frac{1}{2},k+\frac{1}{2}}=\frac{\Delta_j z^n_{i+1,l+\frac{1}{2},k}+\Delta_j z^n_{i,j+\frac{1}{2},k+1}+\Delta_j z^n_{i+1,j+\frac{1}{2},k+1}+\Delta_j z^n_{i,j+\frac{1}{2},k}}{4}\\
&(\frac{\partial z_{i,j,k}}{\partial k})^n_{i+\frac{1}{2},j+\frac{1}{2},k+\frac{1}{2}}=\frac{\Delta_k z^n_{i+1,j,k+\frac{1}{2}}+\Delta_k z^n_{i,j+1,k+\frac{1}{2}}+\Delta_k z^n_{i+1,j+1,k+\frac{1}{2}+}\Delta_k z^n_{i,j,k+\frac{1}{2}}}{4}
\end{cases}
\end{eqnarray}

\item {\bf main\_LC.cpp}\\
The main procedure of the Lagrangian fluid code. If you want to artificially add non-uniformity in the $\theta$ and $\phi$ directions without using the OK3 code, change it here. 
\item {\bf outputRMS.cpp}\\
It contains a procedure to output the results for the RMS non-uniformity.
\item {\bf output\_LC.cpp}\\
The result data are stored by this procedure. The time interval of data output is 0.1 ns in the Lagrangian code. The user can adjust the output step in "input\_LC.h". The physical quantity (for example, velocity) defined at the grid points of the mesh is output to outputS1. The physical quantity defined at the center of the mesh (for example, temperature, density) is output to outputS2. 
\item {\bf output\_to\_EulerCode.cpp}\\
This file contains a procedure for outputting the data used in Euler code. After the beam irradiation is completed, the file is output every 0.1ns. 
\item {\bf relax.cpp}\\
The following equation is used as the basic equation for the temperature relaxation\cite{Tahir}. 
	\begin{eqnarray}
		\begin{cases}
			C_{V_i}\frac{dT_i}{dt}=-K_{ie}\\
			C_{V_e}\frac{dT_e}{dt}=K_{ie}-K_{re}\\
			C_{V_r}\frac{dT_r}{dt}=K_{re}
		\end{cases}
	\end{eqnarray}

Here, $K_{ie}$ is the energy exchange rate between the ions and the electrons,  and $K_{re}$ the energy exchange rate between the radiation and the electrons.
\begin{eqnarray}
	\begin{cases}
		K_{ie}=C_{V_i}\omega_{ie}(T_i-T_e)\\
		K_{re}=C_{V_r}\omega_{re}(T_e-T_r)
	\end{cases}
	\end{eqnarray}
	$ \omega_{ie}$ and $ \omega_{re} $ are the collision frequencies between the ions and the electrons and between the radiation and the electrons, respectively. They are obtained by the following formulae: The Compton effect  between the radiation and the electrons is included. Each expression and the solution method are found in Refs. \cite{Tahir, CPC-O-SUKI}. 

I

\end{enumerate}


\subsection{Conversion code}
The Euler meshes are constructed based on the size of the small Lagrange mesh in the conversion code. The 3D conversion process is performed after setting the upper limit of the Euler total mesh number. In order to meet the computer resource limitation, this prescription is employed in the O-SUKI-N 3D code. 

\begin{enumerate}
\item{\bf boundary\_set.cpp}
The function makes the boundary point data between DT and Al layer. 
\item{\bf check\_quantities.cpp}
The function outputs the data of the transformed Euler mesh as a text (csv) file. 
\item{\bf define\_convert.h}
Define the variables necessary for the conversion code.

\item{\bf GenerateEulerMesh.cpp}
The procedure determines the number of the Euler meshes and to secure the necessary memory, just before the data conversion.
\item{\bf Interpolation.cpp}
The function interpolates the data on the Lagrangian mesh to those on the Euler meshes. Figure \ref{interpolation_speed} shows the interpolation method from the Lagrange data to the Euler data. The "MeshSearch.cpp" provides the relation between the Lagrangian mesh location and the Euler mesh location. The following interpolation equation is used to obtain each physical quantity on the Euler meshes. For example, here $\bm u$ shows a velocity. 

\begin{eqnarray}\label{eq:interpolation}
	\bm u(P)&=&\frac{1}{sumR}\nonumber\\		
		&\times&\left[\Bigl(\frac{1}{r_{i,j,k}}\Bigr)^2\bm u_{i,j,k}+\Bigl(\frac{1}{r_{i+1,j,k}}\Bigr)^2\bm u_{i+1,j,k}+\Bigl(\frac{1}{r_{i+1,j+1,k}}\Bigr)^2\bm u_{i+1,j+1,k}\right.\nonumber\\
		&+&\Bigl(\frac{1}{r_{i,j+1,k}}\Bigr)^2\bm u_{i,j+1,k}+\Bigl(\frac{1}{r_{i,j,k+1}}\Bigr)^2\bm u_{i,j,k+1}+\Bigl(\frac{1}{r_{i+1,j,k+1}}\Bigr)^2\bm u_{i+1,j,k}\nonumber\\
		&+&\left.\Bigl(\frac{1}{r_{i+1,j+1,k}}\Bigr)^2\bm u_{i+1,j+1,k}+\Bigl(\frac{1}{r_{i,j+1,k+1}}\Bigr)^2\bm u_{i,j+1,k+1}\right]
	\end{eqnarray}

\begin{eqnarray}
	sumR&=&\left(\frac{1}{r_{i,j,k}}\right)^2+\left(\frac{1}{r_{i+1,j,k}}\right)^2+\left(\frac{1}{r_{i+1,j+1,k}}\right)^2+\left(\frac{1}{r_{i,j+1,k}}\right)^2\nonumber\\
	&+&\left(\frac{1}{r_{i,j,k+1}}\right)^2+\left(\frac{1}{r_{i+1,j,k+1}}\right)^2+\left(\frac{1}{r_{i+1,j+1,k+1}}\right)^2+\left(\frac{1}{r_{i,j+1,k+1}}\right)^2\nonumber
\end{eqnarray}

\begin{figure}[H]
	\centering
	\includegraphics[width=12cm]{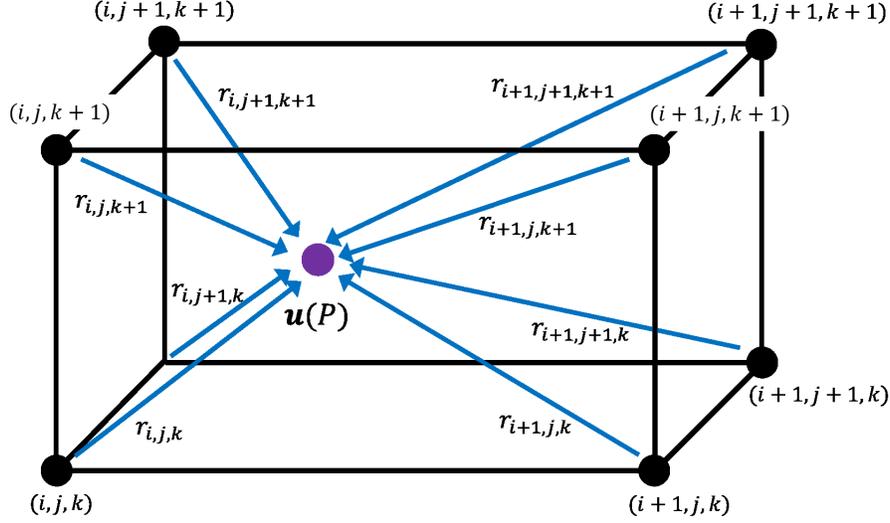}
	\caption{Interpolation of velocity}\label{interpolation_speed}
\end{figure}

Usually the Euler mesh size is small compared with the size of the Lagrange mesh. For the interpolation of physical quantities other than velocity, the physical quantity of the corresponding Lagrange mesh acquired by the "MeshSearch.cpp" is interpolated by the 0th order method. On the other hand, if the corresponding Lagrange mesh is smaller than the Euler mesh, it is done in the same way shown in Fig. \ref{interpolation_speed} and Eq. (\ref{eq:interpolation}) for the example velocity interpolation. This is the special treatment in 3D to optimize the required memory size. 

\item{\bf main\_convert.cpp}
This is the main procedure of the conversion code. The procedure selects the output Lagrangian data transferred to the Euler code among the Lagrangian data sets obtained in the Lagrangian code. The Lagrangian meshes are deformed along with the fluid motion. The Lagrangian code stops, before no mesh is crushed. The conversion range is the all DT layer and a part of the Aluminum region. The volume of the Al region is 2.5 times larger than the thickness of the DT layer. The required number of the Euler meshes is calculated. The Lagrange data sets are examined from the data set from the time of 2ns earlier than the last output data set. The function selects the conversion date, which has the smallest number of the Euler mesh required. If the number of Euler meshes exceeds the number of Euler meshes set in the ''input\_LC.h'', the Euler mesh total number is forced to set to the upper limit defined beforehand.

\item{\bf MeshSearch.cpp}
This procedure examines the location of each Euler mesh among the Lagrangian meshes. The MeshSearch function divides a Lagrange mesh into 12 triangular tetrahedra as shown in Fig. \ref{meshcut}, and examines if the definition point of an Euler mesh is contained in the specific Lagrange mesh. 

\begin{figure}[H]
	\centering
	\includegraphics[width=10cm]{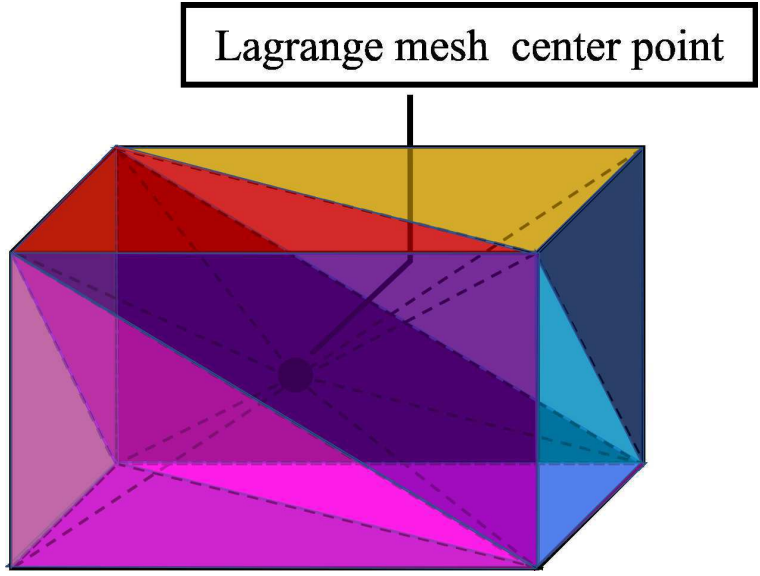}
	\caption{One Lagrange mesh and 12 triangular tetrahedra}\label{meshcut}
\end{figure}

In Fig. \ref{meshsearch}, $\vec P$ represents a coordinate vector of a specific Euler mesh and $\vec R$ represents a coordinate vector of the Lagrangian mesh. The $point1$, $point2$, $point3$, $point4$ and $\vec V_P$ are the positions specified in Fig. \ref{meshsearch}, and vectors composed of the position vectors are as follows: 
\begin{eqnarray*}
	\begin{cases}
		{\vec V}_{11}={\vec R}_{point2}-{\vec R}_{point1}\\
		{\vec V}_{12}={\vec R}_{point3}-{\vec R}_{point1}\\
		{\vec V}_{P1}={\vec P}-{\vec R}_{point1}
	\end{cases}
	\begin{cases}
		{\vec V}_{21}={\vec R}_{point3}-{\vec R}_{point1}\\
		{\vec V}_{22}={\vec R}_{point4}-{\vec R}_{point1}\\
		{\vec V}_{P2}={\vec P}-{\vec R}_{point1}
	\end{cases}\\
	\begin{cases}
		{\vec V}_{31}={\vec R}_{point4}-{\vec R}_{point1}\\
		{\vec V}_{32}={\vec R}_{point1}-{\vec R}_{point1}\\
		{\vec V}_{P3}={\vec P}-{\vec R}_{point1}
	\end{cases}
\begin{cases}
		{\vec V}_{41}={\vec R}_{point4}-{\vec R}_{point2}\\
		{\vec V}_{42}={\vec R}_{point3}-{\vec R}_{point2}\\
		{\vec V}_{P4}={\vec P}-{\vec R}_{point2}
	\end{cases}
\end{eqnarray*}

\begin{figure}[H]
	\centering
	\includegraphics[width=10cm]{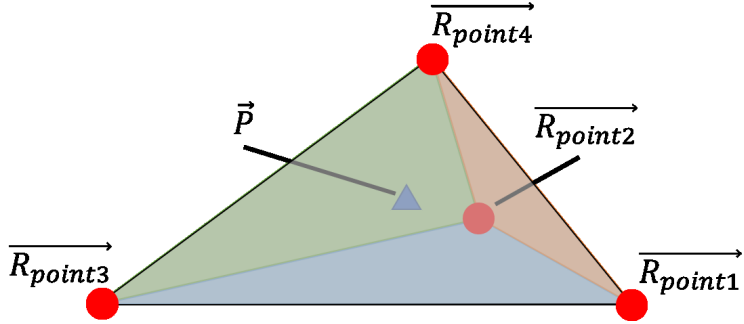}
	\caption{An Euler mesh point $P$ in a tetrahedron of the Lagrange mesh. }\label{meshsearch}
\end{figure}

If the point $P$ is in the triangular pyramid, the following conditions are met. 

\begin{eqnarray}
		\begin{cases}
			\left({\vec V}_{11}\times{\vec V}_{12}\right)\cdot{\vec V}_{P1}>0\\
			\left({\vec V}_{21}\times{\vec V}_{22}\right)\cdot{\vec V}_{P2}>0\\
			\left({\vec V}_{31}\times{\vec V}_{32}\right)\cdot{\vec V}_{P3}>0\\
			\left({\vec V}_{41}\times{\vec V}_{42}\right)\cdot{\vec V}_{P4}>0
		\end{cases}
\end{eqnarray}

\item{\bf output.cpp}
In this procedure the converted data is output. 
\item{\bf read\_variable.cpp}
This procedure reads the file data output by the Lagrange code, after the Lagrangian data set selection. 
\end{enumerate}


\subsection{Eulerian code}
\begin{enumerate}
\item {\bf BoundaryTracking.cpp}\\
It is a function to track the material boundary surfaces. Each boundary point is specified by the coordinates of the three variables: $( BoundaryMesh\_i$,  $BoundaryMesh\_j$, $BoundaryMesh\_k )$. The function interpolates the velocities $u$, $v$ and $w$ at the coordinates by the volume interpolation, and tracks the position of each boundary point. In Fig. \ref {BP} dotted lines represent the material boundaries. When the boundary point exists at the position shown in Fig. \ref {BMV}, the boundary point velocity ($u_b, \ v_b, \ w_b$) is calculated by the volume interpolation method and is obtained by the following equations:
	\begin{eqnarray}
		u_b&=&V_{u1}u_{i+1,j+1,k+1}+V_{u2}u_{i,j+1,k+1}+V_{u3}u_{i,j,k+1}+V_{u4}u_{i+1,j,k+1} \\ \nonumber 
		&+&V_{u5}u_{i+1,j+1,k}+V_{u6}u_{i,j+1,k}+V_{u7}u_{i,j,k}+V_{u8}u_{i+1,j,k}\\
		v_b&=&V_{v1}v_{i+1,j+1,k+1}+V_{v2}v_{i,j+1,k+1}+V_{v3}v_{i,j,k+1}+V_{v4}v_{i+1,j,k+1} \\ \nonumber
		&+&V_{v5}v_{i+1,j+1,k}+V_{v6}v_{i,j+1,k}+V_{v7}v_{i,j,k}+V_{v8}v_{i+1,j,k}\\
		w_b&=&V_{w1}w_{i+1,j+1,k+1}+V_{w2}w_{i,j+1,k+1}+V_{w3}w_{i,j,k+1}\\ \nonumber
	 &+&V_{w4}w_{i+1,j,k+1} +V_{w5}w_{i+1,j+1,k}+V_{w6}w_{i,j+1,k}+V_{w7}w_{i,j,k}\\ \nonumber
	 &+&V_{w8}w_{i+1,j,k}
	\end{eqnarray}

	\begin{figure}[H]
		\centering
		\includegraphics[height=7cm]{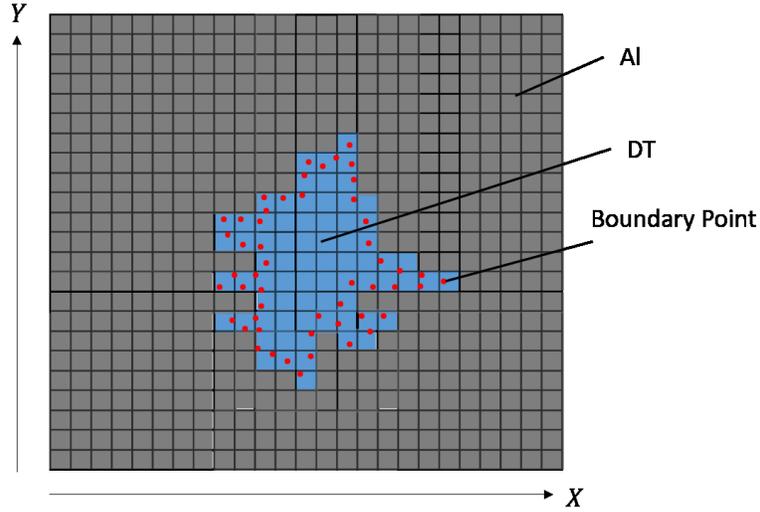}
		\caption{Material boundary points.}\label{BP}
	\end{figure}
	\begin{figure}[H]
		\centering
		\includegraphics[width=10cm]{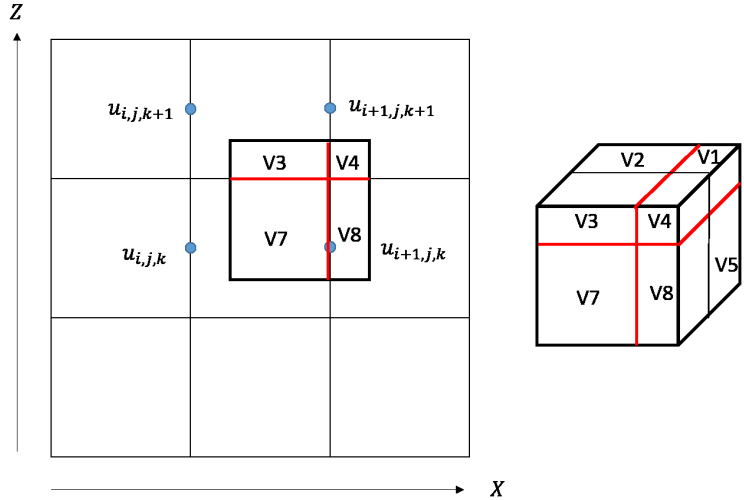}
		\caption{Velocity interpolation by the volume interpolation.}\label{BMV}
	\end{figure}
	
\item {\bf GenerateMatrix.cpp}\\
The mesh total numbers of $(im,\ jm,\ km)$ are loaded from the converted data in GenerateMatrix(). all the variables required  in the Euler code are defined based on the number $(im,\ jm,\ km)$. 

\item {\bf MS\_TDMA.cpp}\\
A function to solve matrixes by TDMA (TriDiagonal-Matrix Algorithm).
\item {\bf MaterialRecognation.cpp}\\
A function to discriminate each material by the material boundary lines. 
\item {\bf PaintMaterial.cpp\\}
The material is specified between the two material boundary lines in the procedure. 
\item {\bf RMFP\_ECSH.cpp}\\
A procedure to calculate the Rosseland mean free path (see Ref. \cite{Zeldovich}). 
\item {\bf ScanLine.cpp}\\
A procedure that specifies the material on each Euler mesh. 
\item {\bf artv\_ECSH.cpp}\\
This file contains a procedure to calculate the artificial viscosity. The three-dimensional artificial viscosity is written as follows: 
\begin{eqnarray}
q_x = \rho C^2_Q \left( \frac{\partial u}{\partial i} \right)^2 + \rho C_L C_s |\frac{\partial u}{\partial i}| \\
q_y = \rho C^2_Q \left( \frac{\partial v}{\partial j} \right)^2 + \rho C_L C_s |\frac{\partial v}{\partial j}| \\
q_z = \rho C^2_Q \left( \frac{\partial w}{\partial k} \right)^2 + \rho C_L C_s |\frac{\partial w}{\partial k}| 
\end{eqnarray}
Here, the discretized artificial viscosities are shown below: 
\small
\begin{eqnarray}
{q_x}^n_{i+\frac{1}{2},j+\frac{1}{2},k+\frac{1}{2}} &=& \rho^n_{i+\frac{1}{2},j+\frac{1}{2},k+\frac{1}{2}} C_Q^2 (u^n_{i+1,j+\frac{1}{2},k+\frac{1}{2}} - u^n_{i,j+\frac{1}{2},k+\frac{1}{2}})^2 \\\nonumber
&& + \rho^n_{i+\frac{1}{2},j+\frac{1}{2},k+\frac{1}{2}} C_L C_s \left| u^n_{i+1,j+\frac{1}{2},k+\frac{1}{2}}-u^n_{i,j+\frac{1}{2},k+\frac{1}{2}} \right| \\
{q_y}^n_{i+\frac{1}{2},j+\frac{1}{2},k+\frac{1}{2}} &=& \rho^n_{i+\frac{1}{2},j+\frac{1}{2},k+\frac{1}{2}} C_Q^2 (u^n_{i+\frac{1}{2},j+1,k+\frac{1}{2}} - u^n_{i+\frac{1}{2},j,k+\frac{1}{2}})^2 \\ \nonumber
&& + \rho^n_{i+\frac{1}{2},j+\frac{1}{2},k+\frac{1}{2}} C_L C_s \left| u^n_{i+\frac{1}{2},j+1,k+\frac{1}{2}}-u^n_{i+\frac{1}{2},j,k+\frac{1}{2}} \right| \\
{q_z}^n_{i+\frac{1}{2},j+\frac{1}{2},k+\frac{1}{2}} &=& \rho^n_{i+\frac{1}{2},j+\frac{1}{2},k+\frac{1}{2}} C_Q^2 (u^n_{i+\frac{1}{2},j+\frac{1}{2},k+1} - u^n_{i+\frac{1}{2},j+\frac{1}{2},k})^2 \\ \nonumber
&& + \rho^n_{i+\frac{1}{2},j+\frac{1}{2},k+\frac{1}{2}} C_L C_s \left| u^n_{i+\frac{1}{2},j+\frac{1}{2},k+1}-u^n_{i+\frac{1}{2},j+\frac{1}{2},k} \right| \\
\end{eqnarray}
\normalsize

\item {\bf define\_ECSH.h}\\
It contains the constant values, the normalization factors and the procedure declarations required. 
\item {\bf dif\_ECSH.cpp}\\
The following equations of motion are used.
\begin{eqnarray}
\label{dmeEu}
\frac{\partial \tilde{u}}{\partial \tilde{t}} &=& -  \left( \tilde{u} \frac{\partial \tilde{u}}{\partial \tilde{x}} + \tilde{v} \frac{\partial \tilde{u}}{\partial \tilde{y}} +\tilde{w} \frac{\partial \tilde{u}}{\partial \tilde{z}} \right) -  \frac{1}{\tilde{\rho}}  \frac{\partial \left(\tilde{p} + \tilde{q}\right)}{\partial \tilde{x}}   \\
\frac{\partial \tilde{v}}{\partial \tilde{t}} &=& -  \left( \tilde{u} \frac{\partial \tilde{v}}{\partial \tilde{x}} + \tilde{v} \frac{\partial \tilde{v}}{\partial \tilde{y}} +\tilde{w} \frac{\partial \tilde{v}}{\partial \tilde{z}} \right) -  \frac{1}{\tilde{\rho}} \frac{\partial \left(\tilde{p} + \tilde{q}\right)}{\partial \tilde{y}}   \\
\frac{\partial \tilde{w}}{\partial \tilde{t}} &=& -  \left( \tilde{u} \frac{\partial \tilde{w}}{\partial \tilde{x}} + \tilde{v} \frac{\partial \tilde{w}}{\partial \tilde{y}} +\tilde{w} \frac{\partial \tilde{w}}{\partial \tilde{z}} \right) -  \frac{1}{\tilde{\rho}} \ \frac{\partial \left(\tilde{p} +  \tilde{q}\right)}{\partial \tilde{z}}   
\end{eqnarray}

Equations (\ref{dmeEu}) are discretized as follows: 

\begin{eqnarray}
u^{n+\frac{1}{2}}_{i,j+\frac{1}{2},k+\frac{1}{2}} &=& u^{n-\frac{1}{2}}_{i,j+\frac{1}{2},k+\frac{1}{2}} - Dt^n \Biggl[ \left\{ \left(u \frac{\partial u}{\partial x} \right)^{n-\frac{1}{2}}_{i,j+\frac{1}{2},k+\frac{1}{2}} + \left(v \frac{\partial u}{\partial y} \right)^{n-\frac{1}{2}}_{i,j+\frac{1}{2},k+\frac{1}{2}} \right. \\  \nonumber
&& \left.  + \left(w \frac{\partial u}{\partial z} \right)^{n-\frac{1}{2}}_{i,j+\frac{1}{2},k+\frac{1}{2}} \right\} + \left\{ \frac{1}{\rho} \frac{\partial (p+q)}{\partial x} \right\}^{n-\frac{1}{2}}_{i,j+\frac{1}{2},k+\frac{1}{2}} \Biggr] 
\end{eqnarray}
Here, 
\begin{eqnarray*}
&& u \frac{\partial u}{\partial x} \bigg|^{n-\frac{1}{2}}_{i,j+\frac{1}{2},k+\frac{1}{2}} \\
&& \ \ \ \ \ \  = \left\{ 
\begin{array}{ll}
u^{n-\frac{1}{2}}_{i,j+\frac{1}{2},k+\frac{1}{2}} \frac{u^{n-\frac{1}{2}}_{i,j+\frac{1}{2},k+\frac{1}{2}} - u^{n-\frac{1}{2}}_{i-1,j+\frac{1}{2},k+\frac{1}{2}}}{Dx^{n-\frac{1}{2}}_{i-\frac{1}{2},j+\frac{1}{2},k+\frac{1}{2}}} &\left( u^{n-\frac{1}{2}}_{i,j+\frac{1}{2},k+\frac{1}{2}} \geq 0 \right) \\
	u^{n-\frac{1}{2}}_{i,j+\frac{1}{2},k+\frac{1}{2}} \frac{u^{n-\frac{1}{2}}_{i+1,j+\frac{1}{2},k+\frac{1}{2}} - u^{n-\frac{1}{2}}_{i+1,j+\frac{1}{2},k+\frac{1}{2}}}{Dx^{n-\frac{1}{2}}_{i-\frac{1}{2},j+\frac{1}{2},k+\frac{1}{2}}} &\left( u^{n-\frac{1}{2}}_{i,j+\frac{1}{2},k+\frac{1}{2}} <0 \right)
\end{array} 
\right. \\
&& v \frac{\partial u}{\partial y} \bigg|^{n-\frac{1}{2}}_{i,j+\frac{1}{2},k+\frac{1}{2}} \\
&& \ \ \ \ \ \  = \left\{ 
\begin{array}{ll}
	v^{n-\frac{1}{2}}_{i,j+\frac{1}{2},k+\frac{1}{2}} \frac{u^{n-\frac{1}{2}}_{i,j+\frac{1}{2},k+\frac{1}{2}} - u^{n-\frac{1}{2}}_{i,j-\frac{1}{2},k+\frac{1}{2}}}{Dy^{n-\frac{1}{2}}_{i-\frac{1}{2},j+\frac{1}{2},k+\frac{1}{2}}} &\left( v^{n-\frac{1}{2}}_{i,j+\frac{1}{2},k+\frac{1}{2}} \geq 0 \right) \\
	v^{n-\frac{1}{2}}_{i,j+\frac{1}{2},k+\frac{1}{2}} \frac{u^{n-\frac{1}{2}}_{i,j+\frac{3}{2},k+\frac{1}{2}} + u^{n-\frac{1}{2}}_{i,j+\frac{1}{2},k+\frac{1}{2}}}{Dy^{n-\frac{1}{2}}_{i-\frac{1}{2},j+\frac{1}{2},k+\frac{1}{2}}} &\left( v^{n-\frac{1}{2}}_{i,j+\frac{1}{2},k+\frac{1}{2}} < 0 \right) 
\end{array} 
\right. \\
&&  w \frac{\partial u}{\partial z} \bigg|^{n-\frac{1}{2}}_{i,j+\frac{1}{2},k+\frac{1}{2}} \\
&& \ \ \ \ \ \  = \left\{ 
\begin{array}{ll}
	w^{n-\frac{1}{2}}_{i,j+\frac{1}{2},k+\frac{1}{2}} \frac{u^{n-\frac{1}{2}}_{i,j+\frac{1}{2},k+\frac{1}{2}} - u^{n-\frac{1}{2}}_{i,j+\frac{1}{2},k-\frac{1}{2}}}{Dz^{n-\frac{1}{2}}_{i-\frac{1}{2},j+\frac{1}{2},k+\frac{1}{2}}} &\left( w^{n-\frac{1}{2}}_{i,j+\frac{1}{2},k+\frac{1}{2}} \geq 0 \right) \\
	w^{n-\frac{1}{2}}_{i,j+\frac{1}{2},k+\frac{1}{2}} \frac{u^{n-\frac{1}{2}}_{i,j+\frac{1}{2},k+\frac{3}{2}} - u^{n-\frac{1}{2}}_{i,j+\frac{1}{2},k+\frac{1}{2}}}{Dz^{n^\frac{1}{2}}_{i-\frac{1}{2},j+\frac{1}{2},k+\frac{1}{2}}} &\left( w^{n-\frac{1}{2}}_{i,j+\frac{1}{2},k+\frac{1}{2}} < 0 \right) 
\end{array} 
\right. 
\end{eqnarray*}

\begin{eqnarray*}
 &&\frac{1}{\rho} \frac{\partial(p+q)}{\partial x}\bigg|^{n-\frac{1}{2}}_{i,j+\frac{1}{2},k+\frac{1}{2}} = \frac{2}{\rho^{n-\frac{1}{2}}_{i+\frac{1}{2},j+\frac{1}{2},k+\frac{1}{2}} + \rho^{n-\frac{1}{2}}_{i-\frac{1}{2},j+\frac{1}{2},k+\frac{1}{2}}}  \\
&& \left\{ \frac{p^{n-\frac{1}{2}}_{i+\frac{1}{2},j+\frac{1}{2},k+\frac{1}{2}} + q^{n-\frac{1}{2}}_{i+\frac{1}{2},j+\frac{1}{2},k+\frac{1}{2}} - \left(p^{n-\frac{1}{2}}_{i-\frac{1}{2},j+\frac{1}{2},k+\frac{1}{2}} + q^{n-\frac{1}{2}}_{i-\frac{1}{2},j+\frac{1}{2},k+\frac{1}{2}} \right) }{Dx^{n-\frac{1}{2}}_{i,j+\frac{1}{2},k+\frac{1}{2}}} \right\}
\end{eqnarray*}

\item {\bf eod\_ECSH.cpp}\\
The following continuity equation is used.
\begin{equation}
\label{eodE}
\frac{\partial \rho}{\partial t} = - \rho \left( \frac{\partial u}{\partial x} +\frac{\partial v}{\partial y} + \frac{\partial w}{\partial z} \right) - \left( u \frac{\partial \rho}{\partial x} + v \frac{\partial \rho}{\partial y} + w \frac{\partial \rho}{\partial z}\right)
\end{equation}
Equation (\ref{eodE}) is discretized as follows: 
\begin{eqnarray}
\rho^{n+1}_{i+\frac{1}{2},j+\frac{1}{2},k+\frac{1}{2}} &=& \rho^{n}_{i+\frac{1}{2},j+\frac{1}{2},k+\frac{1}{2}} - Dt^n \Biggl[ \rho^n_{i+\frac{1}{2}, j+\frac{1}{2}.k+\frac{1}{2}} \left\{ \frac{\partial u}{\partial x} \bigg|^n_{i+\frac{1}{2},j+\frac{1}{2},k+\frac{1}{2}} \right. \\ \nonumber
&& \left. + \frac{\partial v}{\partial y} \bigg|^n_{i+\frac{1}{2},j+\frac{1}{2},k+\frac{1}{2}}+ \frac{\partial w}{\partial z} \bigg|^n_{i+\frac{1}{2},j+\frac{1}{2},k+\frac{1}{2}} \right\} + \left\{  u \frac{\partial \rho}{\partial x} \bigg|^n_{i+\frac{1}{2},j+\frac{1}{2},k+\frac{1}{2}} \right. \\ \nonumber
&& \left. + v \frac{\partial \rho}{\partial y} \bigg|^n_{i+\frac{1}{2},j+\frac{1}{2},k+\frac{1}{2}} + w \frac{\partial \rho}{\partial z} \bigg|^n_{i+\frac{1}{2},j+\frac{1}{2},k+\frac{1}{2}} \right\} \Biggr] 
\end{eqnarray}
\begin{eqnarray*}
&&  \frac{\partial u}{\partial x} \bigg|^n_{i+\frac{1}{2},j+\frac{1}{2},k+\frac{1}{2}} = \frac{ u^n_{i+1,j+\frac{1}{2},k+\frac{1}{2}} - u^n_{i,j+\frac{1}{2},k+\frac{1}{2}}}{D x^n_{i+\frac{1}{2},j+\frac{1}{2},k+\frac{1}{2}}} \\
&& \frac{\partial v}{\partial y} \bigg|^n_{i+\frac{1}{2},j+\frac{1}{2},k+\frac{1}{2}} = \frac{ v^n_{i+\frac{1}{2},j+1,k+\frac{1}{2}} - v^n_{i+\frac{1}{2},j,k+\frac{1}{2}}}{D y^n_{i+\frac{1}{2},j+\frac{1}{2},k+\frac{1}{2}}} \\
&& \frac{\partial w}{\partial z} \bigg|^n_{i+\frac{1}{2},j+\frac{1}{2},k+\frac{1}{2}} = \frac{ w^n_{i+\frac{1}{2},j+\frac{1}{2},k+1} - w^n_{i+\frac{1}{2},j+\frac{1}{2},k}}{D z^n_{i+\frac{1}{2},j+\frac{1}{2},k+\frac{1}{2}}} \\
&& u \frac{\partial \rho}{\partial x} \bigg|^n_{i+\frac{1}{2},j+\frac{1}{2},k+\frac{1}{2}} \\
&& \ \ \ \  = \left\{ 
\begin{array}{ll}
	u^n_{i+\frac{1}{2},j+\frac{1}{2},k+\frac{1}{2}} \frac{\rho^n_{i+\frac{1}{2},j+\frac{1}{2},k+\frac{1}{2}} - \rho^n_{i-\frac{1}{2},j+\frac{1}{2},k+\frac{1}{2}}}{Dx^n_{i,j+\frac{1}{2},k+\frac{1}{2}}} &\left( u^n_{i+\frac{1}{2},j+\frac{1}{2},k+\frac{1}{2}} \geq 0 \right) \\
	u^n_{i+\frac{1}{2},j+\frac{1}{2},k+\frac{1}{2}} \frac{\rho^n_{i+\frac{3}{2},j+\frac{1}{2},k+\frac{1}{2}} - \rho^n_{i+\frac{1}{2},j+\frac{1}{2},k+\frac{1}{2}}}{Dx^n_{i+1, j+\frac{1}{2},k+\frac{1}{2}}} &\left( u^n_{i+\frac{1}{2},j+\frac{1}{2},k+\frac{1}{2}} <0 \right)
\end{array} 
\right. \\
&& v \frac{\partial \rho}{\partial y} \bigg|^n_{i+\frac{1}{2},j+\frac{1}{2},k+\frac{1}{2}} \\
&& \ \ \ \  = \left\{ 
\begin{array}{ll}
	v^n_{i+\frac{1}{2},j+\frac{1}{2},k+\frac{1}{2}} \frac{\rho^n_{i+\frac{1}{2},j+\frac{1}{2},k+\frac{1}{2}} - \rho^n_{i+\frac{1}{2},j-\frac{1}{2},k+\frac{1}{2}}}{Dy^n_{i+\frac{1}{2},j,k+\frac{1}{2}}} &\left( v^n_{i+\frac{1}{2},j+\frac{1}{2},k+\frac{1}{2}} \geq 0 \right) \\
	v^n_{i+\frac{1}{2},j+\frac{1}{2},k+\frac{1}{2}} \frac{\rho^n_{i+\frac{3}{2},j+\frac{1}{2},k+\frac{1}{2}} - \rho^n_{i+\frac{1}{2},j+\frac{1}{2},k+\frac{1}{2}}}{Dy^n_{i+\frac{1}{2}, j+1,k+\frac{1}{2}}} &\left( v^n_{i+\frac{1}{2},j+\frac{1}{2},k+\frac{1}{2}} <0 \right)
\end{array} 
\right. \\
&& w \frac{\partial \rho}{\partial z} \bigg|^n_{i+\frac{1}{2},j+\frac{1}{2},k+\frac{1}{2}} \\
&& \ \ \ \  = \left\{ 
\begin{array}{ll}
	w^n_{i+\frac{1}{2},j+\frac{1}{2},k+\frac{1}{2}} \frac{\rho^n_{i+\frac{1}{2},j+\frac{1}{2},k+\frac{1}{2}} - \rho^n_{i+\frac{1}{2},j+\frac{1}{2},k-\frac{1}{2}}}{Dz^n_{i+\frac{1}{2},j+\frac{1}{2},k}} &\left( w^n_{i+\frac{1}{2},j+\frac{1}{2},k+\frac{1}{2}} \geq 0 \right) \\
	w^n_{i+\frac{1}{2},j+\frac{1}{2},k+\frac{1}{2}} \frac{\rho^n_{i+\frac{1}{2},j+\frac{1}{2},k+\frac{3}{2}} - \rho^n_{i+\frac{1}{2},j+\frac{1}{2},k+\frac{1}{2}}}{Dz^n_{i+\frac{1}{2},j+\frac{1}{2},k+1}} &\left( w^n_{i+\frac{1}{2},j+\frac{1}{2},k+\frac{1}{2}} < 0 \right) \\
\end{array} 
\right.
\end{eqnarray*}
\item{\bf eoenergy\_ECSH}\\
The following basic energy equations are used.
\begin{eqnarray}
	\label{eq:EuEoE}
\frac{\partial T_i}{\partial t} &=& - \left( u \cdot \nabla \right) T_i -\frac{k_B}{C_{V_i}} \Biggl[ \left( \rho B_{T_i} + \frac{p_i +q}{\rho} \right) \left( \nabla \cdot u \right)  \Biggr] \\
\frac{\partial T_e}{\partial t} &=& - \left( u \cdot \nabla \right) T_e -\frac{k_B}{C_{V_e}} \Biggl[ \left( \rho B_{T_e} + \frac{p_e}{\rho} \right) \left( \nabla \cdot u \right)  \Biggr] \\
\frac{\partial T_r}{\partial t} &=& - \left( u \cdot \nabla \right) T_r -\frac{k_B}{C_{V_r}} \Biggl[ \left( \rho B_{T_r} + \frac{p_r}{\rho} \right) \left( \nabla \cdot u \right)  \Biggr]
\end{eqnarray}
Here, $B_{T_i} = 0$ in HIF. The discretized energy equation for the ion temperature, for example, becomes as follows: 
\begin{eqnarray}
{T_i}^{n+1}_{i+\frac{1}{2},j+\frac{1}{2},k+\frac{1}{2}} &=& {T_i}^n_{i+\frac{1}{2},j+\frac{1}{2},k+\frac{1}{2}} - Dt^n \Biggl[ \left\{  u \frac{\partial T_i}{\partial x} \bigg|^n_{i+\frac{1}{2},j+\frac{1}{2},k+\frac{1}{2}} \right. \\ \nonumber
&& \left. + v \frac{\partial T_i}{\partial y} \bigg|^n_{i+\frac{1}{2},j+\frac{1}{2},k+\frac{1}{2}} + w \frac{\partial T_i}{\partial z} \bigg|^n_{i+\frac{1}{2},j+\frac{1}{2},k+\frac{1}{2}} \right\} \\ \nonumber
&& + \frac{1}{{C_{V_i}}^n_{i+\frac{1}{2},j+\frac{1}{2},k+\frac{1}{2}}} \Biggl[ \frac{{p_i}^n_{i+\frac{1}{2},j+\frac{1}{2},k+\frac{1}{2}} + q^n_{i+\frac{1}{2},j+\frac{1}{2},k+\frac{1}{2}}}{\rho^n_{i+\frac{1}{2},j+\frac{1}{2}+k\frac{1}{2}}}  \\ \nonumber
&& \left\{ \frac{\partial u}{\partial x} \bigg|^n_{i+\frac{1}{2},j+\frac{1}{2},k+\frac{1}{2}} + \frac{\partial v}{\partial y} \bigg|^n_{i+\frac{1}{2},j+\frac{1}{2},k+\frac{1}{2}}  + \frac{\partial w}{\partial z} \bigg|^n_{i+\frac{1}{2},j+\frac{1}{2},k+\frac{1}{2}} \right\} \Biggr] \Biggr]
\end{eqnarray}
\begin{eqnarray*}
&& u \frac{\partial {T_i}}{\partial x} \bigg|^n_{i+\frac{1}{2},j+\frac{1}{2},k+\frac{1}{2}} \\
&& \ \ \ \  = \left\{ 
\begin{array}{ll}
	u^n_{i+\frac{1}{2},j+\frac{1}{2},k+\frac{1}{2}} \frac{{T_i}^n_{i+\frac{1}{2},j+\frac{1}{2},k+\frac{1}{2}} -  {T_i}^n_{i-\frac{1}{2},j+\frac{1}{2},k+\frac{1}{2}}}{Dx^n_{i,j+\frac{1}{2},k+\frac{1}{2}}} &\left( u^n_{i+\frac{1}{2},j+\frac{1}{2},k+\frac{1}{2}} \geq 0 \right) \\
	u^n_{i+\frac{1}{2},j+\frac{1}{2},k+\frac{1}{2}} \frac{{T_i}^n_{i+\frac{3}{2},j+\frac{1}{2},k+\frac{1}{2}} - {T_i}^n_{i+\frac{1}{2},j+\frac{1}{2},k+\frac{1}{2}}}{Dx^n_{i+1, j+\frac{1}{2},k+\frac{1}{2}}} &\left( u^n_{i+\frac{1}{2},j+\frac{1}{2},k+\frac{1}{2}} <0 \right)
\end{array} 
\right. \\
&& v \frac{\partial {T_i}}{\partial y} \bigg|^n_{i+\frac{1}{2},j+\frac{1}{2},k+\frac{1}{2}} \\
&& \ \ \ \  = \left\{ 
\begin{array}{ll}
	v^n_{i+\frac{1}{2},j+\frac{1}{2},k+\frac{1}{2}} \frac{{T_i}^n_{i+\frac{1}{2},j+\frac{1}{2},k+\frac{1}{2}} - {T_i}^n_{i+\frac{1}{2},j-\frac{1}{2},k+\frac{1}{2}}}{Dy^n_{i\frac{1}{2},j,k+\frac{1}{2}}} &\left( v^n_{i+\frac{1}{2},j+\frac{1}{2},k+\frac{1}{2}} \geq 0 \right) \\
	v^n_{i+\frac{1}{2},j+\frac{1}{2},k+\frac{1}{2}} \frac{{T_i}^n_{i+\frac{1}{2},j+\frac{3}{2},k+\frac{1}{2}} - {T_i}^n_{i+\frac{1}{2},j+\frac{1}{2},k+\frac{1}{2}}}{Dy^n_{i+\frac{1}{2}, j+1,k+\frac{1}{2}}} &\left( v^n_{i+\frac{1}{2},j+\frac{1}{2},k+\frac{1}{2}} <0 \right)
\end{array} 
\right. \\
&& w \frac{\partial {T_i}}{\partial z} \bigg|^n_{i+\frac{1}{2},j+\frac{1}{2},k+\frac{1}{2}} \\
&& \ \ \ \  = \left\{ 
\begin{array}{ll}
	w^n_{i+\frac{1}{2},j+\frac{1}{2},k+\frac{1}{2}} \frac{{T_i}^n_{i+\frac{1}{2},j+\frac{1}{2},k+\frac{1}{2}} - {T_i}^n_{i+\frac{1}{2},j+\frac{1}{2},k-\frac{1}{2}}}{Dz^n_{i\frac{1}{2},j+\frac{1}{2},k}} &\left( w^n_{i+\frac{1}{2},j+\frac{1}{2},k+\frac{1}{2}} \geq 0 \right) \\
	w^n_{i+\frac{1}{2},j+\frac{1}{2},k+\frac{1}{2}} \frac{{T_i}^n_{i+\frac{1}{2},j+\frac{1}{2},k+\frac{3}{2}} - {T_i}^n_{i+\frac{1}{2},j+\frac{1}{2},k+\frac{1}{2}}}{Dz^n_{i+\frac{1}{2}, j+\frac{1}{2},k+1}} &\left( w^n_{i+\frac{1}{2},j+\frac{1}{2},k+\frac{1}{2}} <0 \right)
\end{array} 
\right. \\
&& \frac{\partial u}{\partial x} \bigg|^n_{i+\frac{1}{2},j+\frac{1}{2},k+\frac{1}{2}} = \frac{u^n_{i+1,j+\frac{1}{2},k+\frac{1}{2}} - u^n_{i,j+\frac{1}{2},k+\frac{1}{2}}}{Dx^n_{i,j+\frac{1}{2},k+\frac{1}{2}}} \\
&& \frac{\partial v}{\partial y} \bigg|^n_{i+\frac{1}{2},j+\frac{1}{2},k+\frac{1}{2}} = \frac{v^n_{i+\frac{1}{2},j+1,k+\frac{1}{2}} - v^n_{i+\frac{1}{2},j,k+\frac{1}{2}}}{Dy^n_{i+\frac{1}{2},j,k+\frac{1}{2}}} \\
&& \frac{\partial w}{\partial z} \bigg|^n_{i+\frac{1}{2},j+\frac{1}{2},k+\frac{1}{2}} = \frac{w^n_{i+\frac{1}{2},j+\frac{1}{2},k+1} - w^n_{i+\frac{1}{2},j+\frac{1}{2},k}}{Dz^n_{i+\frac{1}{2},j+\frac{1}{2},k}}
\end{eqnarray*}
	
\item{\bf eos\_ECSH.cpp}\\
The same equation is used as the equation of state in the Lagrangian code.
\item{\bf fusion.cpp}\\
The fusion reactions are calculated in this procedure. The details are shown in Ref. \cite{CPC-O-SUKI}. The fusion reaction formulae for deuterium and tritium are shown below.
	\begin{eqnarray}
	\begin{split}
		\rm D+\rm D&\xrightarrow[50\%]{}\rm T(1.01{\rm MeV})+\rm p(3.02{\rm MeV})\\
		&\xrightarrow[50\%]{}{\rm He}^3(0.82{\rm MeV})+\rm n(2.45{\rm MeV})\\
		\rm D+\rm T&\rightarrow{\rm He}^4(3.5{\rm MeV})+\rm n(14.1{\rm MeV})
		\end{split}
		\label{NucEq}
	\end{eqnarray}
	D decreases due to the DD and DT reactions from the expression (\ref {NucEq}). The number density $n_{\rm D} $ change  is given bellow: 	
	\begin{eqnarray}
	\label{eq:ReacD}
		\frac{\partial n_{\rm D}}{\partial t}&=&-N_{\rm DD}-N_{\rm DT}\nonumber\\
							 &=&-\frac{1}{2}\langle\sigma v\rangle_{\rm DD}n_{\rm D}n_{\rm D}-\langle\sigma v\rangle_{\rm DT}n_{\rm D}n_{\rm T}
	\end{eqnarray}

Considering the diffusion term of $ \alpha $ particles and the term of $ \alpha $ particle absorption, $ n_\alpha $ is described as follows: 
		\begin{equation}
	\label{eq:alphaReac}
		\frac{\partial n_\alpha}{\partial t}=+\langle\sigma v\rangle_{\rm DT}n_{\rm D}n_{\rm T}-\bm\nabla\cdot\bm F-\omega_\alpha n_\alpha
	\end{equation}
	The discretized $\alpha$ particle reaction is written as: 
	\begin{equation}
		{n_\alpha}^{n+1}_{i+\frac{1}{2},j+\frac{1}{2},k+\frac{1}{2}}={n_\alpha}^{n}_{i+\frac{1}{2},j+\frac{1}{2},k+\frac{1}{2}}+\ \Delta t{n_{\rm D}}^{n}_{i+\frac{1}{2},j+\frac{1}{2},k+\frac{1}{2}}{n_{\rm T}}^{n}_{i+\frac{1}{2},j+\frac{1}{2},k+\frac{1}{2}}\langle\sigma v{\rangle_{\rm DT}}^{n}_{i+\frac{1}{2},j+\frac{1}{2},k+\frac{1}{2}}.
	\end{equation}
The D-D and the D-T reaction rates are shown in Refs. (\cite {CPC-O-SUKI, NRLpf}).	
The flux of the $\alpha$ particle is shown below $\mbox{\boldmath $F$}$.
\begin{eqnarray}
\mbox{\boldmath $F$}  = - D_\alpha \mbox{\boldmath $\nabla$} n_\alpha
\end{eqnarray}
Here $D_\alpha$ is the diffusion coefficient and is expressed by the following equation.
\begin{eqnarray}
\label{eq:alpha_kakusan}
D_\alpha = \frac{ \frac{1}{3} v_\alpha \lambda_\alpha}{1+ \frac{4}{3} \lambda_\alpha \frac{|\nabla n_\alpha|}{n_\alpha}}
\end{eqnarray}
Here $v_\alpha$ is the speed of $\alpha$ particle and $\lambda_\alpha$ the mean free path of $\alpha$. The second term of the denominator in Eq. (\ref{eq:alpha_kakusan}) expresses the flux limiting effect, which limits the excess flux by the steep gradient of the $\alpha$ density. The flux $\mbox{\boldmath $F$}$ of the $\alpha$ particles in the $x$, $y$ and $z$ directions are expressed by the following equations:  
\begin{eqnarray}
F_x = - \frac{ \frac{1}{3} n_\alpha v_\alpha \lambda_\alpha }{n_\alpha + \frac{4}{3} \lambda_\alpha \left| \frac{\partial n_\alpha}{\partial x} \right| } \frac{\partial n_\alpha}{\partial x} \\
F_y = - \frac{ \frac{1}{3} n_\alpha v_\alpha \lambda_\alpha }{n_\alpha + \frac{4}{3} \lambda_\alpha \left| \frac{\partial n_\alpha}{\partial y} \right| } \frac{\partial n_\alpha}{\partial y} \\
F_z = - \frac{ \frac{1}{3} n_\alpha v_\alpha \lambda_\alpha }{n_\alpha + \frac{4}{3} \lambda_\alpha \left| \frac{\partial n_\alpha}{\partial z} \right| } \frac{\partial n_\alpha}{\partial z} 
\end{eqnarray}
	
The energy increases by the $ \alpha $ particle energy deposition are shown below: 
	\begin{eqnarray}
	\label{NucEnIon}
		\Delta T_i=\frac{E_\alpha n_\alpha f_i}{\rho C_{v_i}}\\
	\label{NucEnEle}
		\Delta T_e=\frac{E_\alpha n_\alpha f_e}{\rho C_{v_e}}
	\end{eqnarray}
Here $f$ represents the distribution factor of the $\alpha $ particle energy among ions and electrons \cite{Fraley}. 

\begin{eqnarray}
f_i = \frac{1}{1+\frac{32}{T_e(KeV)}}, \hspace{1cm} f_e= 1-f_i
\end{eqnarray}

The discretized energy increases for ions and electrons are described as follows.

	\begin{eqnarray}
		{T_i}^{n+1}_{i+\frac{1}{2},j+\frac{1}{2},k+\frac{1}{2}}={T_i}^{n}_{i+\frac{1}{2},j+\frac{1}{2},k+\frac{1}{2}}+\Delta t\frac{E_\alpha {n_\alpha}^n_{i+\frac{1}{2},j+\frac{1}{2},k+\frac{1}{2}}{f_i}^n_{i+\frac{1}{2},j+\frac{1}{2},k+\frac{1}{2}}}{\rho^n_{i+\frac{1}{2},j+\frac{1}{2},k+\frac{1}{2}}{C_{v_i}}^n_{i+\frac{1}{2},j+\frac{1}{2},k+\frac{1}{2}}}\\
		{T_e}^{n+1}_{i+\frac{1}{2},j+\frac{1}{2},k+\frac{1}{2}}={T_e}^{n}_{i+\frac{1}{2},j+\frac{1}{2},k+\frac{1}{2}}+\Delta t\frac{E_\alpha {n_\alpha}^n_{i+\frac{1}{2},j+\frac{1}{2},k+\frac{1}{2}}{f_e}^n_{i+\frac{1}{2},j+\frac{1}{2},k+\frac{1}{2}}}{\rho^n_{i+\frac{1}{2},j+\frac{1}{2},k+\frac{1}{2}}{C_{v_e}}^n_{i+\frac{1}{2},j+\frac{1}{2},k+\frac{1}{2}}}
	\end{eqnarray} 
\item{\bf init\_ECSH.cpp}\\
The file initializes the Eulerian code.
\item{\bf load\_convert.cpp}\\
A procedure to read the converted data.
\item{\bf main\_ECSH.cpp}\\
The main function of the Eulerian code.
\item {\bf output\_ECSH.cpp}\\
The results are stored in this procedure.

\end{enumerate}

\section{Shell script files for setup and postprocessing}
A Shell file is prepared for the integrated run throughout from the Lagrange, conversion and Euler codes. However, each code can be also run manually one by one.
After finishing all the simulation process, users may need to visualize the simulation data. Some of the data computed are visualized by the following shell scripts. All shell files require gnuplot 4.6 or later.

\subsection{Calculation set up shell}
\begin{enumerate}
\item{\bf setup\_fusion.h}
The shell file remove the calculation output date and makes the output file.
\end{enumerate}

\subsection{Visualization for the Lagrange code data}
All the visualized data images are stored in the "pic\_La" directory. 
\begin{enumerate}
\item{\bf adiabat.sh}\\
The visualized graph for the time history of the adiabat $\alpha$ calculated in "Insulation.cpp" in the Lagrangian code. 

\item{\bf Animation\_Ti\_MODE.sh}\\
The shell file visualizes the mode analysis results of the ion temperature calculated by "Legendre.cpp" in the Lagrangian code. 

\item{\bf ImplosionVelocity.sh}\\
The shell plots the time histories of the implosion speed averaged over the azimuthal direction for the DT inner surface, the DT outer surface and the averaged DT speed. 
\item{\bf RMSoutput.sh}\\
The shell file plots the time histories of the root-mean-square (RMS) for the ion temperature and the mass density in the DT layer and Al layer. The RMS data is calculated by "RMS.cpp" in the Lagrangian code. 
\item{\bf SLC\_t\_r.sh}\\
The shell file outputs the images of the $r-t$ diagrams representing the time history of the Lagrangian meshes at $\theta =$30, 60, 120 and 150 degrees and at $\phi =$10, 100, 190 and 280 degrees. To execute the shell file, users need to specify the boundary mesh number of each material in the Lagrangian code. 
\end{enumerate}

\subsection{Visualization for the Euler code data}
All visualized data files are stored in the "pic\_Eu" directory. 
\begin{enumerate}
\item{\bf Animation\_atomic\_XY.sh}\\
The shell file visualizes the distributions of the atomic number on the XY plane for each output data in the Euler code. 
\item{\bf Animation\_atomic\_YZ.sh}\\
The shell file visualizes the distributions of the atomic number on the YZ plane for each output data in the Euler code. 
\item{\bf Animation\_rho\_XY.sh}\\
The shell file visualizes the distributions of the mass density on the XY plane for each output data in the Euler code.  
\item{\bf Animation\_rho\_YZ.sh}\\
The shell file visualizes the distributions of the mass density on the YZ plane for each output data in the Euler code.  
\item{\bf Animation\_Ti\_XY.sh}\\
The shell file visualizes the distributions of the ion temperature on the XY plane for each output data in the Euler code. 
\item{\bf Animation\_Ti\_YZ.sh}\\
The shell file visualizes the distributions of the ion temperature on the YZ plane for each output data in the Euler code. 
\item{\bf Fusiongain.sh}\\
The shell file plots the history of the fusion energy gain. 
\item{\bf rhoR.sh}\\
The shell file plots the history of the $\rho R$. 
\end{enumerate}

\section{Instructions for the user}\par
Before running the O-SUKI code, the user must set the target pellet and HIB parameters accordingly as follows:   \\
\begin{small}
{(a)\em OK3 code calculation type:} In 3D O-SUKI-N code, one can select the OK3 illumination code calculation type. The $OK\_Swich = 1$ is the full calculation with OK3. The $OK\_Swich = 5$ is the 1D uniform energy distribution type, and the HIB's energy distribution changes only in the radius direction. The $OK\_Swich = 10$ is the 1D energy distribution with the illumination non-uniformity in the $\theta$ and $\phi$ directions. One can add artificially non-uniformity in the $\theta$ and $\phi$ directions in ''main\_LC.cpp''\\
{(b)\em Projectile ion type:} Five projectile ion types are included in OK3—Pb, U, Cs, C and p. Users can choose one of them or add other species expanding the arrays aZb and aAb in "Input\_LC.h".\\
{(c)\em Ion beam parameters:} The user can specify the HIB radii on the target surface changing the parameter $tdbrc$ in "input\_LC.h". The design of the beam input pulse is also done in the same file. The pulse rise start time, rise time, and beam power are set by variables $t\_beamj$, $del\_t\_beamj$, and $Powerj (j=1\sim5)$, respectively. Users should also input the total input beam energy into $input\_energy$ in the "define\_ECSH.h" manually, when the users want to run the Euler code independetly. As the parameter value of the wobbling beam, the maximum radius of the beam axis trajectory in the rotation and the oscillating frequency should be specified. Users can set the desirable values for the maximum beam trajectory radius $rRot$ in the "InputOK3.h" and the rotational number $rotationnumber$ in the "input\_LC.h".\\
{(d)\em The beam irradiation position:} The file HIFScheme.h contains 1, 2, 3, 6, 12, 20, 32, 60 and 120-beam irradiation schemes. Users can choose one of them or add other HIB irradiation schemes supplementing the file.  \\
{(e)\em The reactor chamber:} Users can specify the chamber radius by changing the parameter of $Rch$. The parameter $dz$ fixes the pellet displacement from the reactor chamber center in the Cartesian PS coordinates (see Fig. \ref{misalign}). In OK3 the target alignment errors of $dx, dy$ and $dz$ can be specified. One can change this setting in the "input\_LC.h".\\
{(f)\em The target pellet structure and mesh number:} The parameter values of target are set in "input\_LC.h" and "init\_LC.h". In "input\_LC.h", users can change the boundary radius of each layer, the total mesh number and the mesh number for each layer. The present O-SUKI-N 3D includes an example DT-Al-Pb structure target defined by target layer-thickness parameters: $Rin, Rbc1, Rbc2$ and $Rout$. Users can add other target materials by expanding the arrays of $aZt0, aZtm, aAt, aUi, aro$ and $SC$ in "InputOK3.h". \\
{(g)\em The maximal Euler mesh number:} The maximal Euler mesh number is set in ''input\_LC.h''. The upper limit of the mesh number should be defined depending on the resource limitation of the workstation used.

\begin{figure}[H]
\centering
\includegraphics[width=5cm]{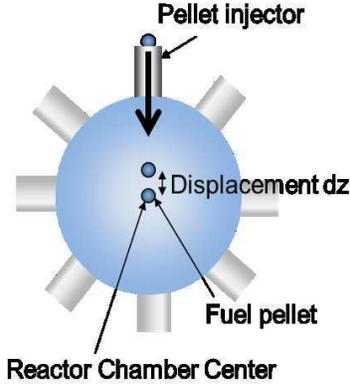}
\caption{Schematic diagram for the target misalignment}\label{misalign}
\end{figure}

If users want to employ a new substance for target structure, usera also need to add the solid density and the atomic mass in "CONSTANT.h". Users can control the Lagrangian radial mesh number for each layer by changing the value $MWC$ in "Input\_LC.h". When $MWC$ = 0, the radial mesh width ($dR1 = dR2 = \cdots$) in all layers becomes equal. When the MWC is large, the radial mesh number for each layer ($num\_k1 = num\_k2 = \cdots$) becomes close to the same number.\\
\end{small}
\par
Users can run "CodeO-SUKI-N-fusion-start.sh" to start running the O-SUKI-N 3D code simulations. When the shell script is executed, the Lagrange fluid code, the data conversion code and the Euler fluid code are sequentially activated. The results of the Lagrangian simulation are saved in the "output" directory, and the results of the Eulerian simulation are saved in "output\_euler". 

\section{Testing the program O-SUKI-N 3D}
The several tests are shown below to present the target fuel implosion dynamics. In the example cases, the HIBs and the target fuel have the following common parameters, which are the same values employed in Ref. \cite{CPC-O-SUKI}: the beam radius at the entrance of a reactor chamber $R_{en}$ = 35 mm, the beam particle density distribution is in the Gaussian profile and all projectile Pb ions have 8 GeV. The target is a multilayered pellet, in which the pellet outer radius is 4 mm, a Pb layer thickness is 0.029 mm, the Al thickness is 0.460 mm, and the DT thickness is 0.083 mm; the Pb, Al and DT layers have the radial mesh numbers of 4, 46 and 30 in these example cases, respectively, and the total mesh number in the theta direction is 90. The input beam pulse is shown in Fig. 12 in Ref. \cite{CPC-O-SUKI}. The beam radius is 3.8mm on the target surface. However, $R_b$ = 3.8mm changes at $\tau_{wb}$ to 3.7mm for the wobbling beam irradiation. Here $\tau_{wb}$ is the rotational period of the beam axis. The rotational frequency is 424MHz ($rotaionnumber$ = 11).


First the 3D Langrange code was run without the OK3 illumination code. This is the case for $OK\_Switch=10$, and we added the artificial non-uniformity $Y_3^2$ (the spherical harmonics) with the amplitude of $30.0\%$. In Fig. \ref{NoOK3_23_Ti} the ion temperature distribution is shown at $t$=35ns, and in Fig. \ref{NoOK3_23_rho} the mass density distribution is presented at $t$=35ns. The target shape is largely distorted due to the non-uniformity of the HIBs deposition energy distribution.

\begin{figure}[H]
		\centering
		\includegraphics[width=10cm]{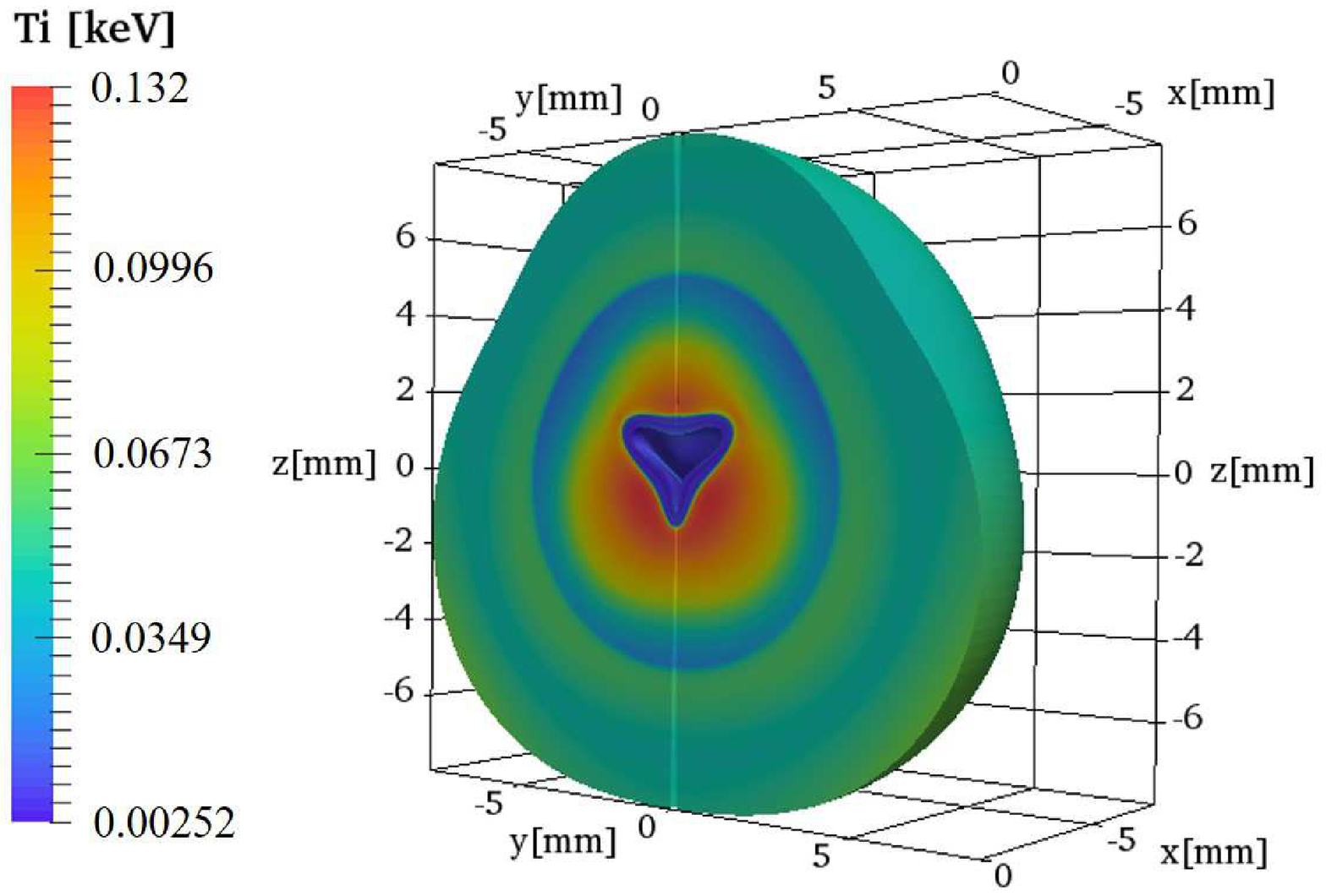}
		\caption{Ion temperature in the 3D Lagrange code without OK3 code at $t$=35ns. The non-uniformity distiribution is $Y_3^2$ with the amplitude of $30\%$.}\label{NoOK3_23_Ti}
\end{figure}
\begin{figure}[H]
		\centering
		\includegraphics[width=10cm]{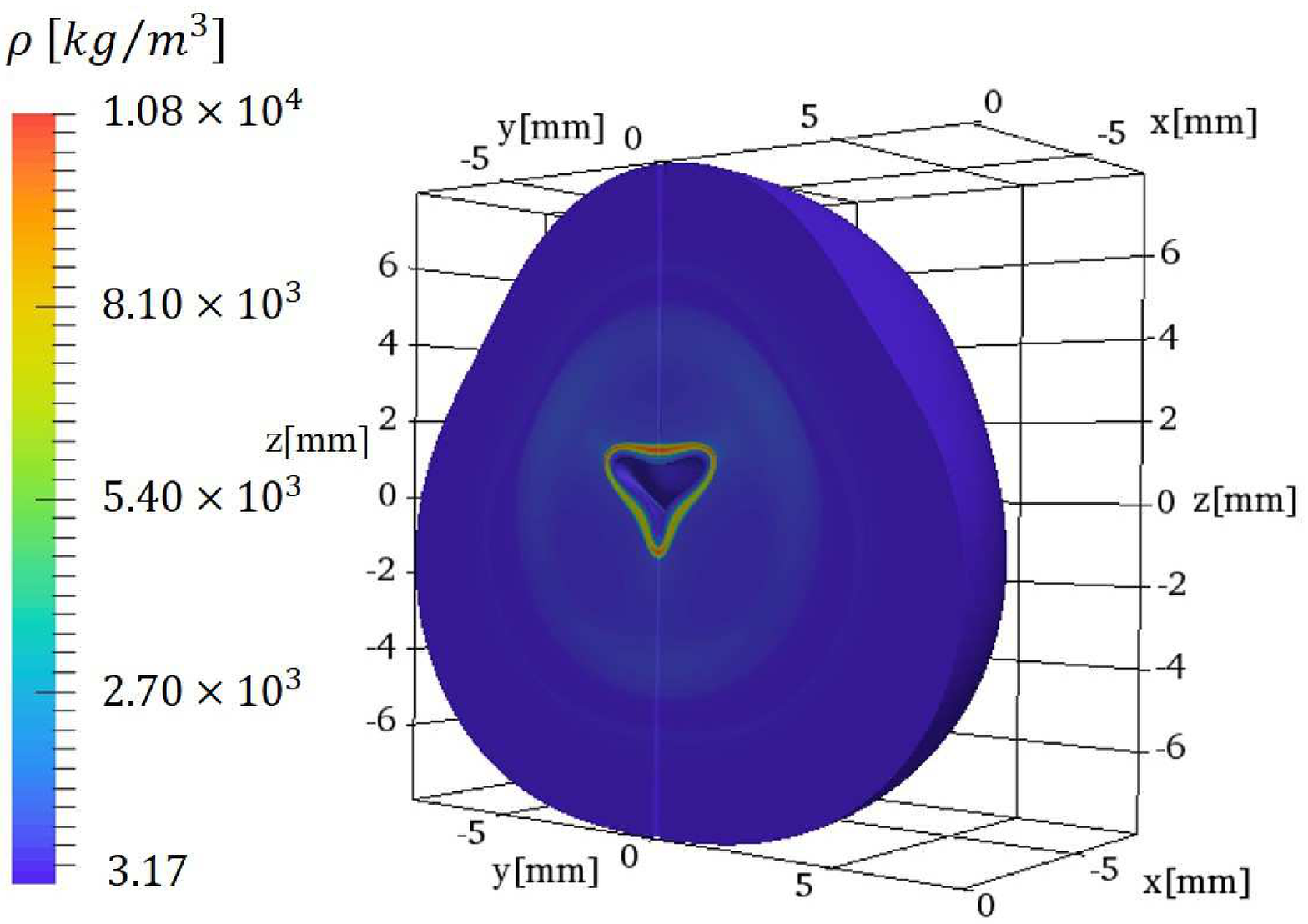}
		\caption{Mass density in the 3D Lagrange code without OK3 code at $t$=35ns. The non-uniformity distriution is $Y_3^2$ with the amplitude of $30\%$.}\label{NoOK3_23_rho}
\end{figure}

We also performed run-through simulation tests. In the example cases, the OK3 code was coupled with the run-through simulations. The implosion data were obtained by the Lagrange code, and the data just before the void closure time were transferred to the Euler code through the data Conversion code. Two cases are computed for the target fuel implosion dynamics with the spiral wobbling or without the oscillating HIBs. These examples are the run-through simulations with the OK3 illumination code ($OK\_Switch = 1$). The input beam pulse, employed in the run-through tests, is shown in Fig. \ref {Beam}. This beam input energy is 5.4MJ. We show the $r-t$ diagram for the case without the HIBs wobbling in Fig. \ref{rt}. The Lagrange-code test results stored in the output directory are visualized in Figs. \ref {fusion_Ti} for the target ion temperature ($T_i$) distributions at $t$ = 0.0, 15.0, 20.0 and 22.5 ns for the case with the HIBs wobbling behavior.  The RMS non-uniformity results are shown in Figs. \ref{fusion_RMS} (a) for DT layer's Ion temperature($T_i$), (b) for DT layer's Mass density($\rho$), (c) for Al layer's Ion temperature($T_i$) and (d) for Al layer's Mass density($\rho$). 
When the HIBs have the wobbling motion during the implosion with the wobbling frequency of 424MHz, the radius acceleration distributions are shown in Figs. \ref{Vr_tp} (a) in the $\theta$ direction and (b) in the $\phi$ direction at $t=6.25t_w=11.2ns$ (solid lines) and at $t=6.75t_w=12.2ns$ (dotted lines). Here $t_w$ shows the one rotation time. Figures \ref{Vr_tp} present that the non-uniformity phase of the implosion acceleration is controlled externally by the HIBs wobbling behavior \cite{CPC-O-SUKI, RSato2}.  

\begin{figure}[H]
		\centering
		\includegraphics[width=7.5cm]{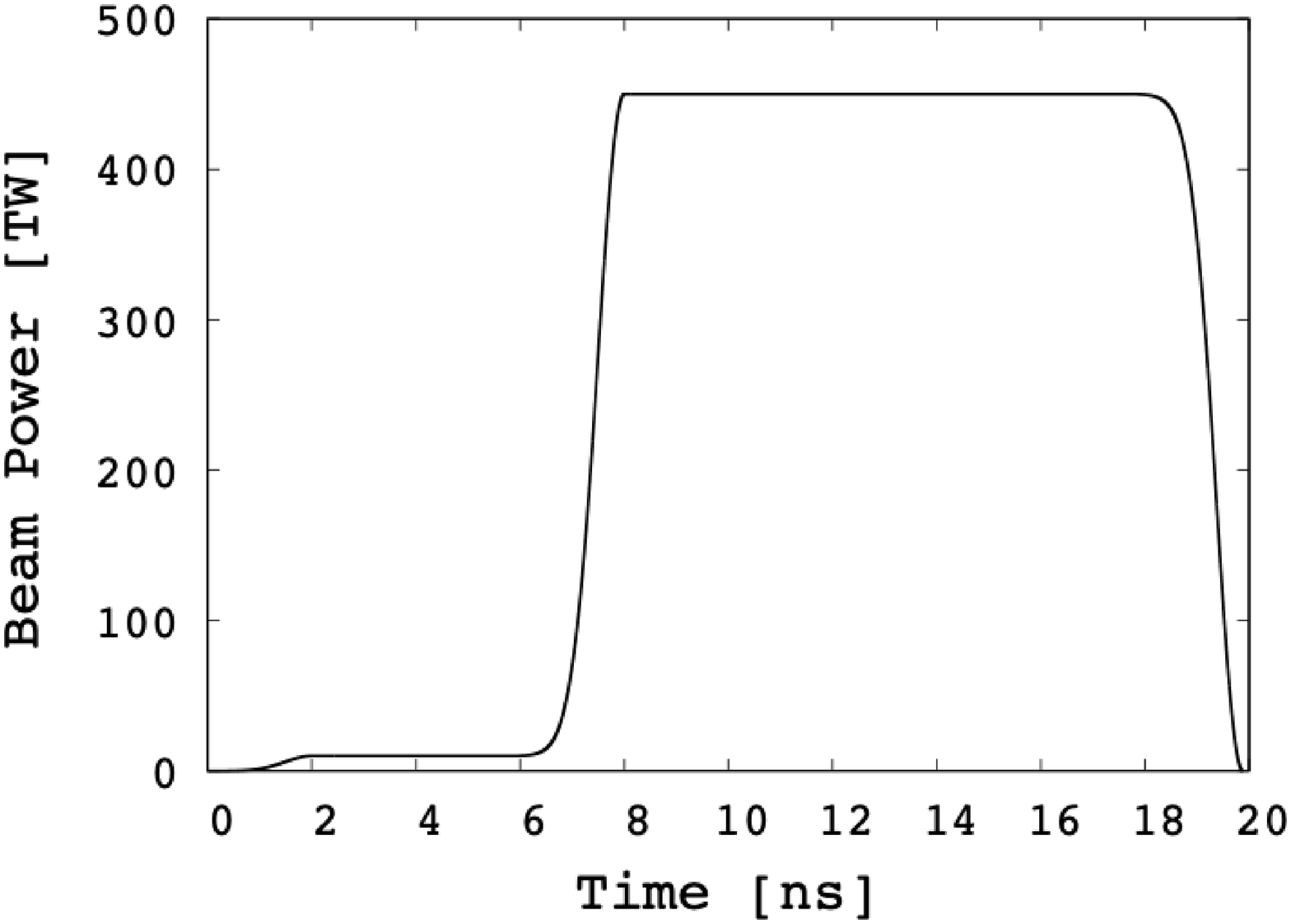}
		\caption{Input beam pulse shape used in the example run-through tests.}\label{Beam}
\end{figure}
\begin{figure}[H]
		\centering
		\includegraphics[width=8cm]{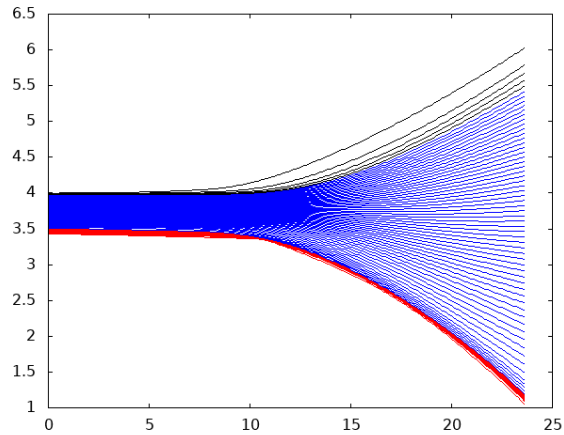}
		\caption{The $r-t$ diagram for the implosion with the HIBs wobbling illumination. The black line area shows the Pb layer, the blue line area Al and the red line area is DT.}\label{rt}
\end{figure}
\begin{figure}[H]
		\centering
		\includegraphics[width=6.5cm]{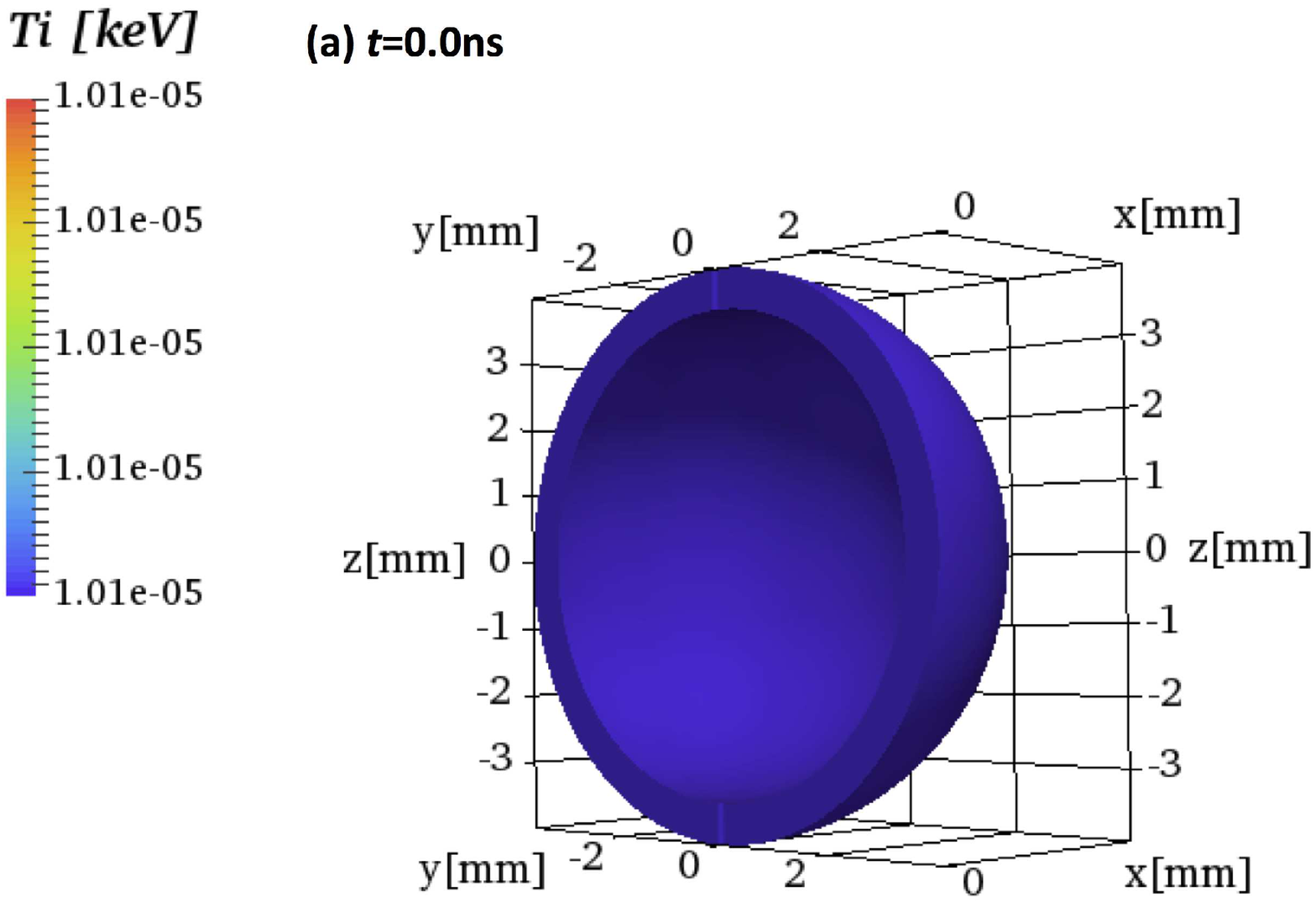}
		\includegraphics[width=6.5cm]{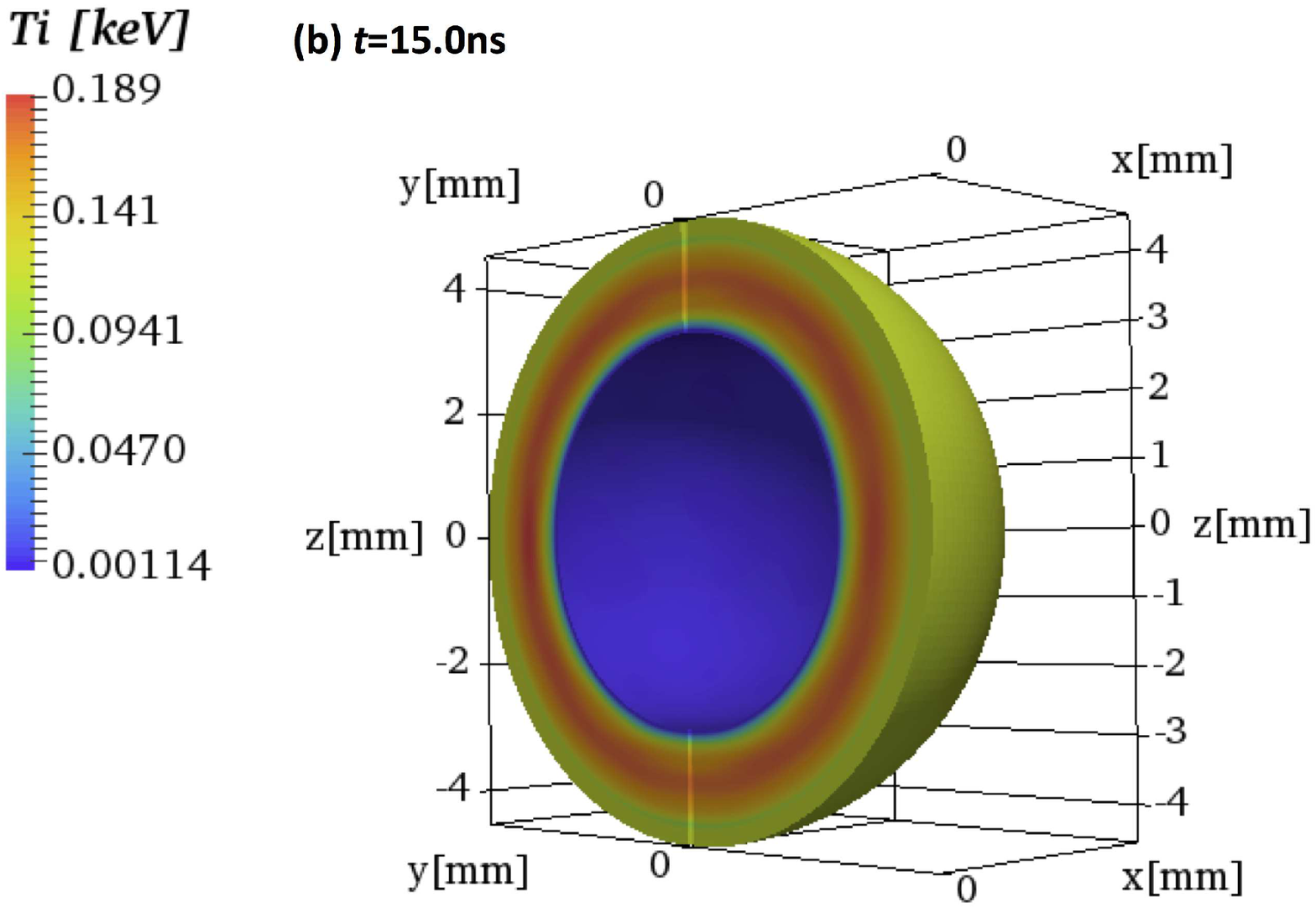}\\
		\includegraphics[width=6.5cm]{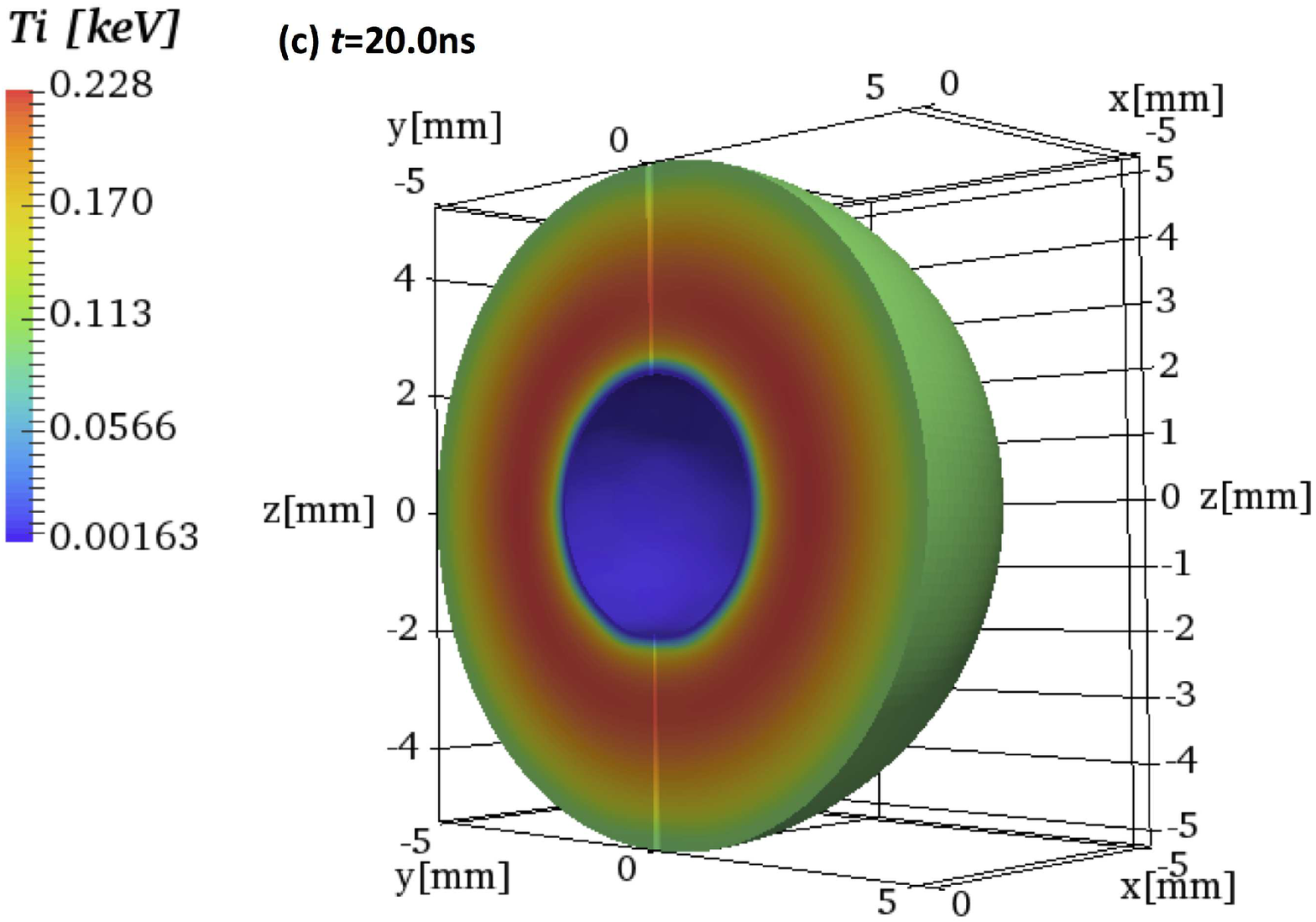}
		\includegraphics[width=6.5cm]{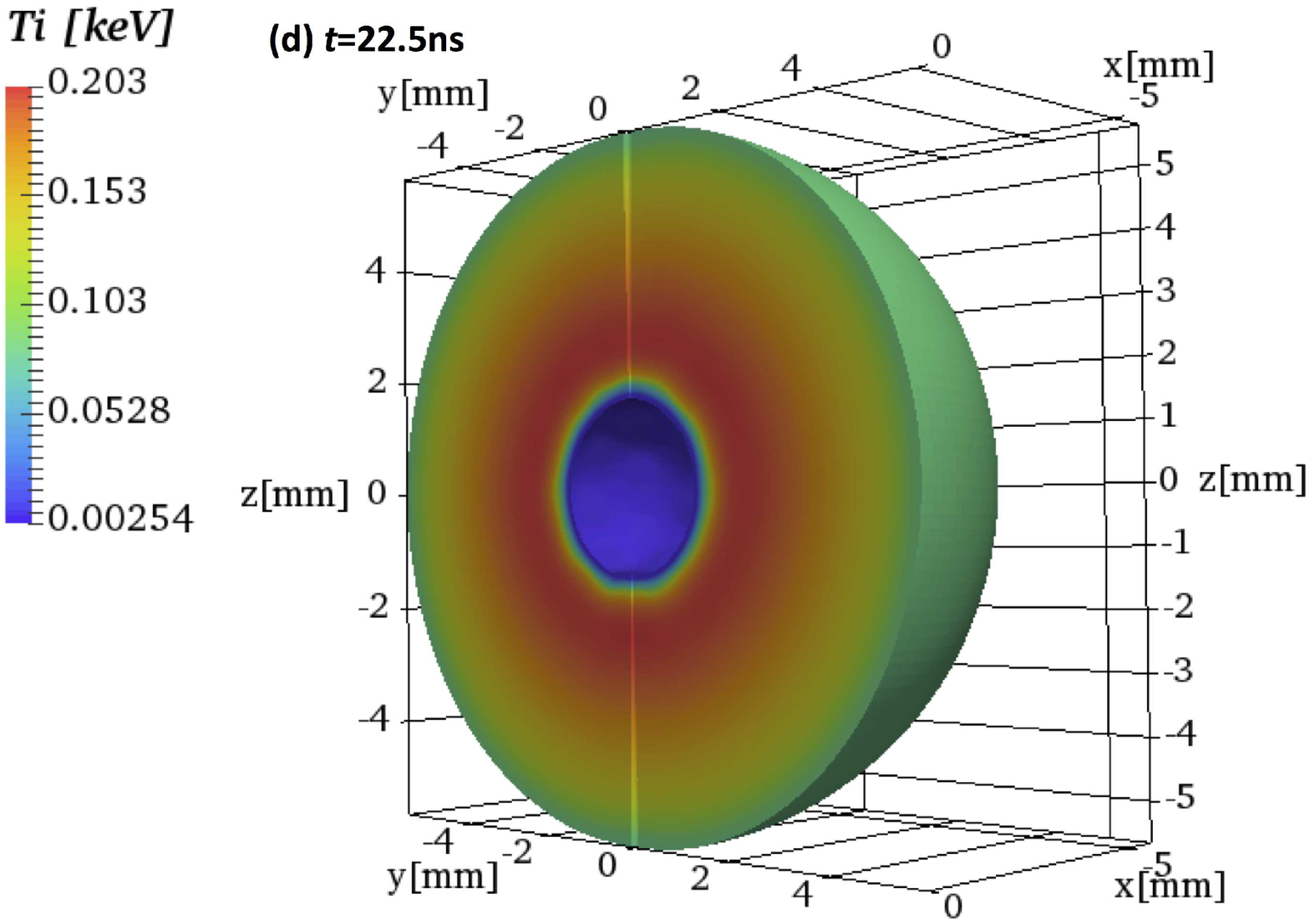}\\
		\caption{Ion temperature distributions in the example run-through test with the HIBs wobbling illumination at (a) $t$=0.0ns, (b) 15.0ns, (c) 20.0ns and (d) 22.5ns.}\label{fusion_Ti}
\end{figure}
\begin{figure}[H]
		\centering
		\includegraphics[width=6.5cm]{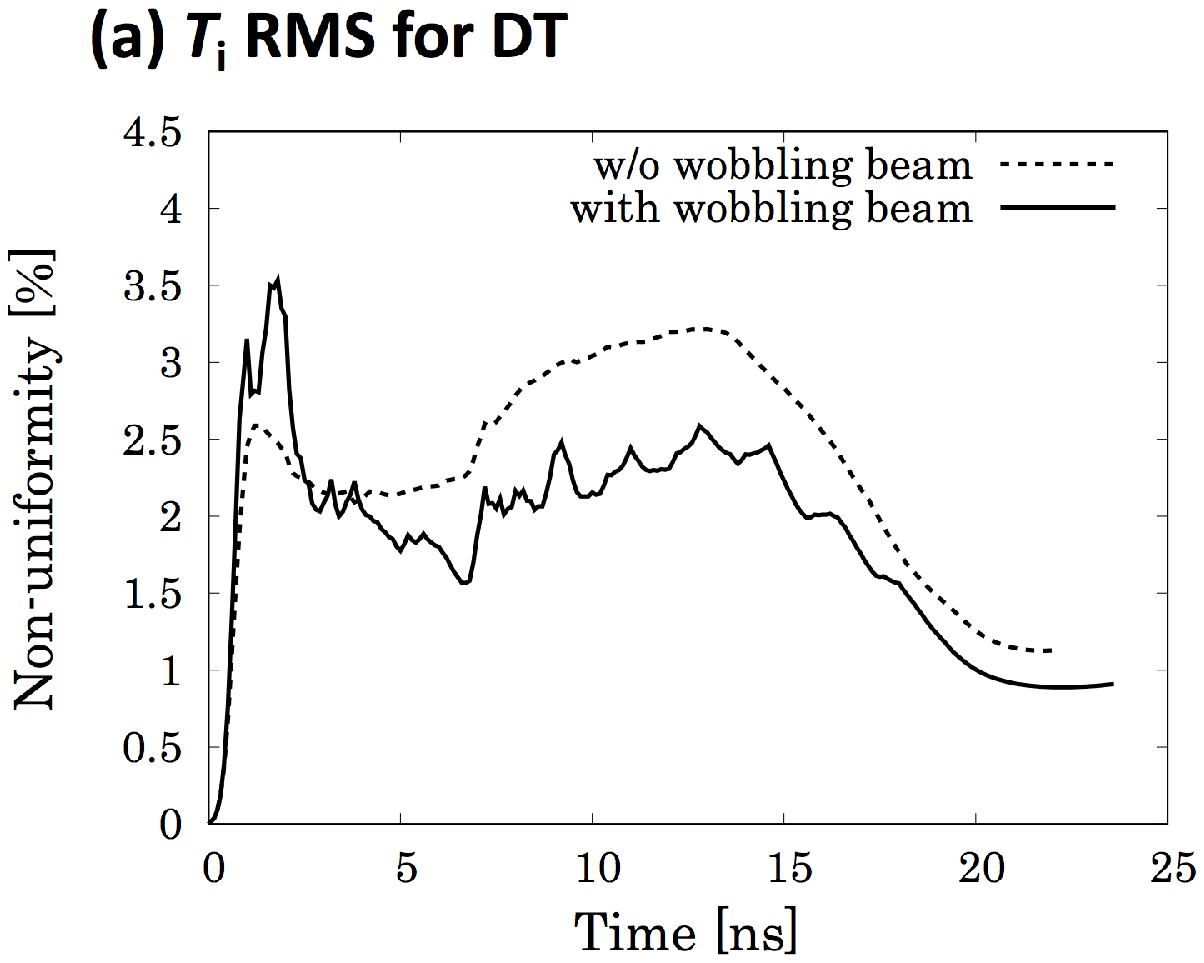}
		\includegraphics[width=6.5cm]{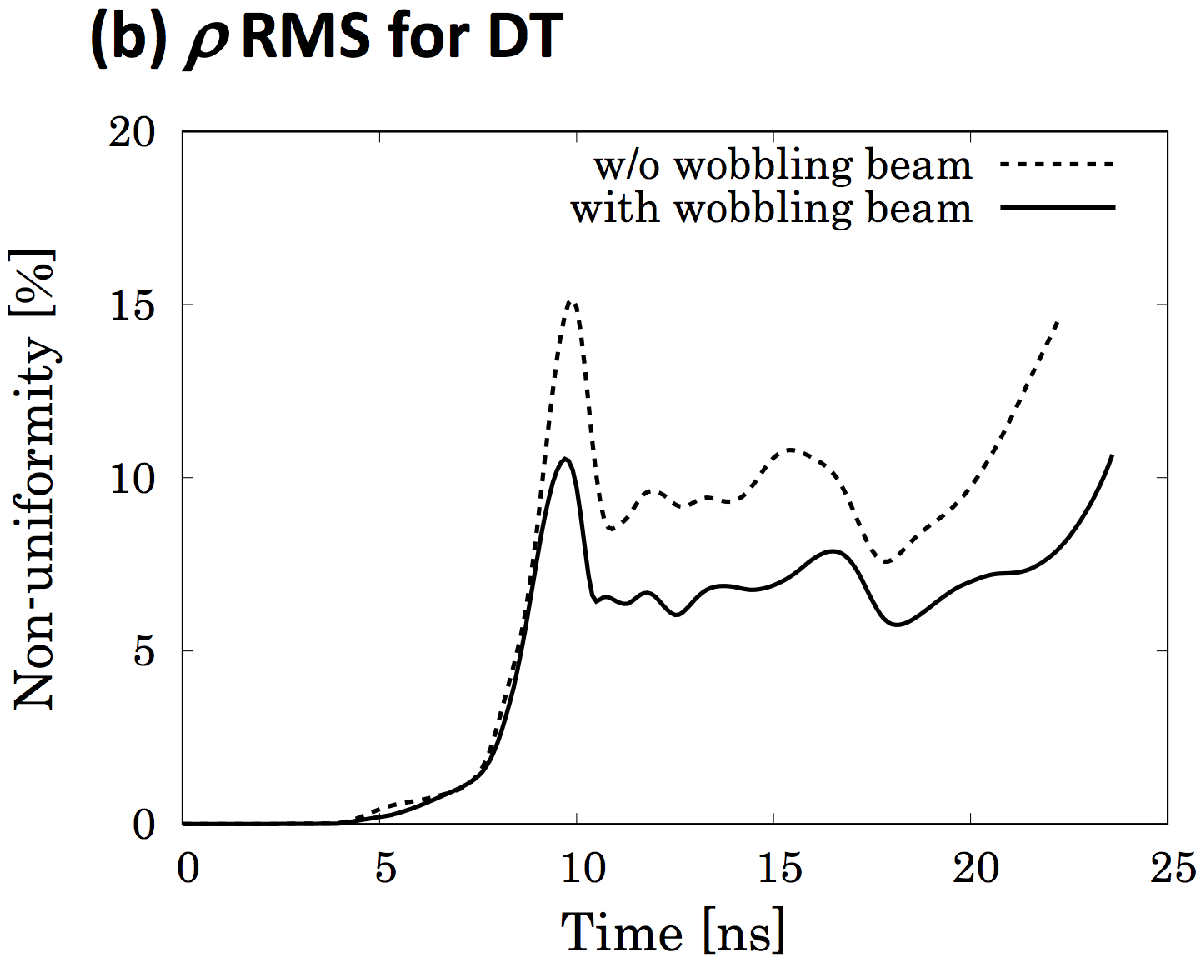}\\
		\includegraphics[width=6.5cm]{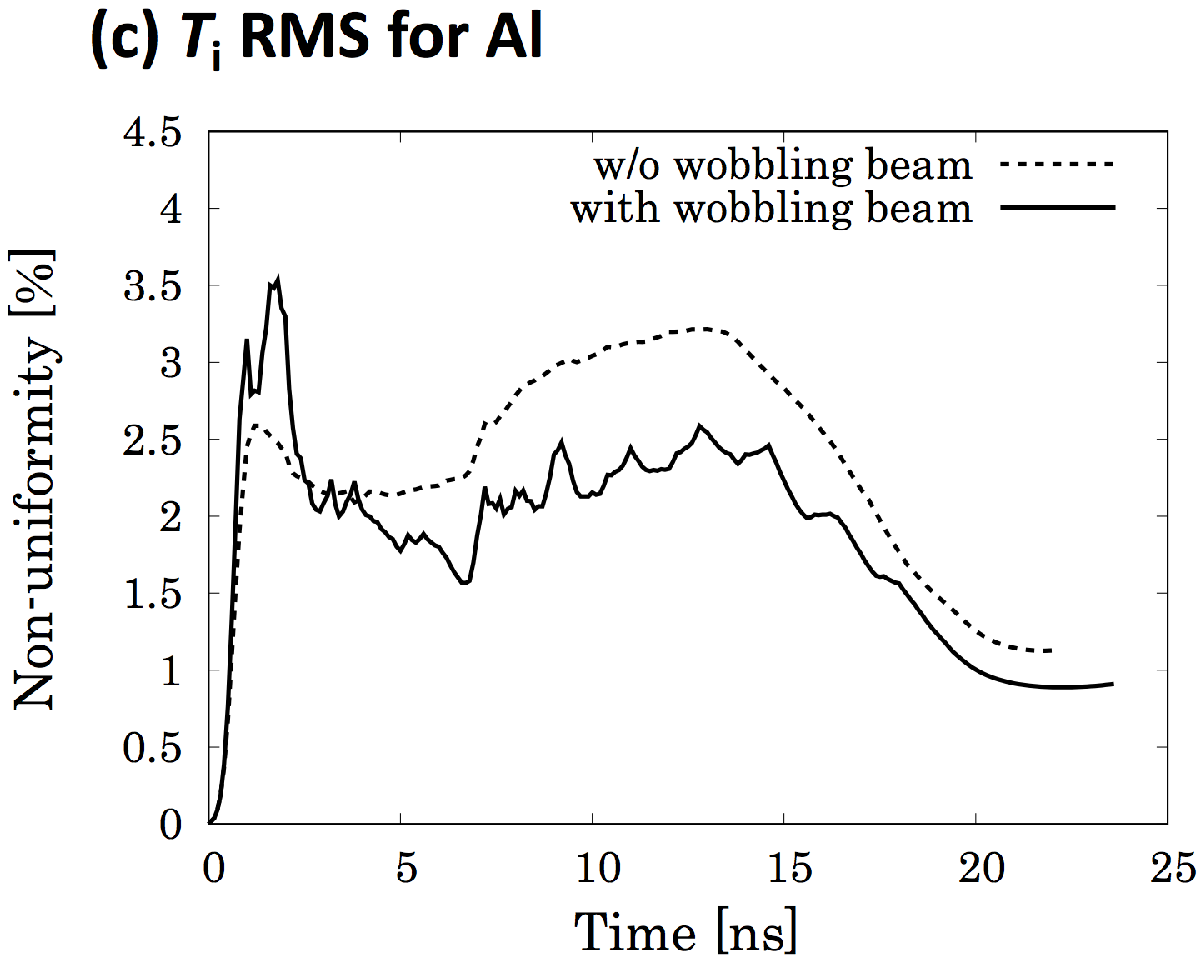}
		\includegraphics[width=6.5cm]{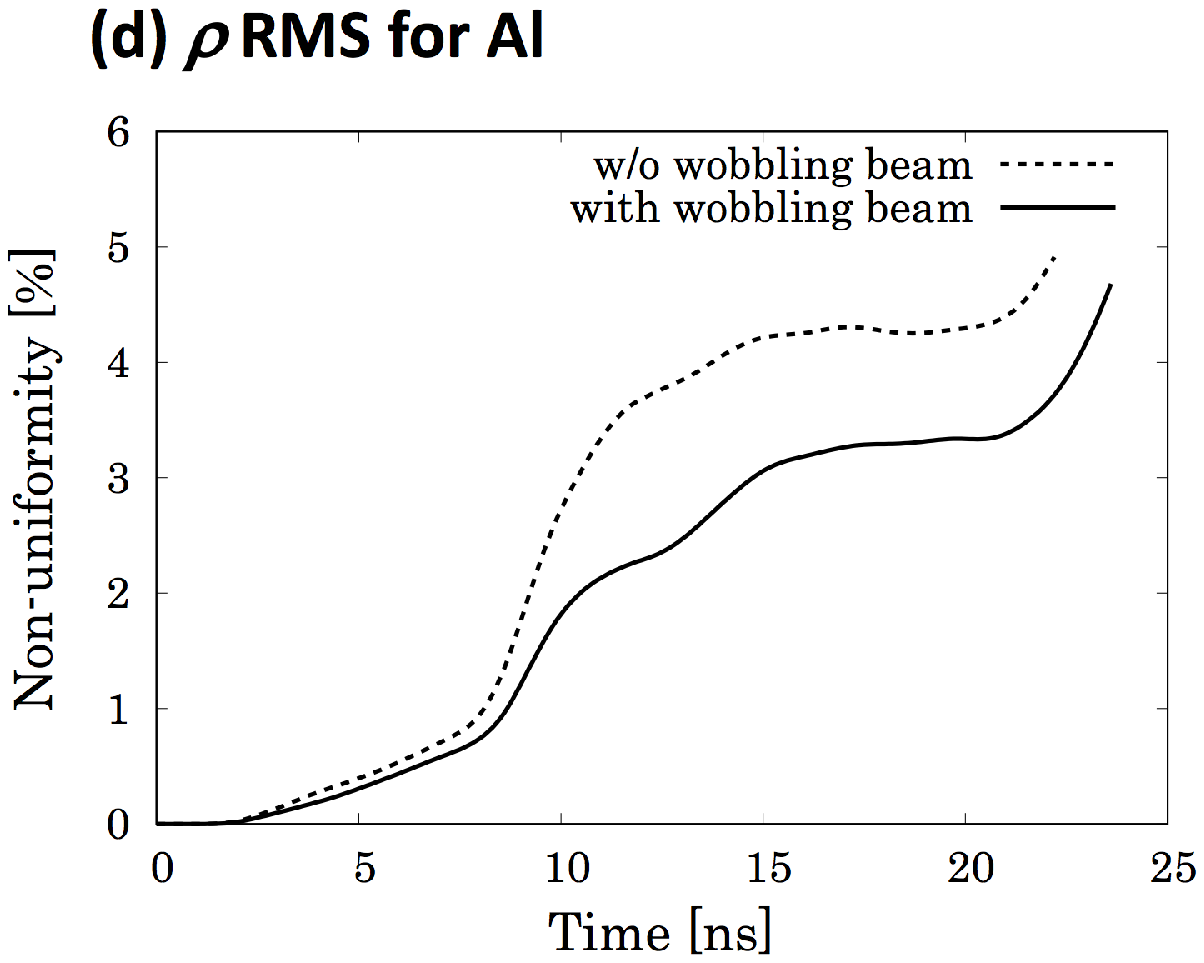}\\
		\caption{RMS non-uniformity histories of (a) the DT ion temperature, (b) the DT mass density, (c) the Al ion temperature and (d) the Al mass density for the cases with the wobbling HIBs (solid lines) and without the wobbling HIBs (dotted lines).}\label{fusion_RMS}
\end{figure}
\begin{figure}[H]
		\centering
		\includegraphics[width=6.5cm]{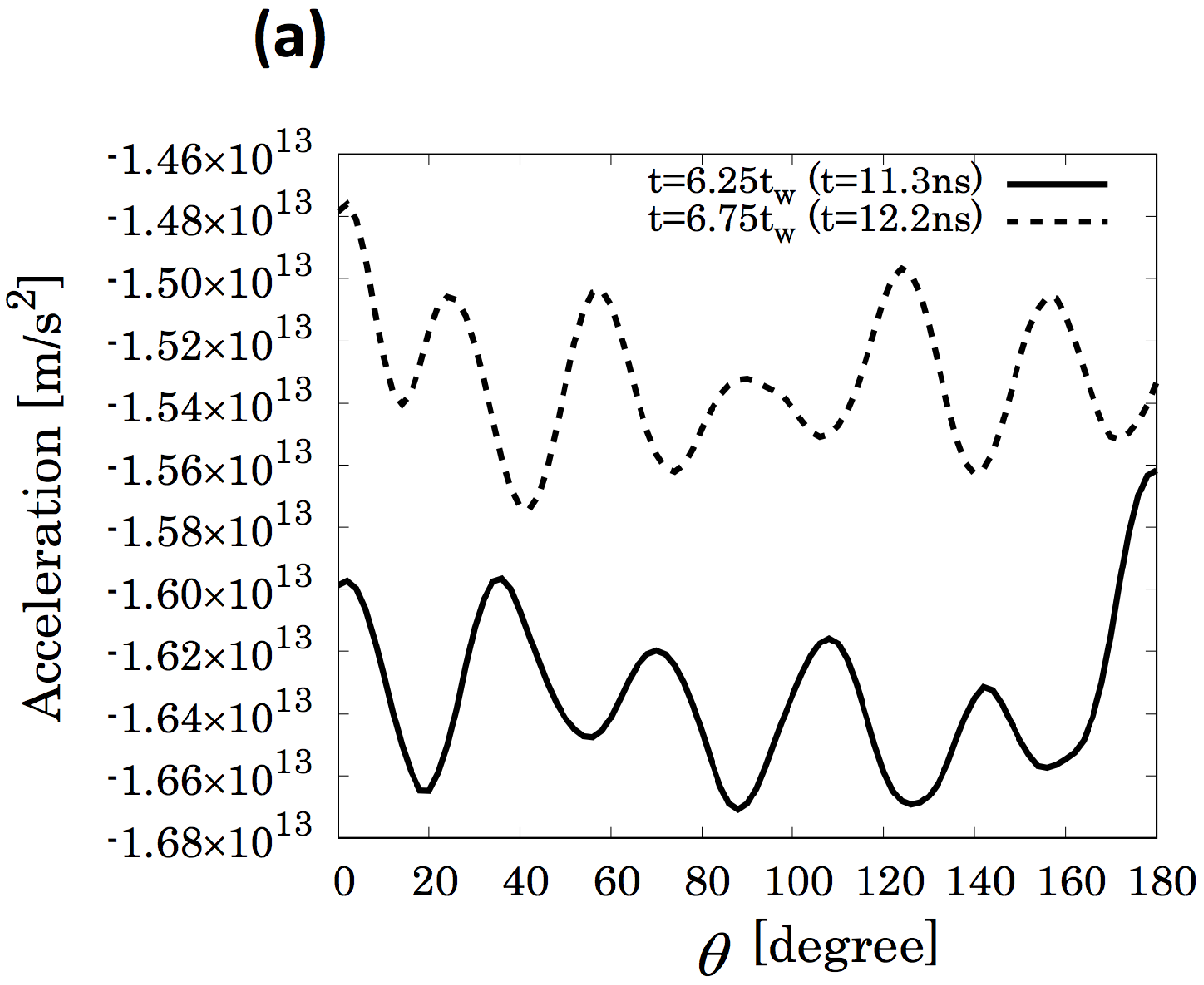}
		\includegraphics[width=6.5cm]{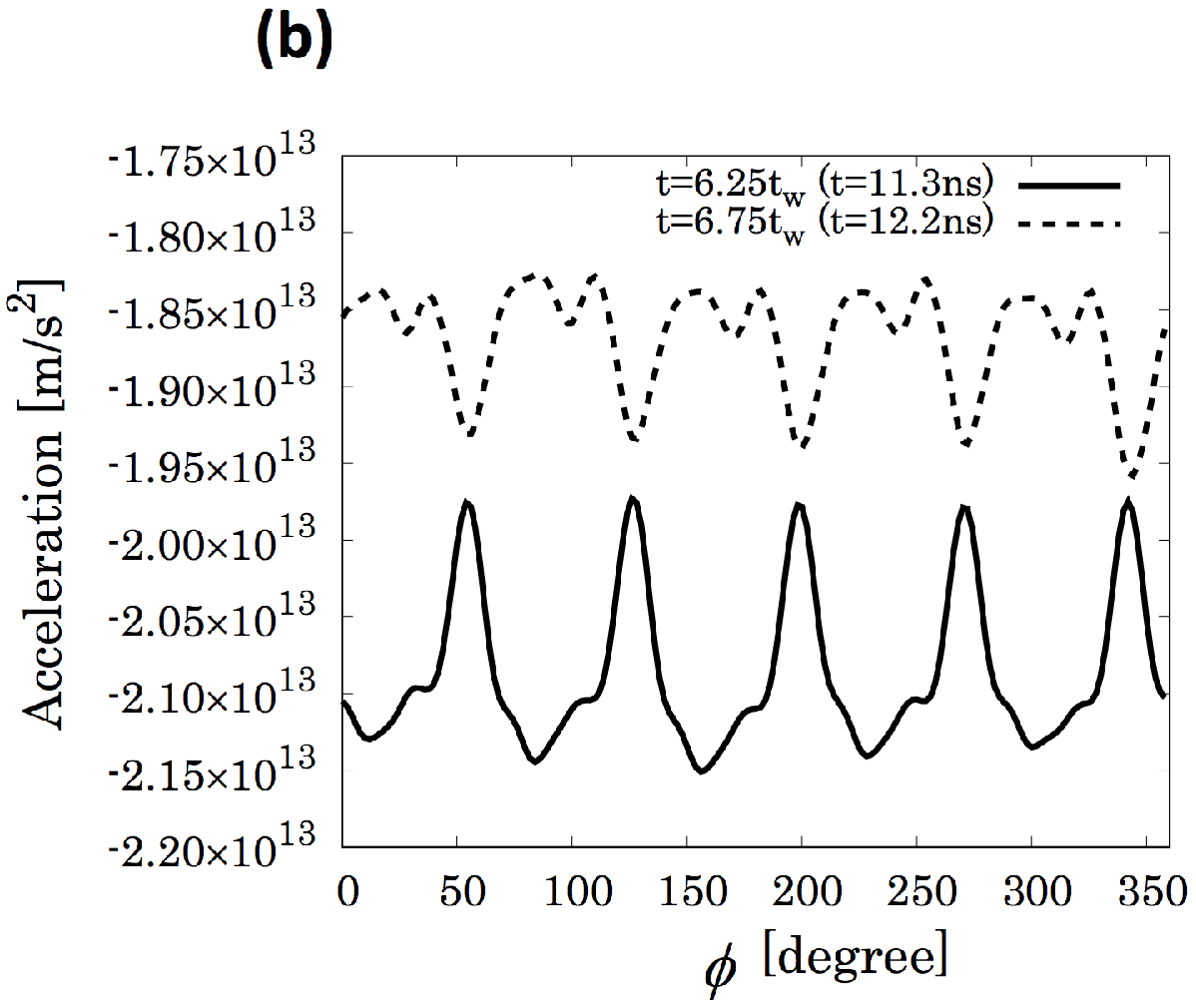}\\
		\caption{Radial acceleration distributions in (a) $\theta$ and (b) $\phi$. The solid lines show the acceleration ditributions at $t=6.25t_w=11.3ns$, and the dotted lines at $t=6.75t_w=12.2ns$.}\label{Vr_tp}
\end{figure}

After the Lagrange code computation, the implosion data are converted and transferred to the Euler code. Figures \ref{Ti_EuWobblIgnited} show the ion temperature distributions by the Euler code. Figures \ref{Ti_EuWobblIgnited} show that the DT fuel is ignited and the gain obtained is about 17.5 in this example case. For a realistic HIF reactor design, the implosion parameters should be further optimized to obtain a sufficient gain, which should be larger than 30$\sim$40 in HIF \cite{CPC-O-SUKI, Kawata1, Kawata2, RSato2}. 

\begin{figure}[H]
		\centering
		\includegraphics[width=13cm]{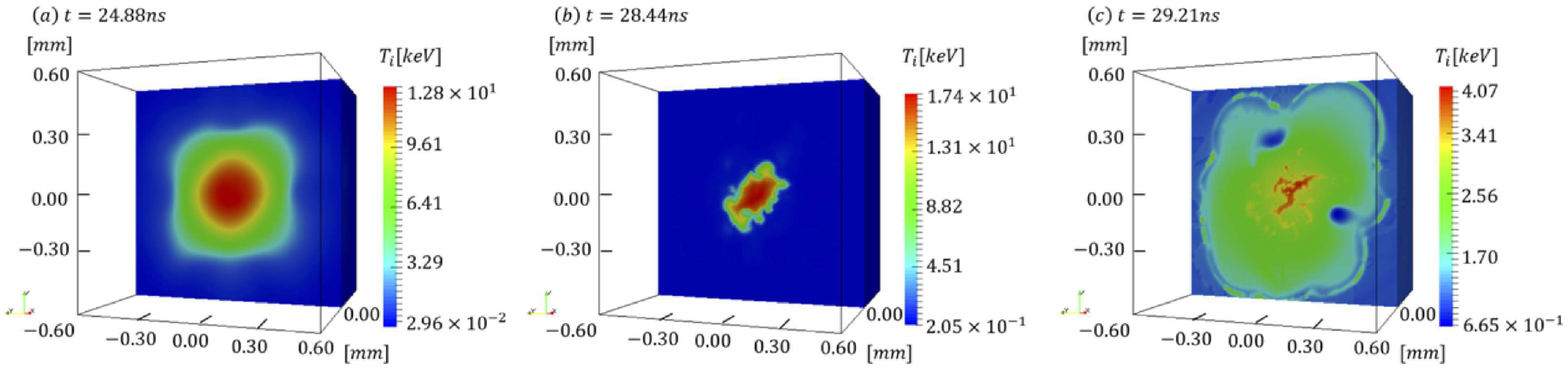}
		\caption{Ion temperature distributions (a) at $t=$24.88ns, (b) at 28.44ns and at 29.21ns.}\label{Ti_EuWobblIgnited}
\end{figure}

In order to check the accuracy of the 3D Euler code, we also performed the Euler code tests, using the initial conditions of the 2D Euler code. The initial conditions in the Euler code are the output of the Lagrangian code.  To this end, the 2D Euler initial conditions were converted into 3D. Therefore, the physical values are uniform in the $\phi$ direction. The Lagrangian test 2D results for the target ion temperature ($T_i$) and the mass density ($\rho$) distribution at $t$ = 29 ns are shown in Figs. 14 and 15 in Ref. \cite{CPC-O-SUKI} for the cases with and without the wobbling HIBs.  The 2D Eulerian test results for the fusion energy gain is shown in Fig. 16 in Ref. \cite{CPC-O-SUKI}.  In Fig. \ref{Ti_Eu_3d} we show the ion temperature distributions by the 3D Euler code. The wobbling HIBs are not used in this simulation. In this case the fuel is ignited at $t \sim $30.1ns. The histories of the fusion gain $G$ of the 2D code and the 3D code are shown in Fig. \ref{FusionGain_Eu}. The fusion gain was 52.5 by the 2D code and 57.6 by the 3D code. In addition, we also did another test for the wobbling HIBs (see Figs. 15 and 16 in Ref. \cite{CPC-O-SUKI}), and the fusion gain was 76.1 in 2D \cite{CPC-O-SUKI} and 67.4 in 3D. The results would confirm that the 3D Euler code reproduces the 2D results successfully for the ignition time and the fusion gain obtained.

\begin{figure}[H]
		\centering
		\includegraphics[width=6.5cm]{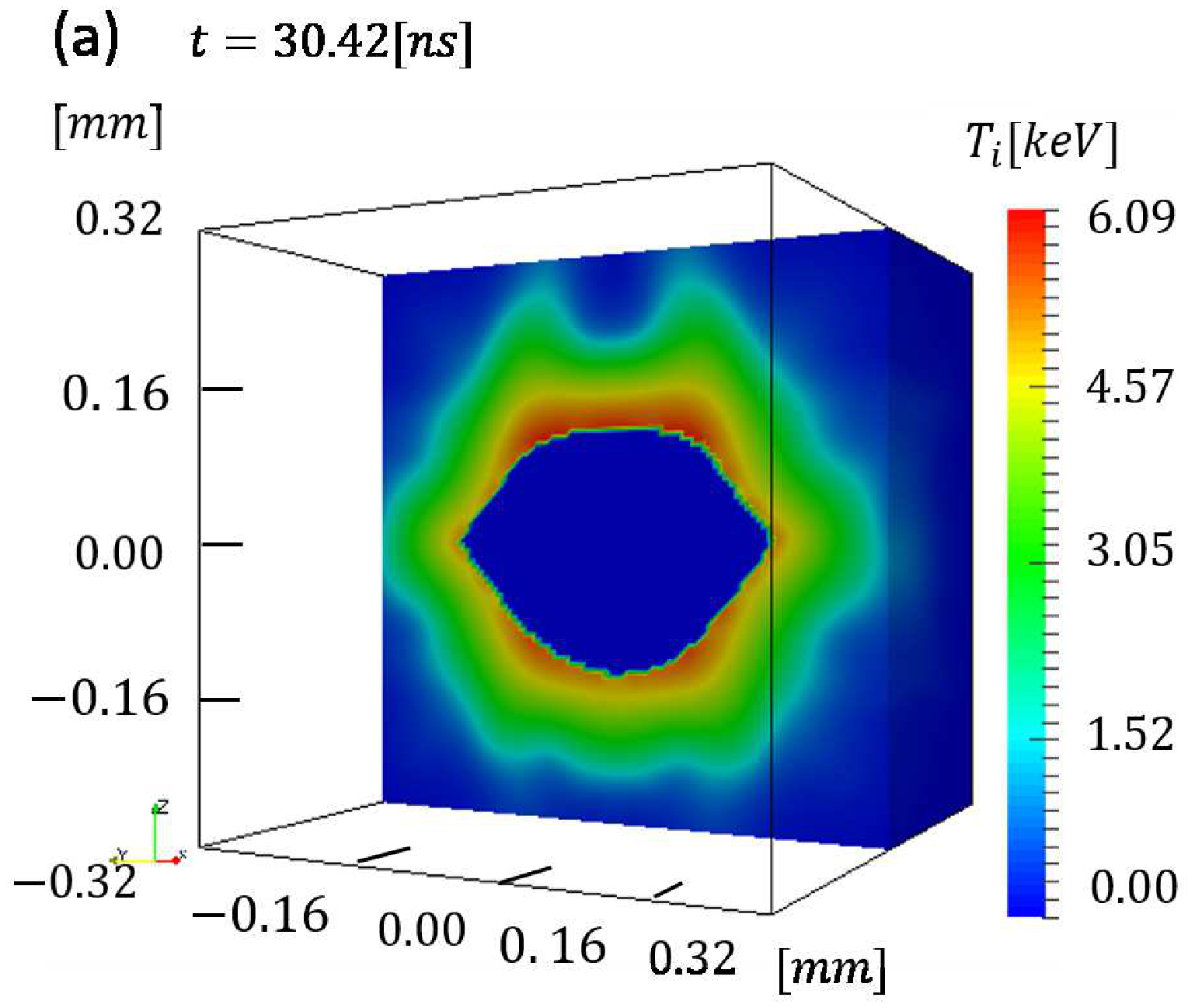}
		\includegraphics[width=6.5cm]{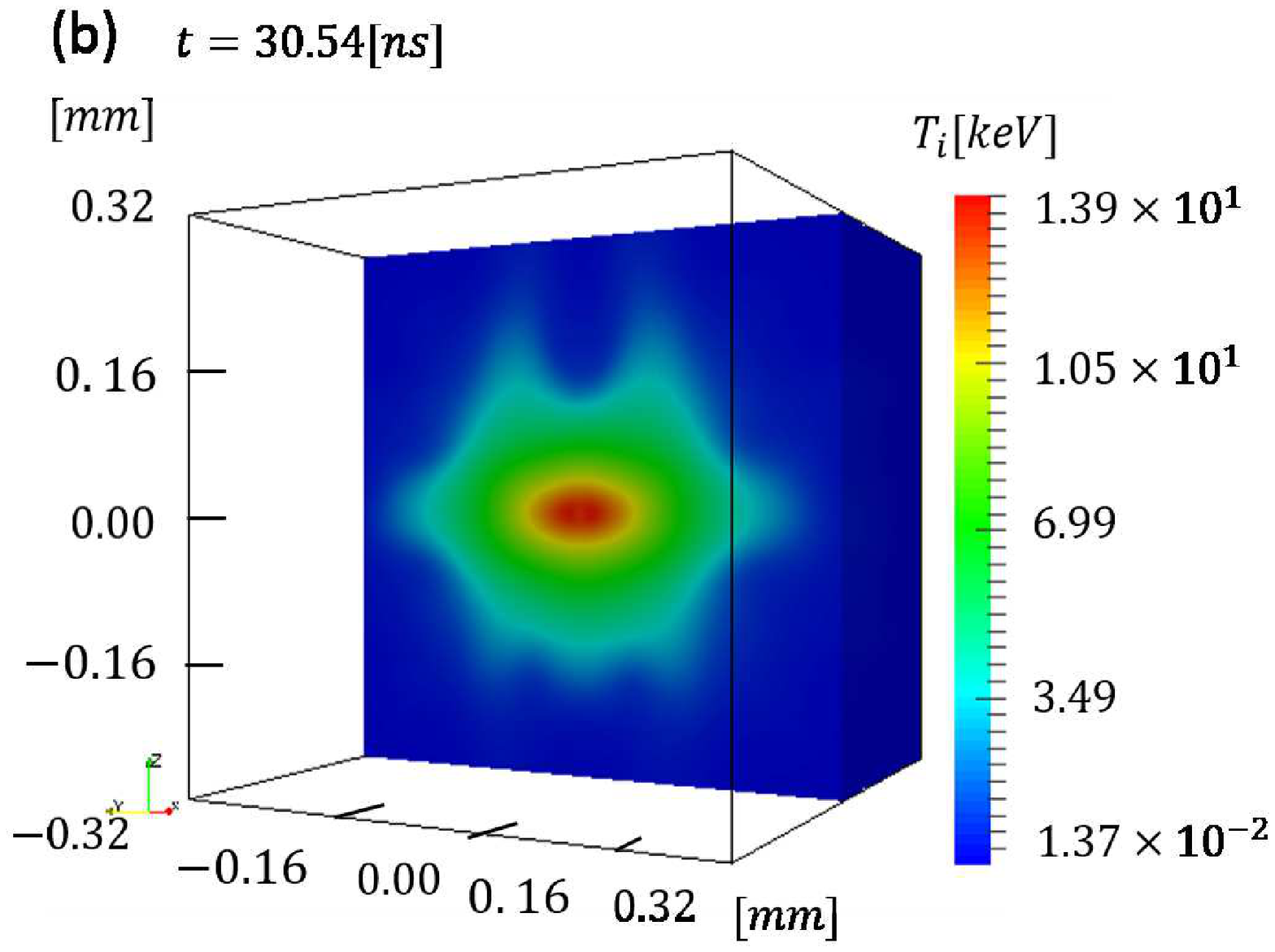} \\
		\includegraphics[width=6.5cm]{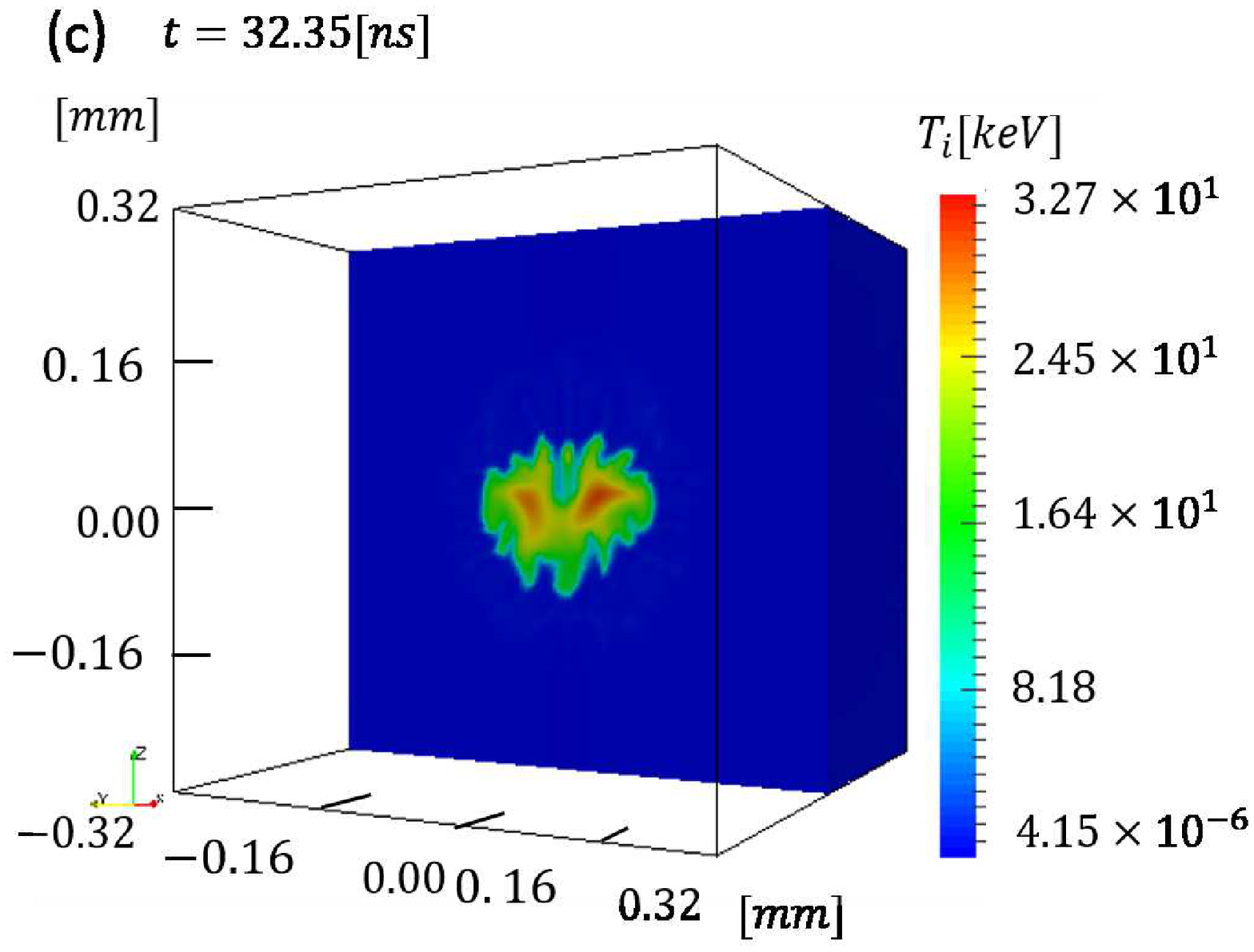}
		\includegraphics[width=6.5cm]{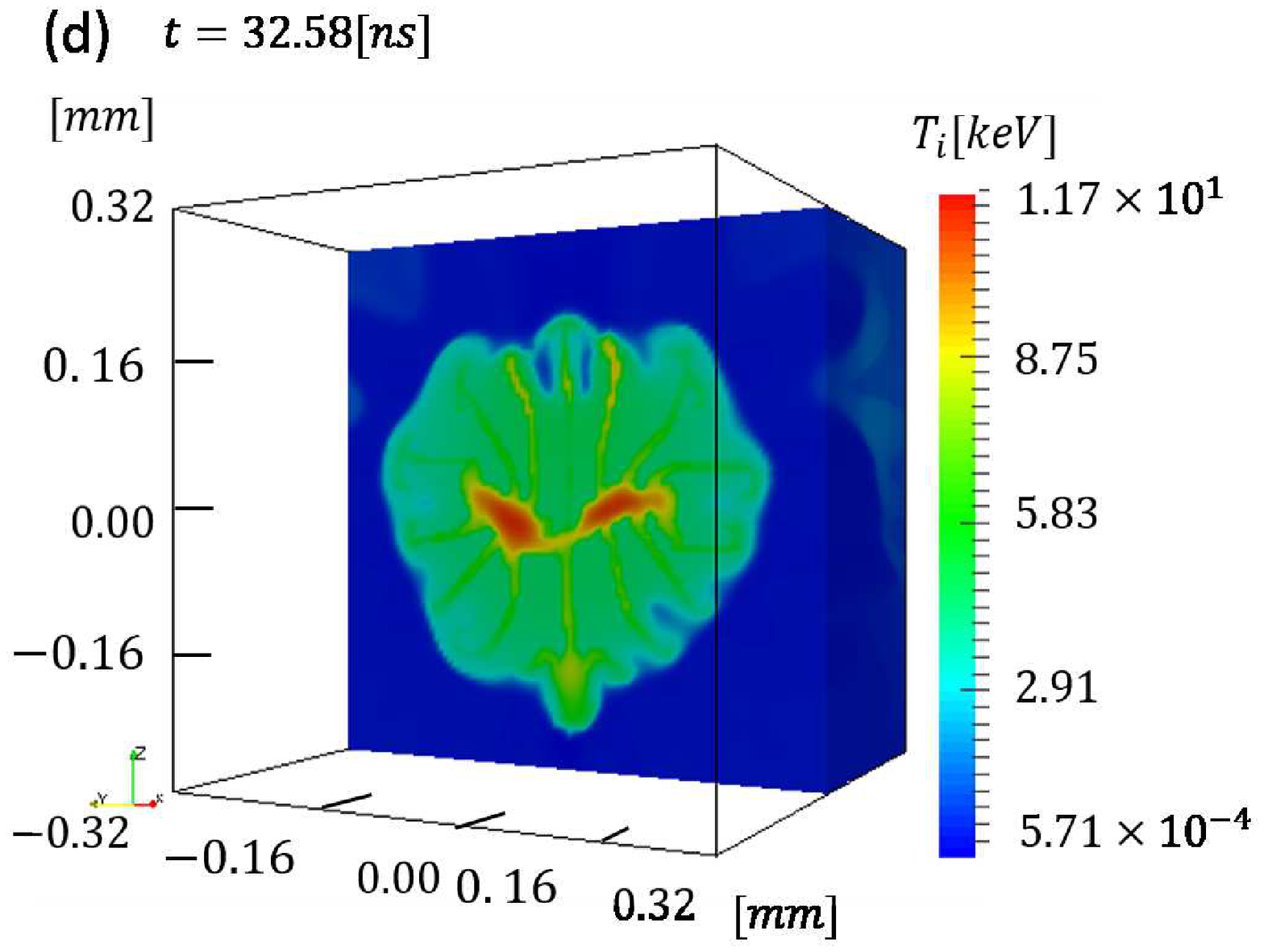} \\
		\caption{Ion temperature distributions by the 3D Euler code without the HIBs wobbling at (a) $t$=30.42ns, (b) 30.53ns, (c) 32.35ns and (d) 32.58ns}\label{Ti_Eu_3d}
\end{figure}

\begin{figure}[H]
		\centering
		\includegraphics[width=11cm]{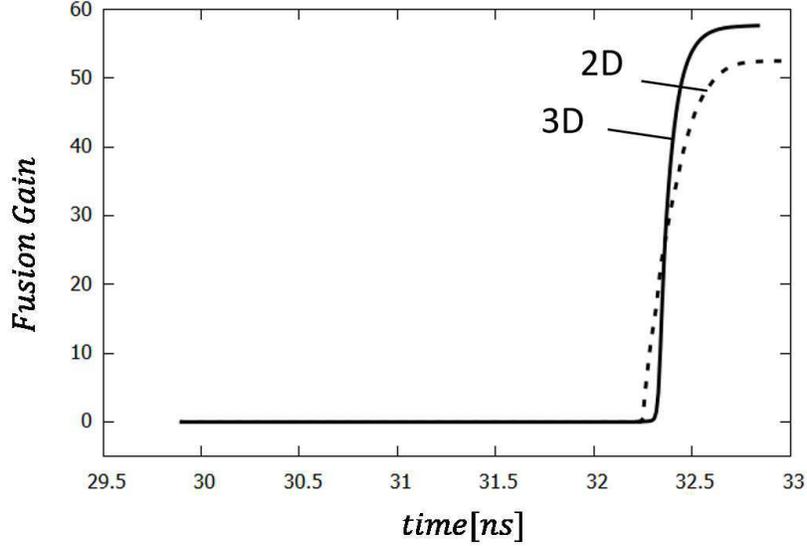}
		\caption{Fusion energy gain curves for the cases with 3D code (a solid line) and with 2D code (a dotted line).}\label{FusionGain_Eu}
\end{figure}

We also simulated the double-cone ignition scheme\cite{Double-cone} using a 3D Euler code. The double-cone ignition scheme was proposed by Prof. Jie Zhang \cite{Double-cone}, and the two compressed DT clouds are created by the gold cones. The two DT spherical clouds collide each other like the impact fusion \cite{Winterberg}. In this example case, the compressed DT maximum density of the DT fuel is set to be $1.0\times 10^5$[kg/m$^3$] with the Gaussian spatial distribution. The DT ignition will be attained by an additional heating, which is not taken into consideration in this example. The ion, electron and radiation temperatures are 10[eV] initially in the Euler code. The radius of the fuel is 92[$\mu$m] and the mass was $0.1$[mg]. We set the colliding speed $w$ of the two DT fuel clouds to $3.0\times10^5$ [m/s]. The ion temperature distributions are shown in Fig. \ref{Double_cone_Ti}.

\begin{figure}[H]
		\centering
		\includegraphics[width=6.5cm]{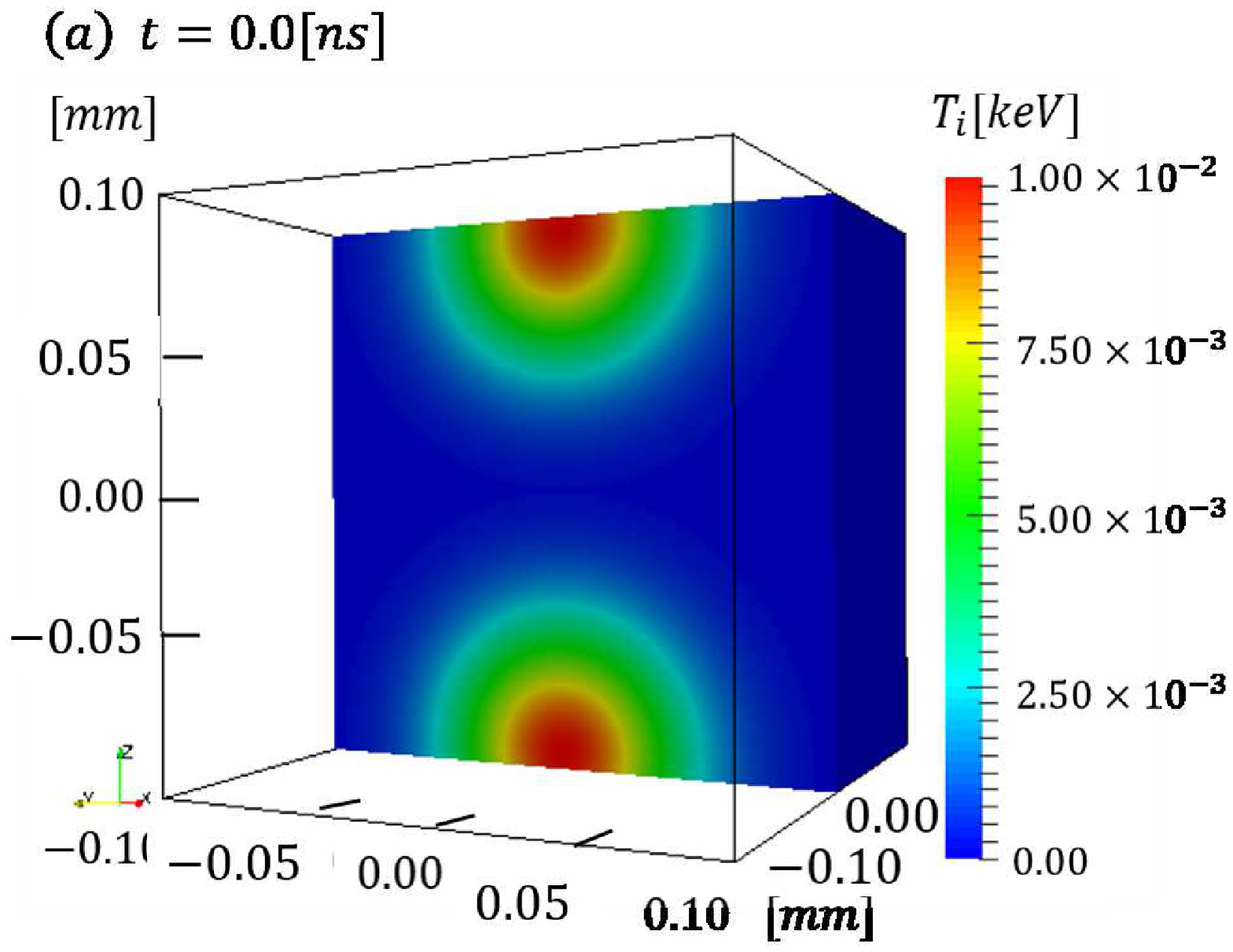}
		\includegraphics[width=6.5cm]{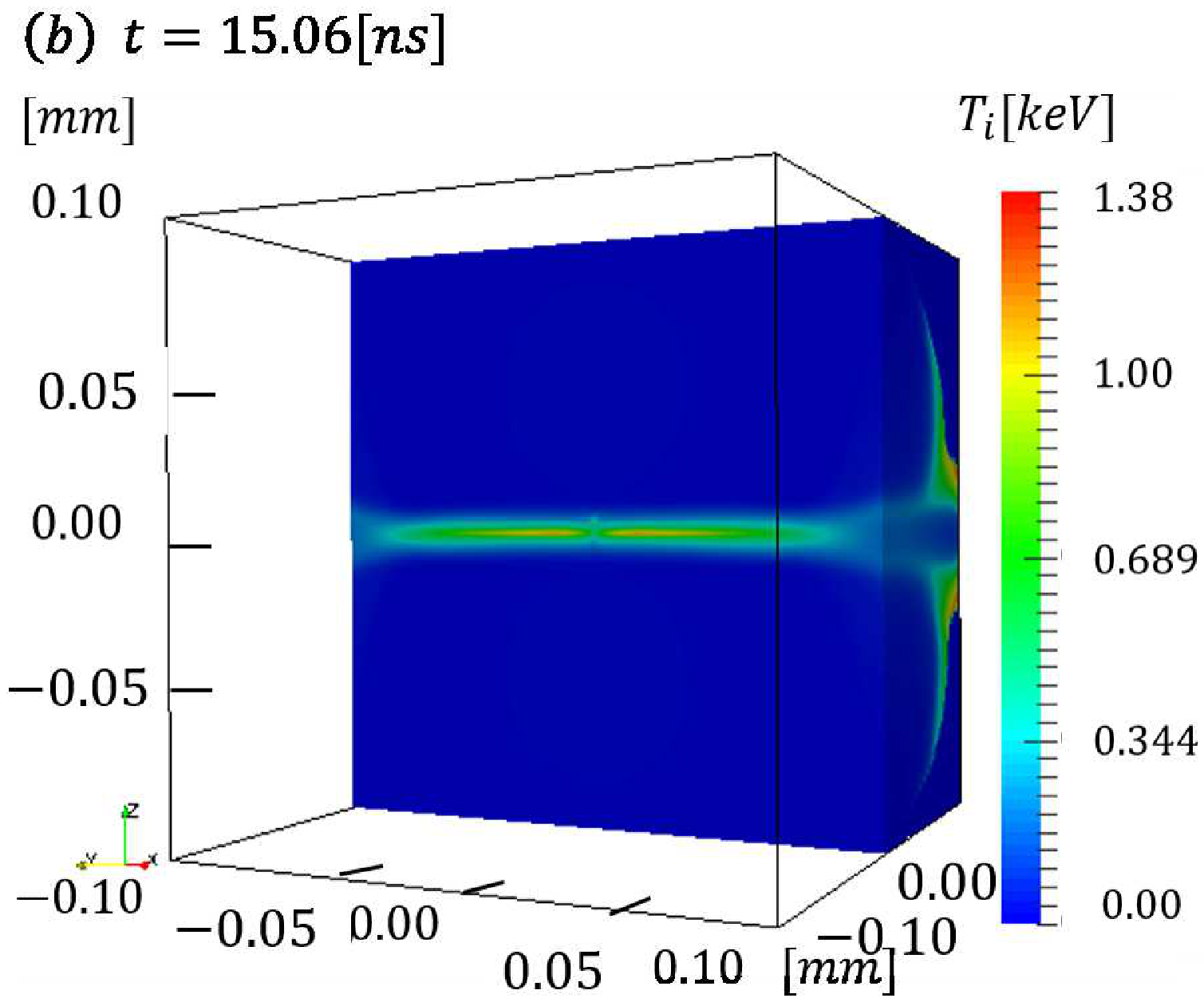} \\
		\includegraphics[width=6.5cm]{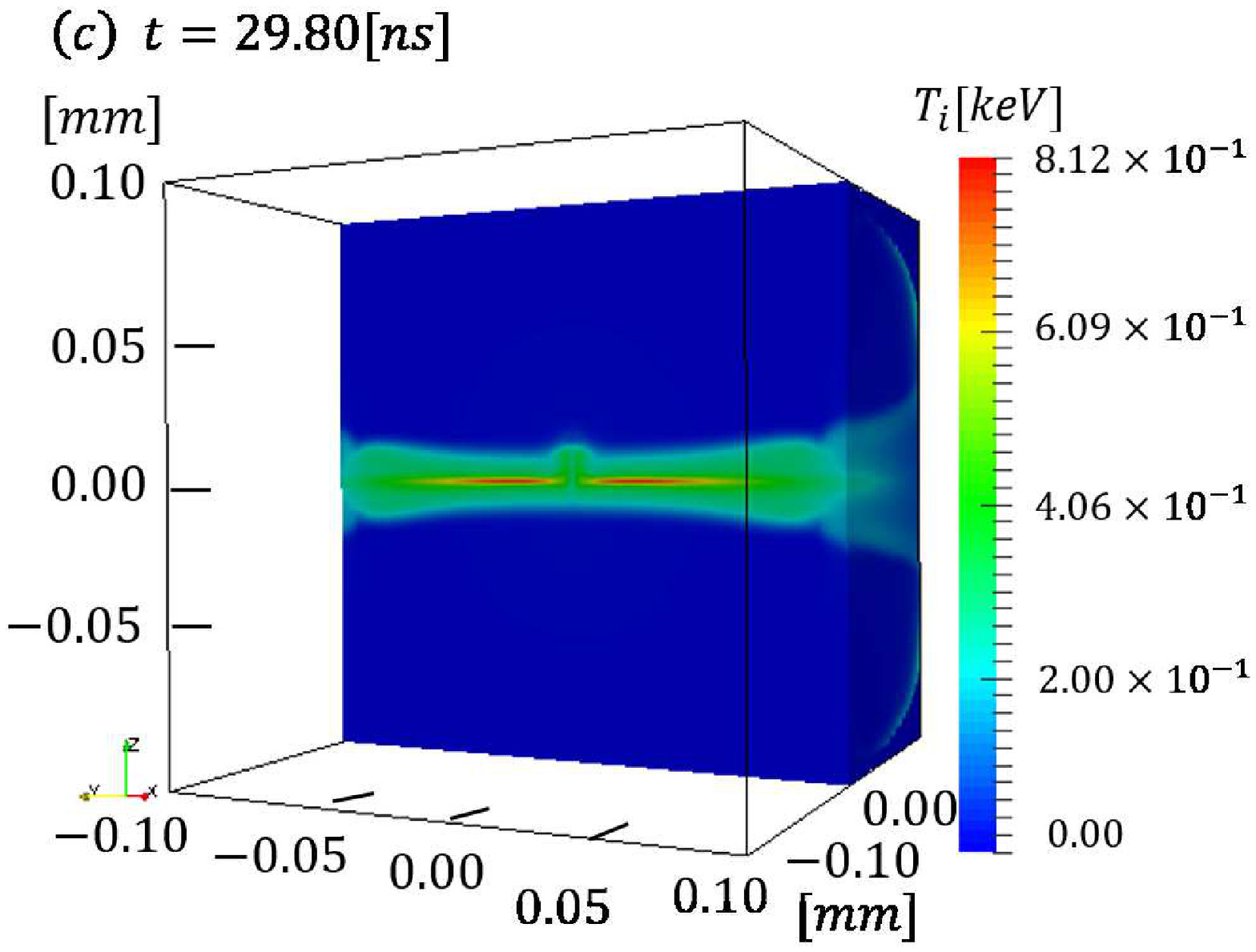}
		\includegraphics[width=6.5cm]{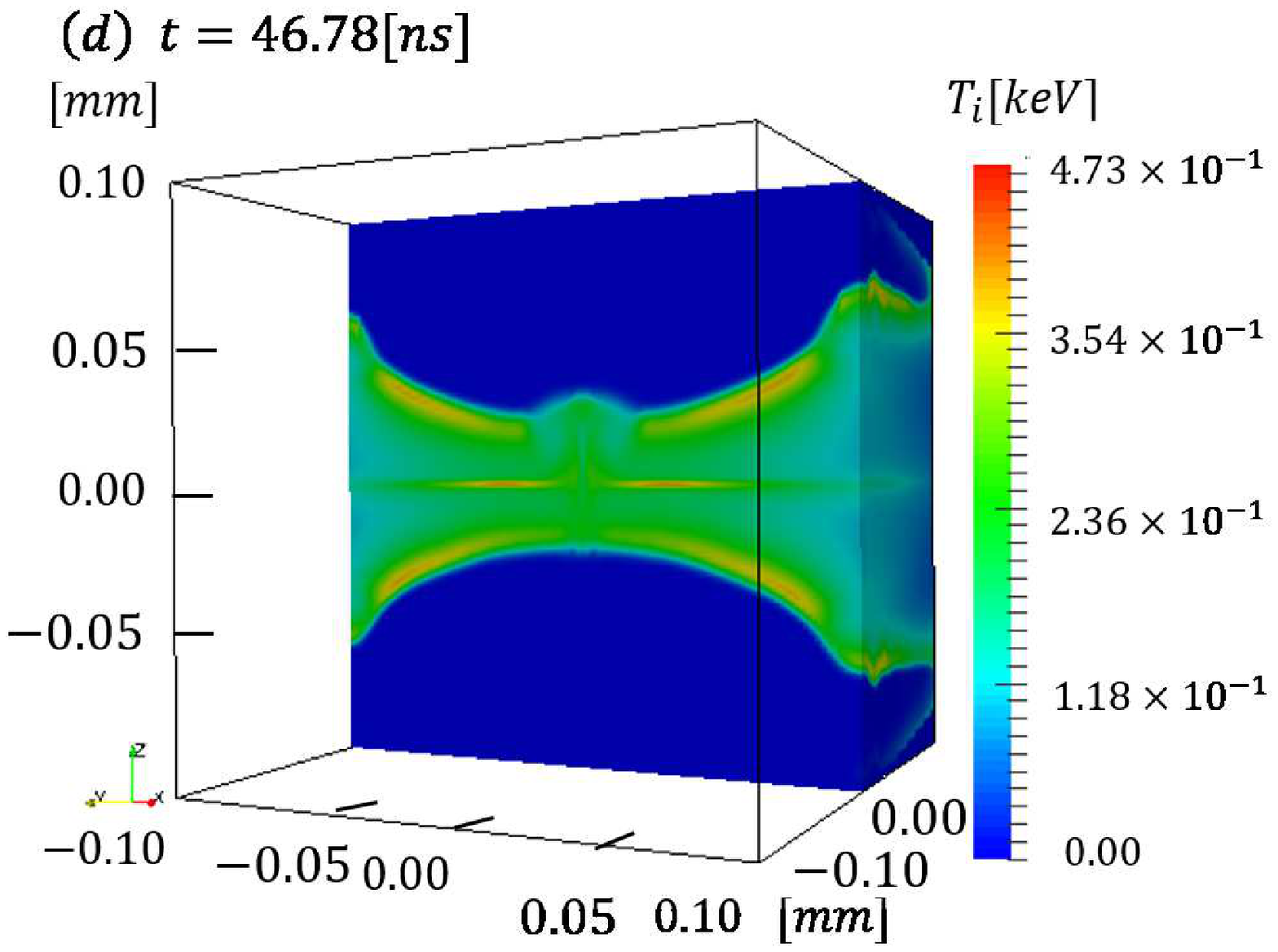} \\
		\caption{Ion temperature distributions for the Double-cone ignition scheme \cite{Double-cone} at (a) $t$=0.0ns, (b) 15.06ns, (c) 29.80ns and (d) 46.78ns.}\label{Double_cone_Ti}
\end{figure}

\section{Conclusions}
We have developed and presented the O-SUKI-N 3D code, which is useful to simulate 3D spherical DT fuel target implosion in HIF. The O-SUKI-N code is an upgraded implosion simulation system from the 2D O-SUKI code, and consists of the Lagrangian fluid code, the data conversion from the Lagrangian code data to the Euler code data, and the Euler code. Near the void closure phase of the DT fuel implosion, the DT fuel spatial deformation is serious. At the stagnation phase the DT fuel is compressed to about a thousand times of the solid density. The O-SUKI-N 3D code would provide a useful tool for the integrated DT fuel target implosion simulation in HIF.

\section*{Declaration of Competing Interest}
     The authors declare that they have no known competing financial interests or personal relationships that could have appeared to influence the work reported in this paper. 
     
\section*{CRediT author statement}
Hiroki Nakamura: Software for Euler code and Conversion code, Validation, Visualization; Ken Uchibori: Software for Lagrange code and Conversion code, Validation, Visualization, Writing draft; Shigeo Kawata: Basic idea, Conceptualization, Methodology, Investigation, Supervision, writing paper; Takahiro Karino: Methodology, Supervision; Ryo Sato: Methodology, Validation; Alexander I. Ogoyski: Software for OK3 code, Validation. 

\section*{Acknowledgments}
	The work was partly supported by JSPS, Japan-U. S.  Exchange Program, MEXT, CORE (Center for Optical Research and Education, Utsunomiya University), Shanghai Jiao Tong University and ILE/Osaka University. The work was also partly done under the collaborations with Xi'an Jia Tong University, Inst. of Modern Physics, Lanzhou, Inst. of Physics, Beijing, Fudan university, Shanghai, Renmin University of China, Beijing, and ELI-Beamlines, Prague. 

\bibliographystyle{elsarticle-num}
\bibliography{<your-bib-database>}

\begin{thebibliography}{99}

\bibitem{CPC-O-SUKI}
R. Sato, S. Kawata, T. Karino, K. Uchibori, T. Iinuma, H. Katoh and A.I. Ogoyski, Comput. Phys. Commun. 240 (2019) 83-100. 
\bibitem{ICFBook}
S. Atzeni and J. Meyer-ter-Vehn, The Physics of Inertial Fusion, Oxford University Press, 2009. 

\bibitem {kwtANDniu}
 S. Kawata and K. Niu, J. Phys. Soc. Jpn. 53 (1984) 3416-3426. 
 
 
 \bibitem{Kawata1}
S. Kawata, T. Karino and A. I. Ogoyski, Matter and Radiation at Extremes 1(2) (2016)  89-113. 

 \bibitem{Kawata2}
S. Kawata, Advances in Physics x 6(1) (2021)  1873860. 


 
\bibitem{ogoyski1}
A. I. Ogoyski, T. Someya and S. Kawata, Comput. Phys. Commun. 157 (2004) 160-172.
 \bibitem{ogoyski2}
A. I. Ogoyski, S. Kawata and T. Someya, Compt. Phys. Commun. 161 (2004) 143-150.
 \bibitem{ogoyski3}
A. I. Ogoyski, S. Kawata, P. H. Popov, Compt. Phys. Commun. 181 (2010) 1332-1333. 

\bibitem{Schulz}
W. D. Schulz, ”Two-Dimensional Lagrangian Hydrodynamic Difference Equations”, University of California Lawrence Radiation Laboratory Livermore, California, UCRL-6776, 1963. 

\bibitem{Tahir}
N. A. Tahir, K. A. Long, E. W. Laing, J. Appl. Phys. 60 (1986) 898. 

\bibitem{Zeldovich}
Ya. B. Zel'dovich, Yu. P. Raizer, Physics of Shock Waves and High-Temperature Hydrodynamic Phenomena, Dover Books on Physics, New York, 2002. 

\bibitem {mehlhorn}
T.A. Mehlhorn, J. Appl. Phys. 52 (1981) 6522-6532. 

\bibitem{artv}
 J. Von Neumann and R. D. Richtmyer, J. Appl. Phys. 21 (1950) 232-237. 
 
 \bibitem{Christiansen}
J. P. Christianen, D. E. T. F. Ashby, and K. V. Roberts, Computer Physics Communications 7 (1974) 271-287. 
 
\bibitem{Bell}
A. R. Bell, Rutherford Laboratory Report, RL-80-091, 1981. 

\bibitem{NRLpf}
A. S. Richardson, 2019 NRL Plasma Formulary, (2019). 

\bibitem{Fraley}
G. S. Fraley, E. J. Linnebur, R. J. Mason, R. L. Morse, Phys. Fluids, 17 (1974) 474-489. 

\bibitem{RSato2}
R. Sato, S. Kawata, T. Karino, K. Uchibori and A. I. Ogoysk, Scientific Reports, 6659 (2019) https://doi.org/10.1038/s41598-019-43221-7. 

\bibitem{Double-cone}
J. Zhang, W. M. Wang, X. H. Yang, D. Wu, Y. Y. Ma, J. L. Jiao, Z. Zhang, F. Y. Wu, X. H. Yuan, Y. T. Li and J. Q. Zhu, Rhilosophical Tran. Royal Soc. A (2020) https://doi.org/10.1098/rsta.2020.0015. 

\bibitem{Winterberg}
F. Winterberg, Z. Naturforschg. 19a (1964) 231-239. 

 \end{thebibliography}



\end{document}